\documentclass[12pt, a4paper, twoside]{book}

\usepackage{amssymb,amsmath,amsthm,verbatim,times}
\usepackage{textcomp,cancel,subfig,anysize,array,rotating,geometry}
\usepackage{feynmp}
\usepackage{slashed}
\usepackage{color}

\usepackage{graphicx}
\usepackage{dcolumn}
\usepackage{float}
\usepackage{epstopdf}

\usepackage{ifpdf}

\ifpdf
            \usepackage[breaklinks=true]{hyperref}
\else
						\usepackage{graphicx}            
            \usepackage[dvips,a4paper,linktocpage]{hyperref}
      \hypersetup{
          pdfauthor = {Aaron Farrell},
          pdftitle = {Topological superconductivity without proximity effect},
          pdfsubject = {M.Sc. Thesis}
      }
\fi

\usepackage[toc,page]{appendix}
\usepackage{dsfont,bbold}

\usepackage[numbers,sort&compress]{natbib}

\usepackage{setspace}
\usepackage{fancyhdr}
\numberwithin{equation}{section}

\renewcommand{\eqref}[1]{Equation \ref{#1}}

\usepackage{sectsty}

\sectionfont{\bfseries\Large\centering}
\subsectionfont{\bfseries\selectfont}

\makeatletter
\newcommand*\std@chapter{}
\let\std@chapter=\@chapter
\renewcommand*\@chapter[2][]{\std@chapter[#2]{#2}\chaptermark{#1}}
\makeatother

\newcommand{\centeredrotatedfigure}[3]{
		\begin{figure}[ht]%
		\begin{center}
		\resizebox{\columnwidth}{!}{\includegraphics[angle=-90]{#1}}
		\end{center}
		\caption{#3} %
		\label{#2}%
}

\newcommand{\nonnumsection}[2]{ \section*{#1}
\thispagestyle{fancyplain}
}

\makeatletter
\def\thickhrulefill{\leavevmode \leaders \hrule height 1ex \hfill \kern \z@}
\def\@makechapterhead#1{%
  \vspace*{10\p@}%
  {\parindent \z@ \centering \reset@font
        {\Huge \scshape \thechapter}
        \par\nobreak
        \vspace*{10\p@}%
        \interlinepenalty\@M
        \thickhrulefill
        \par\nobreak
        \vspace*{2\p@}%
        {\Huge \bfseries #1\par\nobreak}
        \par\nobreak
        \vspace*{2\p@}%
        \thickhrulefill
    \vskip 60\p@
  }}
\def\@makeschapterhead#1{%
  \vspace*{10\p@}%
  {\parindent \z@ \centering \reset@font
        {\Huge \scshape \vphantom{\thechapter}}
        \par\nobreak
        \vspace*{10\p@}%
        \interlinepenalty\@M
        \thickhrulefill
        \par\nobreak
        \vspace*{2\p@}%
        {\Huge \bfseries #1\par\nobreak}
        \par\nobreak
        \vspace*{2\p@}%
        \thickhrulefill
    \vskip 60\p@
  }}

\def\kv{{\bf k}}

\def\beq{\begin{equation}}
\def\eeq{\end{equation}}
\def\rv{{\bf r}}
\def\Qv{{\bf Q}}
\def\Sv{{\bf S}}
\def\Pf{\text{Pf}}

\graphicspath{{./Figures/}}

\begin{document}


\newcommand{\HRule}{\rule{\linewidth}{0.5mm}}

\begin{titlepage}

\begin{center}

\textsc{\LARGE McGill University}\\[1.5cm]

\HRule \\[0.4cm]
{ \huge \bfseries Topological superconductivity without proximity effect}\\[0.4cm]

\HRule \\[1.5cm]

{ \large Aaron Farrell\\ Department of Physics \\ April 2013 \\ McGill University, Montreal }

\vfill

{ A thesis submitted to McGill University in partial fulfillment of the requirements of the degree of Master of Science}\\[0.5cm]

{ \copyright Aaron Farrell, 2013}\\[0.5cm]

\end{center}

\end{titlepage}

\newpage

\pagestyle{empty}

\newpage

\pagenumbering{roman}
\singlespacing
\frontmatter


\pagestyle{fancy}
\nonnumsection{Abstract}

\begin{onehalfspace}
The search for a Majorana Fermion has been an area of intense interest in condensed matter research of late. This elusive particle, predicted to exist in 1937, has been sought after for both fundamental and practical reasons. On the fundamental level, no particle to date has been observed to be a Majorana fermion, meanwhile on the practical level a Majorana fermion, if found, would represent a non-abelian anyon and could thus be used to build a quantum computer. The search for a Majorana Fermion has recently shifted to topological superconductivity. Topological superconductors are categorized by the nontrivial winding of their order parameter phase and for this reason are expected to support Majorana Fermions in their vortex cores. Owing to this, the study of topological superconductors has intensified in recent years. Current proposals for a device that may behave as a topological superconductor are based on semiconductor heterostructures, where the spin-orbit coupled bands of a semiconductor are split by a band gap or Zeeman field and superconductivity is induced by proximity to a conventional superconductor.  In this setup, topological superconductivity is obtained in the semiconductor layer and the proposed heterostructures typically include two or three layers of different materials. In this thesis we propose a simplification to these types of devices, suggesting a way in which the superconducting layer can be replaced. Part of our proposal includes a model Hamiltonian for these types of systems. This thesis will also develop several different methods to analyze this model Hamiltonian in various different parameter regimes with the ultimate goal of classifying its topology.  


\newpage
\nonnumsection{R\'esum\'e}

R\'ecemment, une r\'egion d'int\'er\^et en la recherch\'e de la mati\`ere condens\'ee est le recherche pour les ``Majorana Fermions". Les physiciens sont fascin\'es avec cette particule pour des raisons fondamentales et pratiques. Fondamentalement, une particule se comporte comme un Majorana Fermion n'a jamais \'et\'e trouv\'ee avant. Pratiquement, un Majorana Fermion pourrait \^etre utilis\'e pour la construction d'un ordinateur quantique. Dans les derni\`eres ann\'ees, les chercheurs ont commenc\'e \`a chercher pour des Majorana Fermions dans les supraconducteurs. En particulier, les supraconducteurs topologiques sont crus de supportes les Majorana Fermions dans leur vortex cores et de ce fait des nombreux dispositifs supraconducteurs topologiques ont \'et\'e propos\'ees. Les propositions r\'ecemment sont bas\'ees sur les h\'et\'erostructures de trois ou deux couches. Dans ces h\'et\'erostructures, les bandes dÕun semiconducteur avec le couplage de spin-orbit sont s\'epar\'ees par le champ Zeeman d'une couche ferromagn\'etique (ou un champ appliqu\'e).  Apr\`es cette, supraconductivit\'e topologique est \'etablie dans la couche de semiconductrice en raison de la proximit\'e d'une couche de supraconducteur ordinaire. Dans cette th\`ese nous proposons une simplification des dispositifs d\'ecrits ci-dessus; nous sugg\'erons un moyen d'enlever la couche de supraconductivit\'e. Nous commen\c{c}ons par proposer un Hamiltonian du cette syst\`eme et proc\`ede \`a d\'evelopper des nombreuses m\'ethodes pour analyser cette Hamiltonian avec l'objectif ultime de classifier la topologie de ce syst\`eme.    

\newpage
\nonnumsection{Acknowledgments}

I would like to thank my colleagues, friends, and family for their continued support throughout my studies. Additionally, I would like to acknowledge my supervisor, Dr. Tami Pereg-Barnea, for all of her help and guidance during the completion of this research. The work in this thesis has been funded by the McGill Tomlinson Masters fellowship, the Natural Sciences and Engineering Research Council of Canada and the McGill Schulich Fellowship.   


\end{onehalfspace}

\setcounter{tocdepth}{1}
\tableofcontents

\listoffigures


\mainmatter

\begin{doublespace}



\chapter{Introduction}

\section{Majorana Fermions in Condensed Matter Systems}
A very celebrated result in the early development of quantum theory came in 1928 when Paul Dirac formulated the famous Dirac equation. Complex valued solutions to Dirac's equation provided a description of relativistic, spin-$\frac{1}{2}$ particles such as electrons and protons. One notable (and at the time profound) property of these complex solutions is that they provide a description of particles that have unique anti-particles. 

In 1937 Italian physicist Ettore Majorana\cite{Majorana} proposed a modification to the Dirac equation which would lead to it having purely real solutions. These real valued solutions, since termed Majorana fermions, were truly novel as they described spin-$\frac{1}{2}$ particles ({\it i.e.} fermions) that are their own anti-particles. With a theoretical framework for these curious particles in place, many experimentalists became interested in their possible existence and the search for Majorana fermions began. Efforts were initially focused on finding a fundamental particle behaving as a Majorana fermion, however all fundamental fermions discovered to date have had unique antiparticles. That being said, the verdict is still out on the nature of the neutrino.   

 This ``search" for Majorana fermions has recently shifted to condensed matter physics. Condensed matter is believed to be promising territory for the realization of a Majorana fermion because the excitations of interest in condensed matter systems are not limited to fundamental particles. In condensed matter we are instead interested in quasiparticles; emergent degrees of freedom that arise when a complicated system of many interacting constituents (for example a solid) behaves {\it as if} it is populated by different, weakly interacting particles. The physical properties of these quasiparticles ({\it e.g.} mass, charge and dispersion relation) are determined not only by the medium in which they move but also by the interactions of the fundamental particles in the system. A common example of a quasiparticle occurs in the mathematical description of a semiconductor. In a semiconductor the quasiparticles of interest have the same charge as an electron and a mass determined by the lattice of nuclei and electron-electron interactions present in the material. More exotic examples of quasiparticles include spinons and composite fermions. It is hoped that in some condensed matter systems the conditions are just right for the constituent quasiparticles to behave as Majorana fermions. The next section of this introduction will focus on a condensed matter system where the conditions may be sufficient to see a Majorana fermion.
 
 Before moving on, the reader should note that finding a Majorana fermion is not just of interest because of the fundamental considerations outlined above; there are also important practical applications for this elusive particle. The reason for this is that Majorana fermions are predicted behave as a non-abelian anyons. To motivate why non-abelian anyons are of interest let us first recall some fundamental exchange symmetries of indistinguishable quantum particles. Three dimensional quantum particles must be either bosons of fermions. The exchange of two bosons (fermions) modifies the many-body wave function by a plus (minus) sign. For example, in a two-body system we have $\psi(\rv_1, \rv_2) = \pm \psi(\rv_2, \rv_1)$ with the upper (lower) sign for bosons (fermions) and $\psi$ the two-body, real space wave function. When we make the transition to two dimensions these exchange statistics become more complicated and may be classified as abelian or non-abelian. It is possible for the wave function to pick up a phase $\alpha \ne 0, \pi$ such that under exchange $\psi(\rv_1, \rv_2) = e^{i\alpha} \psi(\rv_2, \rv_1)$, we call such particles {\em abelian} anyons. Instead of the simple phase $e^{i\alpha}$ characteristic of  abelian anyons, particle exchange of non-abelian anyons is represented by matrices acting on the many-body wave function. Further, these matrices do not commute in general and so the net effect of exchanging several non-abelian anyons on the many-body state depends on the order in which this exchange is performed. In this way one can use exchange of non-abelian anyons to store information in a many-body system. This novel property means that non-abelian anyons ({\it e.g.} Majorana fermions) may be used to build a quantum computer\cite{Nayak}. This important practical application gives additional impetus to the search for a Majorana fermion.  
 
 \section{Topological Superconductors}
 
Lately, the most promising route for realizing Majorana fermions in condensed matter systems has been focused on superconductors. In superconductors Gauge symmetry is spontaneously broken and as a result particle number is not conserved. The fundamental excitations of interest, so called Bogoliubons, are then a superposition of particles and holes. The role of antiparticles in condensed matter systems is played by holes and so another way of looking at these Bogoliubons is as a superposition of electrons and (very roughly speaking) ``antielectrons". One can then imagine that if we are careful in creating this superposition we could have elementary excitations that are an {\em equal} superposition of particles and holes/antiparticles. This equal superposition of particles and holes would then describe a particle that is its own antiparticle or, in other words, a Majorana fermion.   

Illustrating this point is easiest in the language of second quantization\cite{franz}. Consider a many-body quantum system with single particle fermionic states labeled by some index $\alpha$ (this may stand for spin, position, momentum {\it etc.}). We can then define the operator $c_{\alpha}^\dagger$ ($c_{\alpha}$) which creates (annihilates) an electron in (from) quantum state $\alpha$. Alternatively, one may think of $c_\alpha$ as creating an anti-fermion. These operators obey the canonical anticommutation relations $\{c_\alpha, c_{\alpha'}\}= \{c_\alpha^\dagger, c_{\alpha'}^\dagger\}=0$, $\{c_\alpha, c_{\alpha'}^\dagger\}=\delta_{\alpha,\alpha'}$ which require $c_\alpha \ne c_{\alpha}^\dagger$. Imagine now that for some problem of interest it becomes more convenient to describe our system in terms of new particles created (destroyed) by $\gamma^\dagger_{\alpha}$ ($\gamma_{\alpha}$). When we write these new particles in terms of the old ones, it is possible to create two equal superpositions of a particle and a hole. These two possible superpositions are $\gamma_{\alpha} \propto c_{\alpha} + c_{\alpha}^\dagger$ and $\gamma_{\alpha} \propto i(c_{\alpha} - c_{\alpha}^\dagger)$. For these special combinations we then have $\gamma_{\alpha} = \gamma_{\alpha}^\dagger$, and the $\gamma$ particles are their own antiparticles.

Not surprisingly such a specific superposition of particles and holes requires a very special type of superconductor. Readers familiar with the theory of superconductors will note that singlet superconductors are described by quasiparticles that are a superposition of electrons and holes with {\em opposite} spin, thus a traditional singlet superconductor will not provide the correct conditions to see quasiparticles that behave as Majorana fermions. This hints that the type of superconductor required to support Majorana fermions should involve a superposition of particles and holes of the same spin, such a superposition occurs in spin-triplet superconductors. It turns out\cite{Read} that the special type of superconductor required in order to support a Majorana fermion is a topological, spin-triplet $p_x+ip_y$ superconductor.   

In order to properly describe a topological, spin-triplet superconductor we must first discuss the meaning of topology in condensed matter. Topology is a relatively new avenue of research in condensed matter physics. By studying the wave function of a system one can classify the system as either trivial or topological by defining objects known as topological invariants. In a superconductor this topological invariant is equivalent to studying the ``winding" of the order parameter phase as we encircle momentum space. In a topological superconductor, the pairing function has a phase that winds by an odd multiple of $2\pi$ in momentum space. A topological, spin-triplet $p_x+ip_y$ superconductor is then a superconductor whose pairing function winds by $2\pi$ in momentum space and whose order parameter has $p_x+ip_y$ symmetry. 

Unfortunately there are, at the time of writing, no known materials that are naturally topological, spin-triplet $p_x+ip_y$ superconductors. Some materials have been found to have triplet $p$-wave pairing, but their topology has yet to be proven non-trivial \cite{Kallin}. The fact that these special superconductors do not seem to occur naturally has lead to theoretical proposals that take an indirect route at creating spin-triplet $p$-wave pairing. These proposals start with several independent solid state systems and combine them in such a way that the net result is a quantum state that is either a topological superconductor, or analogous to one. In the next section we will describe several prominent proposals for inducing an effective topological superconducting state. 

\section{Proposals Using Heterostructures}

The notion of combining several solid state devices in order to create an effective $p$-wave topological superconductor was pioneered by Fu and Kane\cite{Fu}. This group showed that a three dimensional topological insulator layer placed in proximity to a conventional $s$-wave superconductor will develop topological superconductivity. The topology of the system is inherited from the topological insulator while the pairing necessary for superconductivity is induced by proximity to the superconductor. The pairing function of the superconductor is projected onto one of the spin-orbit coupled bands native to the topological insulator. This unique spin structure forces the order parameter induced by proximity to the superconductor to wind its phase by $2\pi$ in momentum space and therefore the superconductivity is topological. 

\begin{figure}[tb]
\begin{center}
  \setlength{\unitlength}{1mm}

   \includegraphics[scale=.85]{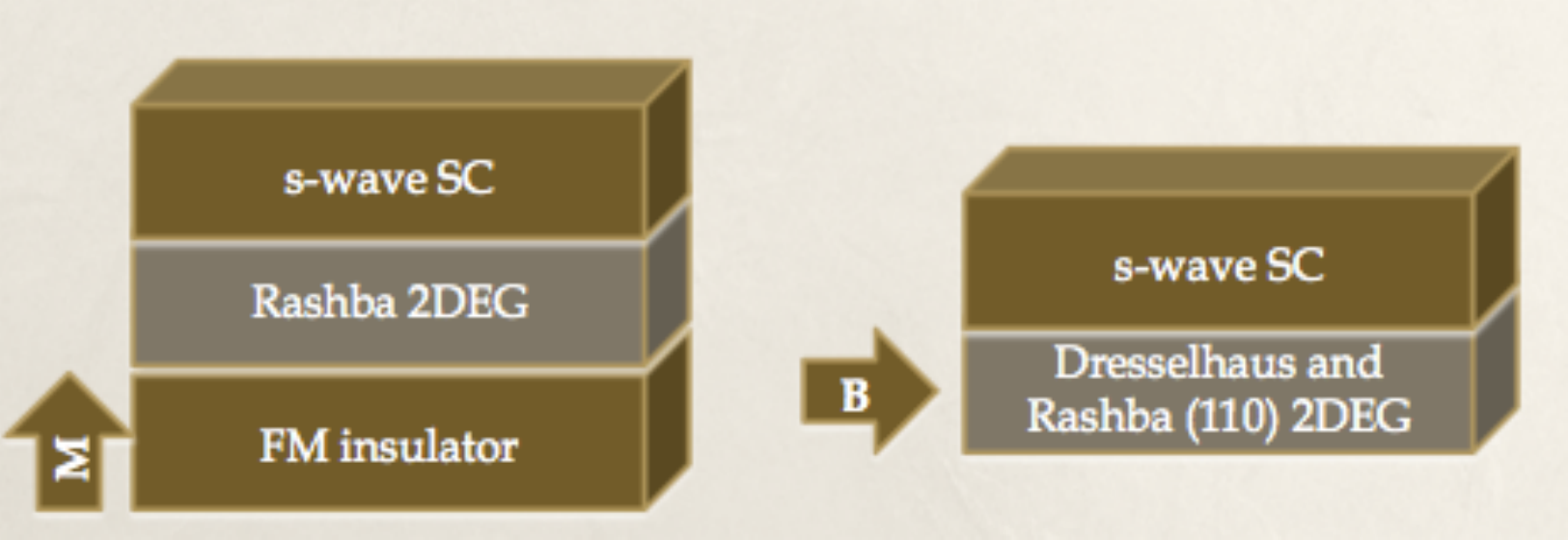}
   \end{center}
\caption{{\small
Heterostructure devices as proposed in References \cite{Sau} (left) and \cite{Alicea} (right).
     }
     }\label{fig:cartoon1}
\end{figure}

\begin{figure}[tb]
\begin{center}
  \setlength{\unitlength}{1mm}

   \includegraphics[scale=.43]{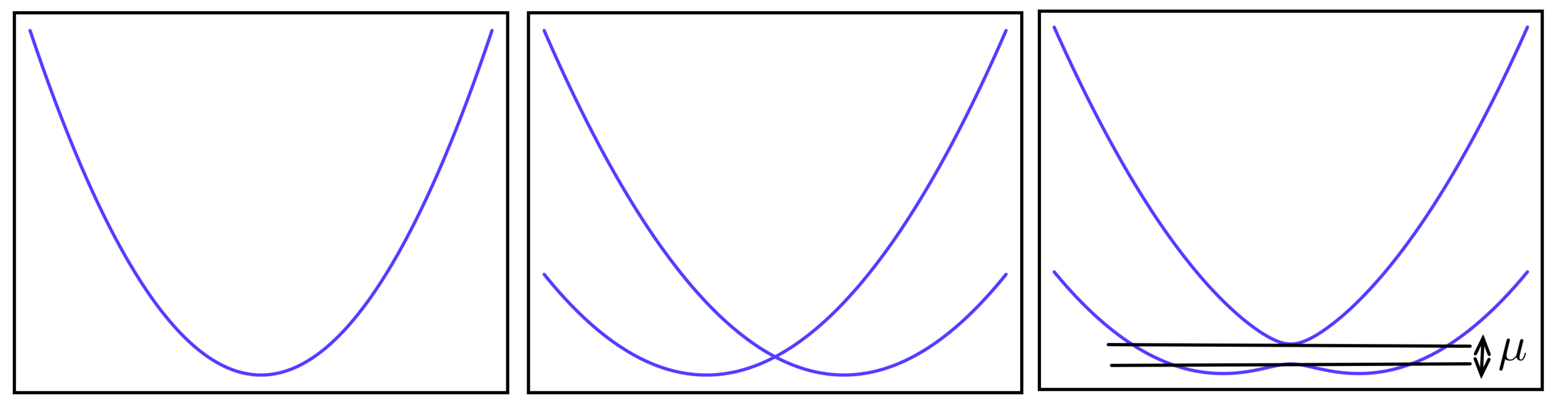}
   \end{center}
\caption{{\small
Cartoon plot of energy dispersion versus wave vector $k$ in arbitrary units. This figure shows, from left to right, how the dispersion of a 2DEG changes as we add spin-orbit coupling and then Zeeman splitting. If the chemical potential of the 2DEG lies in the gap opened by the Zeeman field, then proximity induced pairing from the superconductor will drive the system into a topological superconducting phase expected to support a Majorana fermion. 
     }
     }\label{fig:cartoon2}
\end{figure}

The idea of using several components to induce a state of topological superconductivity was further developed and refined by the group of Sau, Lutchyn, Tewari and Das Sarma\cite{Sau}. This group proposed the three layered device pictured on the left of Figure \ref{fig:cartoon1}. This heterostructure begins with a two-dimensional electron gas (2DEG) with Rashba spin-orbit coupling. One side of the 2DEG is connected to a normal $s$-wave superconductor (SC) and on the other side it is attached to a ferromagnetic (FM) insulator. The magnetization of the FM insulator points perpendicular to the interface between it and the 2DEG. Each of these pieces plays an important role in inducing a topological superconducting phase. With the help of the cartoon picture in Fig \ref{fig:cartoon2} we will discuss these roles. Starting in a 2DEG {\em without} spin-orbit coupling (far left of Fig. \ref{fig:cartoon2})  we have electrons with the typical parabolic dispersion. The Rashba spin-orbit coupling then ``splits" this parabolic band into two separate parabolic bands both offset from $k=0$ (middle panel of Fig. \ref{fig:cartoon2}). The magnetic field of the FM insulator in turn opens up a gap between these two spin-orbit split bands  (far right of Fig. \ref{fig:cartoon2}). Finally, the superconducting layer induces pairing between the electrons in the 2DEG causing them to become superconducting as well. The work in Ref. \cite{Sau} shows that if the chemical potential of the system lies within the gap in the spin-orbit split bands then the resulting superconductor should be topological in nature and support a Majorana fermion.

Motivated by simplicity and practicality Alicea\cite{Alicea} later offered an alternative to the model of Sau and coworkers.  The main success of Alicea's alternative set-up was the elimination of the FM insulating layer from the previous device. In theory one may think this would be an easy task, simply eliminate the FM insulator by applying a magnetic field perpendicular to the interface of the SC and 2DEG. This approach, unfortunately, introduces undesirable orbital effects that are harmful to the topological phase. A second simple fix might be to apply the magnetic field parallel (rather than perpendicular) to the SC-2DEG interface. This strategy also fails, not because of orbital effects but because magnetic fields applied in this direction fail to open the gap in the spin-orbit split bands that is vital for non-trivial topology. In order to both avoid orbital effects and still open a gap Alicea tactfully altered the spin structure of the 2DEG. 

Alicea's work suggested that instead of starting with a 2DEG with only Rashba spin-orbit coupling, one should instead begin with both Rashba and Dresselheus spin-orbit coupling. Such a scenario occurs in quantum wells grown along the (110) direction\cite{Alicea}. A sketch of this device appears on the right of Figure \ref{fig:cartoon1}. The advantage of having both kinds of spin-orbit coupling becomes apparent when we consider what type of spin alignment is induced by each type of spin-orbit coupling. Rashba spin-orbit coupling gives preference for the spins in the 2DEG to align {\em within} the plane of the interface between the 2DEG and the SC. On the contrary, Dresselheus spin-orbit coupling influences electron spins to align out of the plane defined by the component interfaces. Having both of these spin-orbit coupling flavours present tilts the plane in which the electron's spins will tend to align away from the plane of the 2DEG-SC interface. With this spin plane slightly tilted one can apply a magnetic field along the plane of the interface between the 2DEG and SC and open a gap in exactly the same way as the FM insulating layer did in the Sau {\it et al} model. The advantage of this auxiliary set up is two-fold: (1) a gap is opened using a magnetic field instead of FM insulator, so the device design is simplified and (2) the magnetic field (and hence the Zeeman splitting) can easily be tuned in the laboratory, something that is not an option with the fixed field FM insulator.   

 The main motivation for the work appearing in this thesis comes from Alicea's success in simplifying device design by considering several relatively simple changes in device set-up. Our main goal in this work is to further simplify device design by eliminating the SC layer from these current heterostructure proposals. Rather than inducing electron pairing {\em via} the proximity effect, we address the question of whether the necessary pairing can be obtained from electron-electron interactions. The key aspects of realizing an effective topological, $p+ip$-wave superconductor in these past models was the proper combination of spin-orbit coupling and Zeeman splitting. The system we have in mind is then a ``one-layered" device where spin-orbit coupled, two-dimensional electrons interact with each other. Zeeman splitting will then be induced in this system through a magnetic field applied in the proper direction (or {\it via} several other methods that will be discussed in the work to come). 
 
 One possible practical situation we are considering is as follows. Perhaps there is a material that, at zero Zeeman field, is either a normal ({\it i.e.} non-topological) superconductor or in some other phase. Our hope is that this non-topological material becomes a topological superconductor when the proper Zeeman field is applied. We now close the introduction chapter of this thesis with an outline of the chapters to come.

\section{Outline of this Thesis}

 In order to address the question of whether we see topological superconductivity in systems of interacting, spin-orbit coupled electrons we investigate such systems using several different methods. First we address the case of weak interactions between electrons. This is done by using so-called variational mean field theory and gives a general first impression of the system. We then go beyond this mean field study by considering the opposite limit of the problem, that of strongly interacting electrons. To this end, we develop a ``strong-coupling" expansion in order to extract the relevant parts of the Hamiltonian valid in the strong electron coupling limit. We then present two methods to analyze this strong coupling theory, one numerical and one analytical. With both of these limits of the problem investigated we can compare between them in order to find overlaps in their results and therefore extract some intuition about the problem. 
 
 The reader should consider the work in the latter chapters of this thesis as ongoing. In particular we are still awaiting descriptive data from the rather lengthy numerical calculation outlined in Chapter 6. The goal of Chapters 5 and 6 will then be to motivate the work we have done, develop the methods we use and present the preliminary data that we have obtained thus far.
 
 The remainder of this thesis is laid out as follows. In order to set the stage for the rest of the thesis, a short Chapter 2 presents and provides some discussion on the model we will use for spin-orbit coupled, interacting electrons. The following chapter then focuses on our mean field results. It begins by acquainting the reader with the general method of variational mean field theory and moves on to present and discuss results from this analysis. With the weakly interacting analysis presented and discussed, Chapter 4 begins analysis of the strong coupling limit. This chapter presents a method known in the literature as a strong coupling expansion in order to show how we transform our original model Hamiltonian into one valid at strong coupling. With this effective model in place, Chapters 5 and 6 present two different methods of analyzing it. Chapter 5 develops and presents results from a method known as the Gutzwiller approximation, while Chapter 6 uses Variational Monte Carlo to perform a numerical analysis of the model. Finally, Chapter 7 closes with some summary and conclusions.


\chapter{A Model For Spin-Orbit Coupled, Interacting Electrons}

\section{The Extended Hubbard Model}
Introduced in 1963\cite{Hubbard, Gutzwiller, Kanamori}, the Hubbard model has become a major workhorse in theoretical condensed matter physics. This model is regarded by most to be the simplest possible Hamiltonian that captures the essential physics of many-fermion systems with short-range repulsive interactions. Owing to its simplicity, the Hubbard model was been used for decades to describe a variety of condensed matter systems. Its applications have ranged from antiferromagnetism\cite{anderson} to treatment of the metal-insulator transition\cite{Mott, Hubbard2} in addition to high temperature superconductors\cite{param, anderson2}.

The form of the Hubbard model is given by
\begin{equation}
H_{Hub} = T+H_U,
\end{equation}
where $T$ describes the kinetic energy of the electrons and $H_U$ accounts for interaction effects. Using the notation of second quantization where $c^\dagger_{i,\sigma}$ ($c_{i,\sigma}$) creates (destroys) an electron of spin $\sigma$ at lattice site $i$ and defining the occupation operator $n_{i,\sigma} =c^\dagger_{i,\sigma} c_{i,\sigma}$ we have
\begin{equation}
T = -t \sum_{\langle i,j\rangle , \sigma} (c^\dagger_{i,\sigma} c_{j,\sigma} + c^\dagger_{j,\sigma} c_{i,\sigma}),
\end{equation}
where $\langle i,j\rangle$ denotes a sum over nearest neighbours. The above expression for the kinetic energy represents the tight-binding kinetic energy associated with electrons hopping on nearest neighbours. The interaction term in this model is given by   
\begin{equation}
H_U = U_0\sum_i n_{i,\uparrow} n_{i,\downarrow} \ \ \ \ (U_0>0).
\end{equation} 
This term assigns an energy cost $U_0$ for electrons occupying the same lattice site, $i$.

As is easily seen in the mathematical description above, the interaction term $H_U$ assumes that all interactions involving electrons other than those occupying the same lattice site are negligible. One way to improve upon the severity of this assumption is to include a term that allows for interactions between electrons on nearest neighbour lattice sites. When one includes such additional terms the model studied is typically called the ``extended Hubbard model" in the sense that it has been extended to include additional interaction effects. In this thesis we will study a Hamiltonian with an additional term to account for interactions between electrons on nearest neighbour lattice sites. Our attention is then concerned with the following ``extended" Hubbard Hamiltonian
 \begin{equation}
H_{EHM} = T+H_U+H_V.
\end{equation}
where 
\begin{equation}
H_V = V_0 \sum_{\langle i,j\rangle, \sigma, \sigma'} n_{i,\sigma}n_{j,\sigma'} \ \ \ \ \ \ (V_0<0).
\end{equation}     
This expression assigns an energy cost $V_0$ for having electrons (of either spin flavour) on nearest neighbour lattice sites. 

On physical grounds one would argue that we should take the sign of $V_0$ to be positive, after all electrons on nearest neighbour sites should repel each other thereby increasing their energy. In taking $V_0$ to be negative we have followed previous work on the extended Hubbard model without spin-orbit coupling. This work was carried out in the context of the cuprates\cite{Scalapino,Onari} where it was shown that on-site repulsion treated in the Eliashberg formalism leads to $d$-wave pairing on bonds. This same result can be obtained at a mean field level if one introduces an effective nearest-neighbour attraction such as the term $H_V$ introduced above. Thus rather than viewing $H_V$ as an object with a transparent physical origin, we should think of it as an effective interaction introduced to make connection with a more rigorous treatment of the problem. 

\section{Spin-Orbit Coupling on a Lattice}

We begin our description of the lattice model for spin-orbit coupling by simply stating the model and then providing motivation for why we feel it is appropriate (and versatile) afterwards. To this end, our spin-orbit coupling contribution to the Hamiltonian is given by 
\begin{equation}
H_{SO} = \sum_{\kv}\psi^\dagger_{\kv} \mathcal{H}_\kv \psi_{\kv},
\end{equation}
where $\mathcal{H}_\kv = {\bf d}_\kv \cdot \vec{\sigma}$ with ${\bf d}_\kv = (A\sin{k_x}, A\sin{k_y},  2B (\cos{k_x}+\cos{k_y}-2)+M)$, $\vec{\sigma}$ is a vector of Pauli matrices and $\psi_{\kv} = (c_{\kv,\uparrow}, c_{\kv,\downarrow})^T$ where we have introduced the Fourier transformed creation and annihilation operators $c_{\kv,\sigma} =N^{-1/2} \sum_{i} e^{-i\rv_i \cdot \kv} c_{i,\sigma}$. In the above, the parameters $A,B$ and $M$ are material dependent parameters characterizing spin-orbit coupling strengths.    

For discussion's sake we break the above spin-orbit coupling Hamiltonian into three pieces as follows
\begin{eqnarray}
H_{SO} &=& \nonumber \sum_{\kv}\psi^\dagger_{\kv} (d_1(\kv) \sigma_x + d_2(\kv) \sigma_y) \psi_{\kv}+ \sum_{\kv}2B (\cos{k_x}+\cos{k_y})\psi^\dagger_{\kv}  \sigma_z \psi_{\kv}\\ \nonumber &+&(M-4B) \sum_{\kv}\psi^\dagger_{\kv}  \sigma_z \psi_{\kv} \\ &=& H_{xy} + H_{z} + H_{Z}
\end{eqnarray}
Each of these three terms plays an important role in creating our analogue to the systems explored in References \cite{Sau} and \cite{Alicea}. Furthermore, the entire contribution $H_{SO}$ has been chosen to make contact with previous work on spin-orbit coupling in the context of HgTe quantum wells\cite{BHZ}. Below we elaborate on these two properties, beginning with the former. 

The past studies of Sau\cite{Sau} {\it et al} and Alicea\cite{Alicea} have both been performed in the continuum limit and have both made use of Rashba spin-orbit coupling. Rashba spin-orbit coupling is given by
\begin{equation}
H_{{Rashba}} = -i \alpha\int d^2 {\bf r} {\psi}^\dagger({\bf r}) (\sigma_x\partial_y-\sigma_y\partial_x){\psi}({\bf r})
\end{equation}  
where $\sigma_i$ are Pauli matrices, ${\psi}({\bf r}) = (\psi_{\uparrow}({\bf r}) , \psi_{\downarrow}({\bf r}) )^T$ with field operator $\psi_{\sigma}(\rv)$  and $\alpha$ is a strength parameter. This Rashba term can be viewed as an effective magnetic field that aligns the spins of the electrons in a plane normal to their momenta. Writing the above definition in terms of the momentum operators $\tilde{\psi}({\bf k})=(\psi_\uparrow({\bf k}),\psi_\downarrow({\bf k}))^T$ we have
 \begin{equation}
H_{{Rashba}} =  \alpha\int d^2 {\bf k}\tilde{ \psi}^\dagger({\bf k}) (\sigma_xk_y-\sigma_yk_x)\tilde{\psi}({\bf k})
\end{equation}  
We now take the above continuum model to a lattice model by making the following replacements: $k_i \to \sin{k_i}$, $\int d^2\kv \to (2\pi/L)^2\sum_{\kv}$ and $\psi_\sigma(\kv) \to L c_{\kv,\sigma}/2\pi$. This gives
 \begin{equation}
H_{{Rashba}} \to  \alpha\sum_{ {\bf k}} \psi^\dagger({\bf k}) (\sigma_x \sin{k_y}-\sigma_y \sin{k_x})\psi({\bf k}). 
\end{equation}  
where ${\psi}({\bf k})=(c_{\kv, \uparrow},c_{\kv, \downarrow})^T$. We see that (up to a phase) this is very similar to the object $H_{xy}$ in our model of $H_{SO}$ above. Therefore, $H_{xy}$ aligns spins within the plane and plays the role of Rashba spin-orbit coupling in our model Hamiltonian.

Alicea\cite{Alicea} additionally made use of Dresselheus spin-orbit coupling in his proposal for a device that should realize topological superconductivity. Physically, Dressulheus spin-orbit coupling leads to alignment of electron spins {\em out of} the plane of the quantum well. The piece of $H_{SO}$ that will play this role here is $H_{z}$. $H_{z}$  says that the energy cost of an electron of spin $\sigma$ and momentum $\kv$ is $2B\sigma(\cos{k_x}+\cos{k_y})$. This quantity is positive for one value of $\sigma$ and negative for the other. Say for concreteness the quantity $2B(\cos{k_x}+\cos{k_y})$ is positive for the value of $\kv$ we have chosen, then for electrons at wave vector $\kv$ there is an energy savings for having spin down. This behaviour can be reversed at a different point in the Brillouin zone, {\it i.e.} we could choose a different value of $\kv$ for which $2B(\cos{k_x}+\cos{k_y})$ is negative. Overall $H_{z}$ leads to tilting of spins out of the plane and whether this tilting is up or down depends on which part of the Brillouin zone we are in.  

Finally we have the term $H_{Z}$. This term plays the role of the Zeeman splitting used in References \cite{Sau} and \cite{Alicea}. The strength of this splitting is $M-4B$ and it gives preference to one spin type or the other in the entire Brillouin zone. As will be discussed in our connection with work in \cite{BHZ}, this Zeeman term can actually originate from several different sources. In the specific case that it comes from an applied magnetic field we will ignore any orbital effects\cite{Sato1}. 

As stated above, the form of our spin-orbit coupling model has been taken to resemble one of the sectors of the model of Bernevig, Hughes and Zhang (BHZ) developed to describe HgTe quantum wells\cite{BHZ}. To put this statement on more concrete ground let us recall this BHZ model here
\begin{equation}\label{hbhz}
\mathcal{H}_{BHZ} = \left( \begin {array}{cc} \mathcal{M}(k)-\tilde{D}k^2&\tilde{A}k_{-}\\ \noalign{\medskip}\tilde{A}k_{+} &-\mathcal{M}(k)-\tilde{D}k^2\end {array}
\right)
\end{equation}
where $\mathcal{M}(k) = \tilde{M}-\tilde{B}k^2$, $k_{\pm} = k_x\pm i k_y$, $k^2=k_x^2+k_y^2$ and we have used the tilde symbol to differentiate between our model parameters and those of this model. If one discretizes the above model by sending $k_i \to \frac{\sin(k_ia)}{a}$ and $k_i^2 \to \frac{2-2\cos(k_ia)}{a^2}$, (although we have set $a=1$ in our work, we include it here for the sake of making an explicit comparison) we obtain exactly our model $T+H_{SO}$ under the condition that we identify $t = \tilde{D}/a^2$, $B = \tilde{B}/a^2$, $A=\tilde{A}/a$ and $M=\tilde{M}$. The reader should note that the BHZ model was written for hyperfine levels in HgTe\cite{BHZ} and so the BHZ values for $A,B$ and $M$ are much larger than the typical values for traditional semiconductors\cite{Alicea}. 

We feel that the two connections made above really showcase the versatility of the model we have introduced. For example, the parameters $A, B$ may be traditional spin-orbit coupling terms ({\it e.g.} Rashba and Dresselheus) such as those used in \cite{Sau} and \cite{Alicea}. In this case $A$ and $B$ will be small compared to the hopping amplitude $t$\cite{Alicea}. Additionally, we are in the position to treat $A$ and $B$ as if they are BHZ type parameters. In the Hamiltonian in Eq. (\ref{hbhz}) one can have\cite{rothe, lu, guigou} $\tilde{B}, \tilde{A} \sim \tilde{D}$ which translates to $B, A\sim t$ in terms of our parameters. The same versatility is true of the Zeeman/mass like parameter $M$; it may come from an applied field, a band gap in the quantum well\cite{BHZ} or proximity to a FM insulator.      

With these important points made we are ready to proceed with our study of this system. The Hamiltonian we will use to describe our system of interacting, spin-orbit coupled electrons is then given by
\begin{equation}\label{full}
H = H_{EHM} + H_{SO} = T + H_U + H_V + H_{SO}.
\end{equation}
The next chapter will develop and discuss mean field results of this Hamiltonian. Some of the results in the proceeding Chapter can be found in \cite{farrell}.


\chapter{Results from Variational Mean Field Theory}
\section{Variational Mean Field Theory}
\subsection{The Variational Method}
To map the phase diagram of the Hamiltonian in Eq. (\ref{full}) we will utilize (zero-temperature) variational mean field theory. At the very core of this method is the variational theorem of quantum mechanics. The variational theorem  states that for some Hamiltonian $H_v$ and some ``trial" ground state $|\psi_v\rangle$ which in general is not an eigenstate of $H_v$ an upper bound on the ground state energy of $H_v$ is given by
\begin{equation}
E \le \frac{\langle \psi_v|H_v|\psi_v\rangle}{\langle \psi_v|\psi_v\rangle}
\end{equation}
where the equality holds if and only if $|\psi_v\rangle$ is the true ground state of $H$. One beauty of this theorem is that one can construct a trial wave function built out of a set of parameters $\{\lambda\}$, define the trial energy
\begin{equation}
E_t (\lambda) = \frac{\langle  \psi(\lambda)|H_v|\psi (\lambda)\rangle}{\langle \psi(\lambda)|\psi(\lambda)\rangle}
\end{equation} 
and then minimize $E_t(\lambda)$ over the set $\{\lambda \}$. In this case the $\{\lambda\}$ are called ``variational parameters" and the set which minimizes $E_t(\lambda)$, which we will call $\{\lambda_{opt}\}$, can be used to define the wave function $|\psi (\lambda_{opt})\rangle$ and energy $E_t(\lambda_{opt})$  which are the best possible estimates of the true ground state wave function and energy. 

Variational mean field theory makes use of the fact that the trial ground state can be built out of a set of variational parameters in order to extract physical information about the system described by $H_v$. This is done by choosing a set of variational parameters that correspond to various order parameters of possible physical phases. One builds these order parameters into the trial wave function by constructing a quadratic ``auxiliary Hamiltonian" which explicitly breaks the symmetries associated with whatever phase one wishes to search for in a given model Hamiltonian. The variational parameters are contained in this auxiliary Hamiltonian as order parameters that favour the breaking of specific symmetries. Being quadratic the auxiliary Hamiltonian is easily diagonalized to find an auxiliary ground state. One can than appeal to the variational theorem and use this auxiliary ground state as a trial ground state, construct the variational energy $E_v$ and then minimize over the space of order parameters. Finding that a given order parameter is zero at the minimum of $E_v$ signifies lack of a given phase in the original Hamiltonian, while a non-zero value signifies a tendency towards this phase.

\subsection{Defining an Auxiliary Hamiltonian}
For the purposes of this thesis we will be most interested in searching for phases of both superconductivity and antiferromagnetism. This is due to the fact that these are the common phases found in model Hamiltonians like $H$ without spin-orbit coupling as well as the fact that our ultimate goal is to find a phase of topological superconductivity.  We define our auxiliary Hamiltonian by replacing the quartic parts in $H$ by terms which favour superconductivity and antiferromagnetism {\it viz} 
\begin{equation}\label{defaux}
H_{Aux} = T + H_{SO} + H_{SC} + H_{AFM} 
\end{equation}
where $T$ and $H_{SO}$ are the quadratic parts of Eq. (\ref{full}) and $H_{SC}$ and $H_{AFM}$ are the terms which favour superconductivity and antiferromagnetism respectively. Antiferromagnetism is a phase in which spins on opposing sub-lattices are anti-aligned. With this fact in mind, we construct $H_{AFM}$ as follows
\begin{equation}
H_{AFM} = S \sum_i (-1)^{i_x+i_y} (c^\dagger_{i,\uparrow}c_{i,\uparrow}-c^\dagger_{i,\downarrow}c_{i,\downarrow}) = S \sum_i e^{i\Qv\cdot \rv_i} (c^\dagger_{i,\uparrow}c_{i,\uparrow}-c^\dagger_{i,\downarrow}c_{i,\downarrow})
\end{equation}
where $\Qv=(\pi,\pi)$. In the above $S$ is typically called the N\'eel order parameter and will act as a variational parameter as described in the opening of this section. We can see in the real space description above that a non-zero value of $S$ leads to it being energetically favourable for the spins on opposite sites to have anti-aligned spins. 

Next we must determine a suitable form for $H_{SC}$. We recall that superconductivity results from broken Gauge symmetry, something that is favoured by the traditional BCS style contribution
\begin{equation}
H_{SC} = \sum_{\kv} (\Delta_\kv c_{\kv,\uparrow}^\dagger c_{-\kv, \downarrow}^\dagger+\Delta_\kv^* c_{-\kv, \downarrow}c_{\kv,\uparrow})
\end{equation}      
where $\Delta_\kv$ is in general an arbitrary function of $\kv$ to be fixed by minimization. Rather than deal with the variational calculus required in fixing $\Delta_\kv$, we choose to instead build $\Delta_\kv$ out of several different parts with each part favouring certain order parameter ``symmetries". We choose the following
\begin{equation}
\Delta_\kv = \Delta^{(1)}(\cos{k_x}-\cos{k_y})+i\Delta^{(2)}\sin{k_x}\sin{k_y} + \Delta^{(3)}(\cos{k_x}+\cos{k_y})+ \Delta^{(4)}.
\end{equation}
The pairing symmetries descried by the terms above, reading from left to right, are $d$-wave, $id$-wave, extended $s$-wave and $s$-wave. The parameters $\Delta^{(i)}$, $i=1..4$ are all to be treated as variational parameters.
   
   \subsection{Diagonalizing of the Auxiliary Hamiltonian}
Writing all four terms in Eq. (\ref{defaux}) in momentum space the auxiliary Hamiltonian reads
\begin{eqnarray}\label{kaux}
H_{Aux} &=& \sum_{\kv}\xi_\kv c^\dagger_{\kv,\sigma} c_{\kv, \sigma} \\ \nonumber &+& \sum_{\kv} (\psi^\dagger_{\kv} \mathcal{H}_\kv \psi_{\kv}+\Delta_\kv c_{\kv,\uparrow}^\dagger c_{-\kv, \downarrow}^\dagger+\Delta_\kv^* c_{-\kv, \downarrow}c_{\kv,\uparrow}+Sc^\dagger_{\kv+\Qv,\uparrow}c_{\kv,\uparrow}-Sc^\dagger_{\kv+\Qv,\downarrow}c_{\kv,\downarrow}).
\end{eqnarray}  
where $\xi_\kv = \epsilon_\kv -\mu$ with $\epsilon_\kv = -2t(\cos{k_x}+\cos{k_y})$. The above can be written in a form that is readily diagonalized as follows. First, we quadruple count the sums over the first Brillouin zone, this creates four sets of the terms in Eq. (\ref{kaux}) above. One of these sets we leave alone, in one we send $\kv\to-\kv$, in a third we send $\kv\to\kv+\Qv$ and in the final set we perform both a reflection and a shift to send $\kv\to-\kv-\Qv$. Upon making these shifts there are 64 possible combinations of bilinear terms of the form $c c^\dagger$. In order to collect these terms we define the 8-spinor $\Psi_\kv = (c_{\kv, \uparrow}, c_{\kv, \downarrow}, c_{\kv+\Qv, \uparrow}, c_{\kv+\Qv, \downarrow},c^\dagger_{-\kv, \uparrow}, c^\dagger_{-\kv, \downarrow}, c^\dagger_{-\kv-\Qv, \uparrow}, c^\dagger_{-\kv-\Qv, \downarrow})^T$ and make the observation that $H_{Aux}$ may be written as
\begin{equation}
H_{Aux} = \frac{1}{4} \sum_{\kv} \Psi_\kv^\dagger \Lambda_\kv \Psi_\kv
\end{equation}
 where the factor of $1/4$ comes from the quadruple counting described above. The $8\times8$ matrix $\Lambda_\kv$ can be written concisely in terms of sub-blocks as follows
 \begin{equation}
 \Lambda_\kv =    \left(\begin{matrix} 
       h(\kv) & \hat{\Delta}(\kv) \\
       \hat{\Delta}(\kv)^\dagger & -h(-\kv)^* \\
    \end{matrix}\right)
 \end{equation} 
 which represents the Nambu space of particles and holes. The $4\times4$ sub-blocks above are given by
  \begin{equation}
h(\kv) =    \left(\begin{matrix} 
       \hat{\mathcal{H}}(\kv) & S\sigma_z\\
       S\sigma_z & \hat{\mathcal{H}}(\kv+\Qv) \\
    \end{matrix}\right)
   \ \ \ \ \ \   \hat{\Delta}(\kv) =    \left(\begin{matrix} 
      i\Delta_\kv \sigma_y & 0\\
       0 &  -i\Delta_{\kv+\Qv} \sigma_y \\
    \end{matrix}\right)
 \end{equation} 
 where $\hat{H}(\kv) = \xi_\kv+\mathcal{H}_\kv$.
 
 The diagonalization of $H_{Aux}$ can now be done using straightforward linear algebra and by noting some symmetries of the matrix function $\Lambda_\kv$. The first symmetry we exploit is particle-hole symmetry. This symmetry manifests itself in $\Lambda_\kv$ as follows
 \begin{equation}\label{ph}
 \Gamma \Lambda_\kv \Gamma = - \Lambda_{-\kv}^* \ \ \ \text{where  } \Gamma =  \left(  \begin{matrix} 
       0 & I_{4\times4} \\
       I_{4\times4} & 0 \\
    \end{matrix}\right).
 \end{equation} 
 If we define the eigenstates of $\Lambda_\kv$ {\it via} $\Lambda_\kv |\phi^n_\kv\rangle = E_n(\kv) |\phi^n_\kv\rangle$ then this symmetry tells us that if $|\phi^n_\kv\rangle$ is an eigenstate with eigenvalue $E_n(\kv)$ then $(\Gamma|\phi^n_{-\kv}\rangle)^*$ is an eigenstate with eigenvalue $-E_n(-\kv)$. Next we relate results on differing sub-lattices by making use of the symmetry
 \begin{equation}\label{sub}
 \Sigma \Lambda_\kv \Sigma = \Lambda_{\kv+\Qv} \ \ \ \text{where  } \Sigma =  \left(  \begin{matrix} 
       \mathcal{S}_x & 0 \\
       0 & -\mathcal{S}_x \\
    \end{matrix}\right), \ \ \text{and  } \mathcal{S}_x =  \left(  \begin{matrix} 
      0 & I_{2\times2} \\
       I_{2\times2} & 0 \\
    \end{matrix}\right).
 \end{equation}  
Similar to the particle hole-symmetry this tells us that if $|\phi^n_\kv\rangle$ is an eigenstate with eigenvalue $E_n(\kv)$ then $\Sigma \Gamma|\phi^n_{\kv+\Qv}\rangle$ is an eigenstate with eigenvalue $E_n(\kv+\Qv)$. Now between the symmetries in Eqs. (\ref{ph}) and (\ref{sub}) we can write all 8 eigenvalues and eigenvectors of $\Lambda_\kv$ in terms of only two eigenvalues and their associated eigenvectors. We make use of this in order to diagonalize $\Lambda_\kv$.

The diagonalization of $H_{Aux}$ is done by first diagonalizing $\Lambda_\kv$. To do so we begin by simply writing $\Lambda_\kv = W_\kv  \bar{\Lambda}_\kv W_\kv ^\dagger$ where $W_\kv$ is a unitary matrix built out of the eigenvectors of $\Lambda_\kv$ and $\bar{\Lambda}_\kv$ is a diagonal matrix of eigenvalues. Using the symmetries outlined above we can write $\bar{\Lambda}_\kv = \text{diag} (E_{\kv, \uparrow}, E_{\kv, \downarrow}, E_{\kv+\Qv, \uparrow}, E_{\kv+\Qv, \downarrow},-E_{-\kv, \uparrow}, -E_{-\kv, \downarrow}, -E_{-\kv-\Qv, \uparrow}, -E_{-\kv-\Qv, \downarrow})$ where we are free to choose $E_{\kv,\sigma} >0$. Further, using the particle-hole symmetry we are free to define the unitary transformation $W_\kv$ in terms of sub-blocks as follows
\begin{equation}\label{ut1}
W_\kv =  \left(  \begin{matrix} 
      u_\kv &v_{-\kv}^* \\
       v_\kv & u_{-\kv}^* \\
    \end{matrix}\right).
\end{equation} 
while the lattice symmetry defined previously allows the sub-blocks of $W_\kv$ to be written as
\begin{equation}\label{ut2}
u_\kv =  \left(  \begin{matrix} 
      \bar{u}_\kv & \hat{u}_{\kv+\Qv} \\
       \hat{u}_\kv & \bar{u}_{\kv+\Qv} \\
    \end{matrix}\right),
\ \ \ \ \
v_\kv =  \left(  \begin{matrix} 
      \bar{v}_\kv & \hat{v}_{\kv+\Qv} \\
       \hat{v}_\kv & \bar{v}_{\kv+\Qv} \\
    \end{matrix}\right),
\end{equation} 
where the entries above are themselves $2\times2$ matrices. All objects of interest can now be written in terms of the first two columns of $W_\kv$, {\it i.e.} in terms of $\bar{u}_\kv, \hat{u}_\kv, \bar{v}_\kv$ and $\hat{v}_\kv$. 

We now finally diagonalize $H_{Aux}$ by defining the new quasiparticle excitations  $\Phi_\kv = (\gamma_{\kv, \uparrow}, \gamma_{\kv, \downarrow}, \gamma_{\kv+\Qv, \uparrow}, \gamma_{\kv+\Qv, \downarrow},\gamma^\dagger_{-\kv, \uparrow}, \gamma^\dagger_{-\kv, \downarrow}, \gamma^\dagger_{-\kv-\Qv, \uparrow}, \gamma^\dagger_{-\kv-\Qv, \downarrow})^T$ such that $\Phi_\kv = W_\kv ^\dagger \Psi_\kv$ which in turn allows us to write
\begin{equation}
H_{aux} = \frac{1}{4} \sum_\kv  \Phi_\kv^\dagger\bar{ \Lambda}_\kv \Phi_\kv
\end{equation}
Writing the above out explicitly gives $8$ terms of the form $E\gamma^\dagger \gamma$ all of which can have their indices relabeled to give
 \begin{equation}
H_{aux} =\sum_{\kv,\sigma}  E_{\kv,\sigma}\gamma_{\kv,\sigma}^\dagger \gamma_{\kv,\sigma} + E_0
\end{equation}
where $E_0$ is a ground state energy that is unimportant for this discussion. By construction the energies satisfy $E_{\kv,\sigma}>0$ and so it costs the system energy to have $\gamma$-particles. The ground state of the system will then be a vacuum of $\gamma$ operators, that is to say our auxiliary ground state $|\psi_{Aux}\rangle$ must obey the property
\begin{equation}\label{gs}
\gamma_{\kv,\sigma} |\psi_{Aux}\rangle \equiv 0 \ \ \forall(\kv,\sigma).
\end{equation} 
To proceed with our variational mean field theory we need not know the form of the wave function nor the excitation energies (we work at $T=0$). All we need is the property in Equation (\ref{gs}) and the fact that $H_{aux}$ is quadratic.  

The variational energy of the problem is now constructed as follows
\begin{equation}
E_{var} = \langle \psi_{Aux} | H | \psi_{Aux}\rangle 
\end{equation}
where $H$ is from Equation (\ref{full}) and we have taken $|\psi_{Aux}\rangle$ to have unit normalization. The above expectation value can be evaluated by using two simple facts: the auxiliary ground state is the vacuum of $\gamma$ quasi-particles and the auxiliary Hamiltonian is quadratic meaning Wick's theorem applies to all averages we might wish to take. To make use of the first fact we must write all electron operators in terms of quasiparticle operators. This is done most efficiently by way of Eqs. (\ref{ut1}) and (\ref{ut2}) and the fact that $\Psi_\kv = W_\kv \Phi_\kv$ which gives
\begin{equation}
c_{\kv, \alpha} = \sum_{\beta} \left(\bar{u}_{\kv,\alpha,\beta} \gamma_{\kv,\beta}+\hat{u}_{\kv+\Qv,\alpha,\beta} \gamma_{\kv+\Qv,\beta}+\bar{v}^*_{-\kv,\alpha,\beta} \gamma^\dagger_{-\kv,\beta}+\hat{v}^*_{-\kv-\Qv,\alpha,\beta} \gamma^\dagger_{-\kv-\Qv,\beta}\right).
\end{equation}
Making use of the above it straightforward to show the following two central results
\begin{eqnarray}
&& \langle \psi_{Aux} | c^\dagger_{\kv,\alpha}c_{\kv,\alpha'} | \psi_{Aux}\rangle = (\bar{v}_{-\kv}\bar{v}_{-\kv}^\dagger+\hat{v}_{-\kv-\Qv}\hat{v}_{-\kv-\Qv}^\dagger)_{\alpha,\alpha'} \\ \nonumber 
&& \langle \psi_{Aux} | c^\dagger_{\kv,\alpha}c^\dagger_{-\kv,\alpha'} | \psi_{Aux}\rangle = (\bar{v}_{-\kv}\bar{u}_{-\kv}^\dagger+\hat{v}_{-\kv-\Qv}\hat{u}_{-\kv-\Qv}^\dagger)_{\alpha,\alpha'}
\end{eqnarray}
These two results along with straightforward complex conjugation and Wick's theorem allow all possible averages involved in finding $E_{var}$ to be computed.

We are now ready to extract physical information from $E_{var}$. For a given set of order parameters $(\{\Delta^{(i)}\}, S)$ we can numerically diagonalize $\Lambda_\kv$ to find the matrices $\bar{u}_\kv, \hat{u}_\kv, \bar{v}_\kv$ and  $\hat{v}_\kv$. Using these matrices we can find $\langle \psi_{Aux} | H | \psi_{Aux}\rangle$ in order to define $E_{var}(\{\Delta^{(i)}\}, S)$. The function $E_{var}(\{\Delta^{(i)}\}, S)$ is then minimized over the space of order parameters in order to determine what phases, if any, are present. This entire process determines one point in a phase diagram on the 6 dimensional space spanned by the dimensionless parameters $(A/t,B/t,M/t,\mu/t,U_0/t,V_0/t)$. The next section of this chapter will showcase the interesting portions of this phase diagram. For this purpose we will fix $A,B,M$ and $\mu$ in order to study how the ground sate phase depends on the interaction parameters $U_0$ and $V_0$. After this we will fix $U_0$, $V_0$, $M$ and $\mu$ while allowing $A$ and $B$ to vary.  

\section{Results}
\subsection{The mean field phase diagram}
\begin{figure}[tb]
\begin{center}
  \setlength{\unitlength}{1mm}

   \includegraphics[scale=.6]{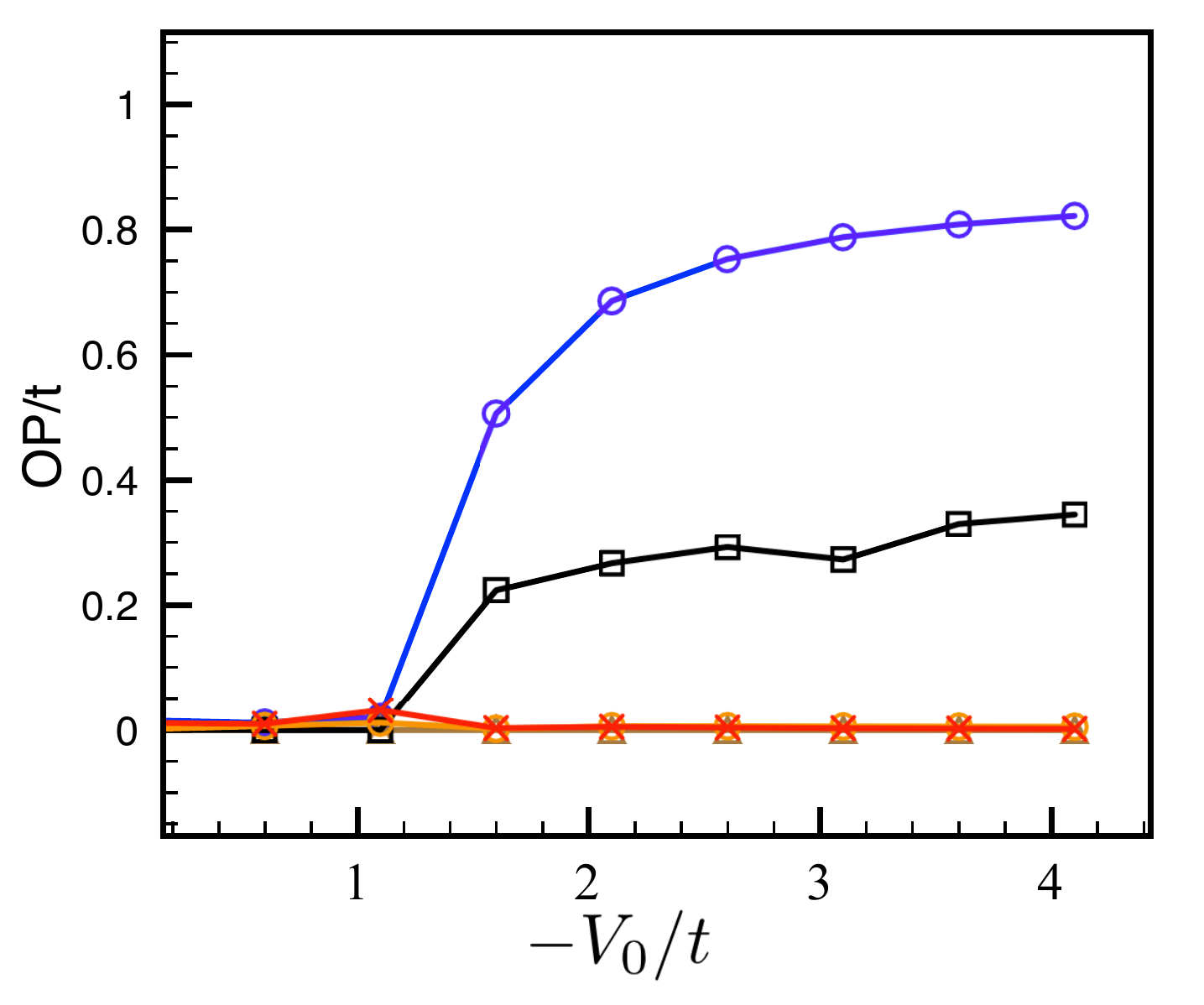}
   \end{center}
\caption{{\small
Representative plot of the order parameters. In this figure all energies are in units of $t$ and we have fixed $A=0.25t$, $B=.5t$, $M=.1t$ and
$U_0=2t$. Blue circles are $\Delta^{(1)}$, black squares are $\Delta^{(2)}$, orange diamonds are $\Delta^{(3)}$, red x's are  $\Delta^{(4)}$ and brown triangles represent $S$. This simulation was done on a $100\times100$ square lattice.  The graph shows the development of $d+id$ order since both $\Delta^{(1)}$ and $\Delta^{(2)}$ become non-zero at the critical coupling. For this figure we have fixed $\mu=0$. This figure also appears in Ref. \cite{farrell}. 
     }
     }\label{fig:OP}
\end{figure}

To continue acquainting the reader with our method we begin by showing an example of what is seen when looking at the order parameters in this study. This is done by plotting the order parameters as a function of the nearest-neighbour interaction strength with all other parameters in the model fixed. This representative plot appears in Fig.~\ref{fig:OP}. We see that $\Delta^{(3)}=\Delta^{(4)}=0$ throughout Fig.~\ref{fig:OP} and so we must conclude that for these parameter values the model does not support $s$-wave nor extended $s$-wave superconductivity. Figure \ref{fig:OP} also shows $S=0$ and so AF is also not a dominant order in this part of parameter space. The lack of $s$ and extended $s$-wave superconductivity is a general characteristic of the phase diagram of this model, however depending on how we tune the spin-orbit coupling parameters it is possible to find a state where AF (and not superconductivity) is the dominant ground state phase. 

\begin{figure*}[tb]
  \setlength{\unitlength}{1mm}

   \includegraphics[scale=.6]{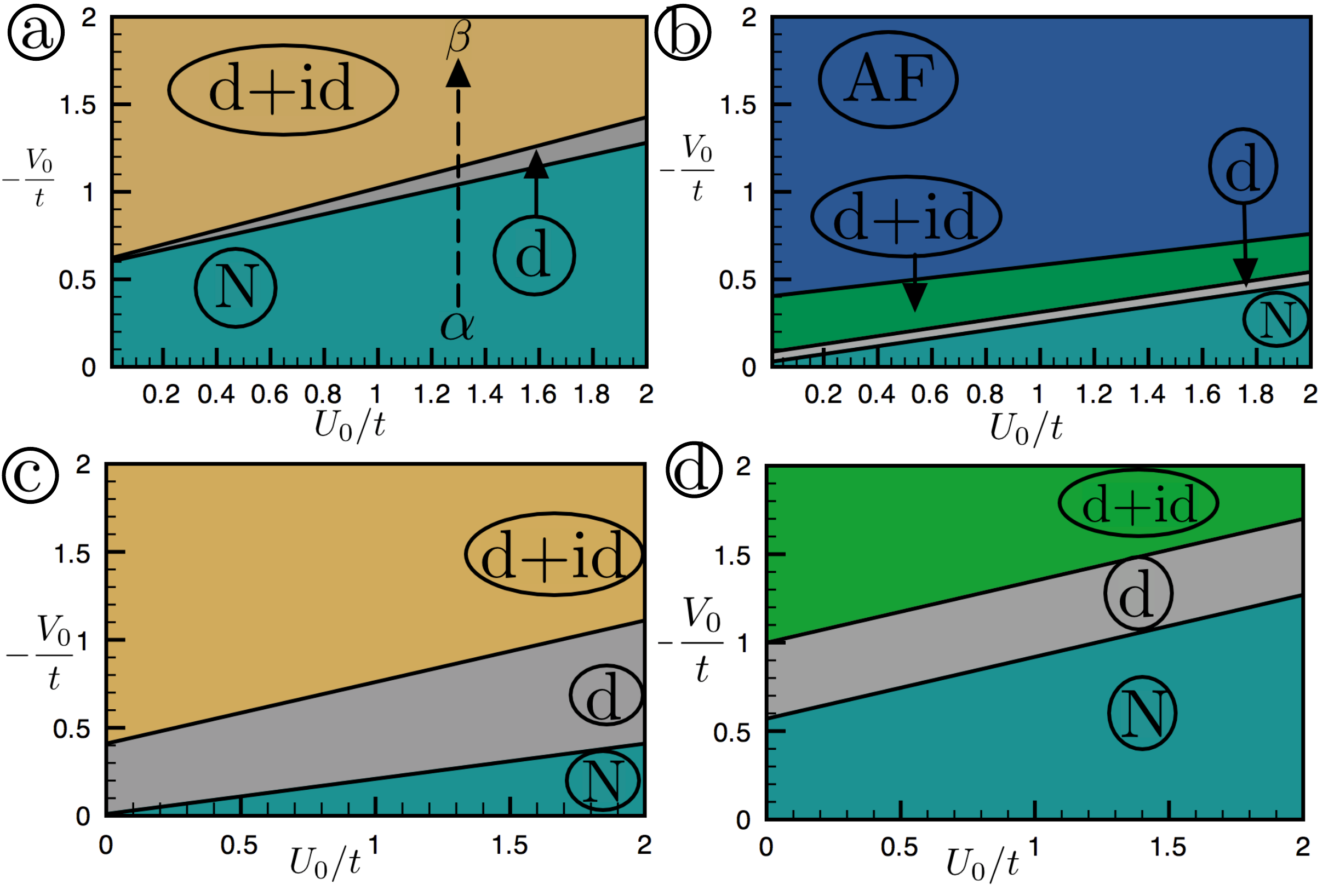}
\caption{
Two-dimensional slices of the phase diagram. The parameters we have chosen are (a) $A=0.25t, B=-0.45t$ and $M=-0.1t$, $\mu=0$, (b) $A=0.25t,B=-0.9t$ and $M=-0.05t$, $\mu=0$,  (c) $A=0.25t, B=0.5t$ and $M=0.1t$, $\mu=0$ and (d) $A=0.25t, B=0.5t$ and $M=0.1t$, $\mu=.1$. The phases are labeled by: ``N" - normal, ``AF" - antiferromagnetic, ``d" - $d$-wave superconductor and ``d+id" - a fully gapped superconductor with order parameter of the form $d_{x^2-y^2}+id_{xy}$. The ``d+id" phase in panels (a) and (c) are topologically trivial and we have coloured it beige while that of panels (b) and (d) is non-trivial and so we have coloured it green. This figure also appears in Ref. \cite{farrell}.}\label{fig:PD11}
\end{figure*}

In order to explore the phase diagram further it is first useful to determine relevant and physical values to use for the spin-orbit coupling and interaction strengths.  Given the various origins of the spin-orbit coupling terms outlined in Chapter 2, estimating realistic values of these parameters is very difficult. In addition, it is also difficult to know what the physically relevant values of $V_0$ and $U_0$ might be. In principle all of these couplings could be determined from {\it ab-initio} calculations but may vary greatly from one material to another.  In view of this, our strategy in plotting the phase diagram will be to explore a large portion of $U_0-V_0$ parameter space while fixing the other model parameters ($A, B, M$) to be similar to the values known\cite{rothe, lu, guigou} for materials such as the two dimensional topological insulators for which the BHZ\cite{BHZ} model was written. Another motivating factor for the values we have chosen is to show the versatility of the model and how this mean field study can have several different ground states depending on the parameters used.  

The differing ground states of our model are presented in Figure \ref{fig:PD11}. This figure gives four plots in a space of the interaction parameters $V_0$ and $U_0$ were we have tuned both from 0 up to $2t$ (half the bandwidth). In panel a we have values of $A,B$ and $M$ where the ground state is found to be superconductivity, while in b the SOC parameters are tuned so that we see a phase with an AF ground state. From a topological standpoint, the $d+id$ phase in Fig.~\ref{fig:PD11} is the most interesting. This is because it is fully gapped and therefore its topological invariant is well defined. In terms of our variational parameters this $d+id$ phase is obtained when both $\Delta^{(1)}$ and $\Delta^{(2)}$ are non-zero.  Later in this chapter we will focus to this $d+id$ region of the phase diagram and investigate its topology.

In the bottom two panels of Fig.~\ref{fig:PD11} show what happens when we tune parameters across a topological phase transition. To do this we have selected $A,B$ and $M$ values where both a $d$-wave and $d+id$-wave phase are seen. In panel c we have set $\mu=0$ and the superconducting phase is topologically trivial. In moving to panel d we have tuned $\mu$ to a value such that non-trivial topology is seen. This also shows that topological superconductivity can be achieved for the relatively low (compared to panel b of Fig.~\ref{fig:PD11} ) value of $B=0.5t$.

\begin{figure*}[tb]
  \setlength{\unitlength}{1mm}

   \includegraphics[scale=.325]{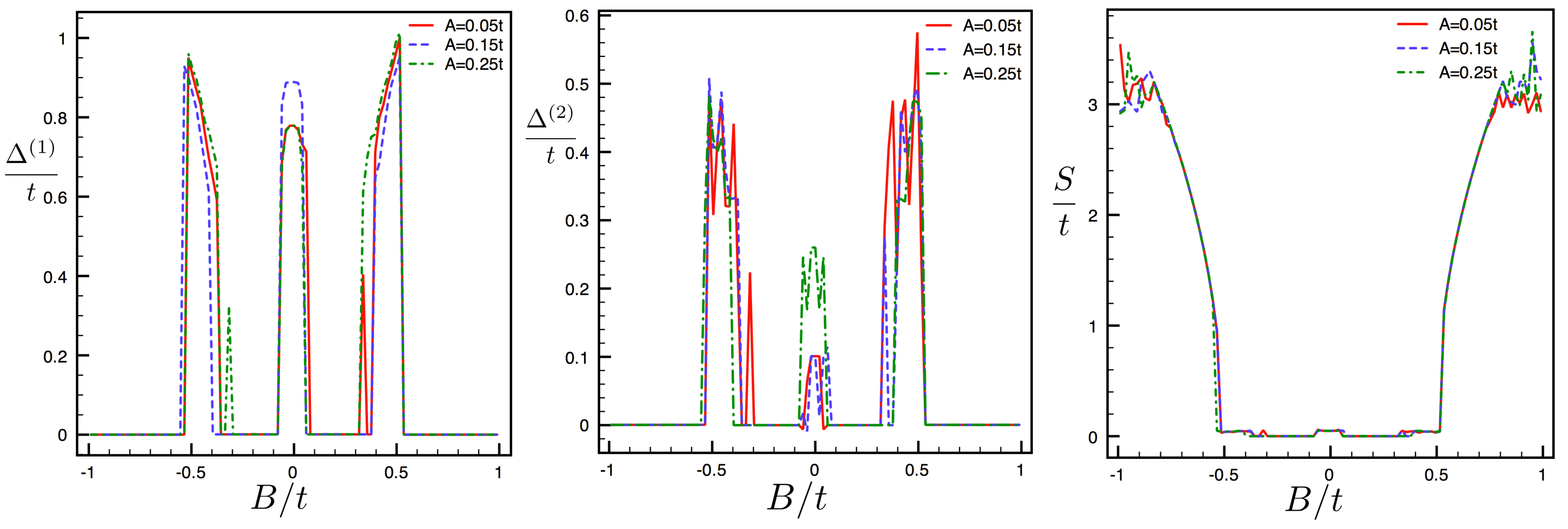}
\caption{{\small
Plot of relevant order parameters as $B$ and $A$ are changed. In this figure we have fixed $V_0=-1.8t$, $U_0=t$, and $M=-.05t$. This figure also appears in Ref. \cite{farrell}.
     }
     }\label{fig:PD2}
\end{figure*}

With a relatively thorough presentation of $U_0-V_0$ slices of the phase diagram complete, we now move on to present a picture of the dependence of various order parameters on the other parameters of the model, namely $A$ and $B$.  To accomplish this we fix the interaction strengths $U_0$ and $V_0$ to be large enough that a phase other than the normal phase can be seen and then scan through $A$ and $B$ values. With this in mind, Fig. \ref{fig:PD2} explores the dependencies of the three order parameters $\Delta^{(1)}, \Delta^{(2)}$ and $S$ on changes in the spin-orbit coupling parameters $A$ and $B$.

\section{Discussion of Results}

Here we provide some discussion and physical motivation for the results of the previous section. We begin in Fig. \ref{fig:PD11}a with the three phases labeled N, d and d+id. For concreteness, we focus on the various phases along a fixed value of $U_0$, for example the dashed-line in the figure starting at point $\alpha$ and ending at point $\beta$. This line begins in the N, or normal, phase. In this context by normal we mean that our mean field theory has found no instability towards superconductivity or AFM here. When we increase $|V_0|$ the system undergoes a transition to a $d_{x^2-y^2}$ superconductor, the d phase. Here the subscript $x^2-y^2$ denotes the form of the $d$-wave symmetry, that is to say we have $\Delta\simeq \cos{k_x}-\cos{ky}$. We can understand this transition to a $d$-wave superconducting phase by realizing that $V_0$ represents the strength of off-site attraction in our model. A significantly strong $V_0$ should therefore lead to pairing on nearest neighbour bonds, just the scenario in a $d_{x^2-y^2}$ superconductor. Continuing along our path $\alpha \to \beta$ one enters the d+id phase, a $d_{x^2-y^2}+id_{xy}$ superconductor. We can also explain this transition {\it via} the increase in $V_0$; at first $V_0$ is strong enough to induce pairing on nearest neighbours (as in the d phase) but as it is increased further it reaches a strength that will effectively lead to pairing on next-nearest neighbour bonds. Such next-nearest neighbour pairing is exactly what a $id_{xy}$-wave superconductor favours. 

To understand why $d_{x^2-y^2}$-superconductivity leads to pairing on nearest neighbour bonds while $id_{xy}$ leads to pairing on next-nearest neighbour bonds we consider transforming the contributions these terms make to the Hamiltonian into real space. We illustrate this here with the $d_{x^2-y^2}$ term and insist that a similar argument holds for the $id_{xy}$ term. The contribution of a tendency towards $d_{x^2-y^2}$ superconductivity enters the Hamiltonian as follows

\beq
H_{x^2-y^2} =\Delta^{(1)}\sum_{\kv} \left((\cos{k_x}-\cos{k_y})c_{\kv,\uparrow} ^\dagger c_{-\kv,\downarrow}^\dagger +\text{h.c.}\right)
\eeq
writing the cosines above in terms of complex exponentials and transforming the electron creation operators to real space yields 
\beq
H_{x^2-y^2} =\frac{\Delta^{(1)}}{2}\sum_{i} \left(c_{i,\uparrow} ^\dagger c_{i+\hat{x},\downarrow}^\dagger+c_{i,\uparrow} ^\dagger c_{i-\hat{x},\downarrow}^\dagger-c_{i,\uparrow} ^\dagger c_{i+\hat{y},\downarrow}^\dagger-c_{i,\uparrow} ^\dagger c_{i-\hat{y},\downarrow}^\dagger +\text{h.c.}\right)
\eeq
From this we clearly see $d_{x^2-y^2}$ leads to a tendency towards pairing on nearest neighbour bonds (recall we are working with a square lattice). 

Continuing with our interpretation of the phase diagram we see that increasing (decreasing) $U_0$ in Fig. \ref{fig:PD11}a causes the $\alpha\to\beta$ line in $U_0-V_0$ space to shift to the right (left) and increases (decreases) the critical values of $V_0$ at which the transition into the d and d+id phase are observed. We can understand this by recalling that $U_0$ represents on-site repulsion, therefore a larger $U_0$ should lead to a reduction in the probability that a site on the lattice is occupied by two electrons. Given that the superconducting order parameters we are interested in are proportional to $\langle c_{i+\delta, \downarrow} c_{i,\uparrow}\rangle$, with $\delta$ a vector to nearest neighbours for  $d_{x^2-y^2}$ and a vector to next-nearest neighbours for $id_{xy}$ superconductivity, a reduced probability for a doubly occupied lattice site should, quite roughly, lead a decrease in this quantity as it will be less likely that site $i+\delta$ is occupied by a spin down electron and site $i$ occupied by a spin up electron. This reduction means that a larger value of $V_0$ is required to realize the same superconducting phases as for smaller $U_0$. 

We now move on to panel b of Fig. \ref{fig:PD11}. We see that the same general pattern is observed as in Fig. \ref{fig:PD11}a,  at small values of $|V_0|$ no superconductivity is observed and as $|V_0|$ is increased  $d_{x^2-y^2}$ and then $d_{x^2-y^2}+id_{xy}$ superconductivity is observed. Again in this phase diagram a stronger value of $U_0$ inhibits the superconductivity and a larger value of $|V_0|$ is required to drive pairing on nearest and next-nearest neighbours. There are two main differences between the phase diagrams in Fig. \ref{fig:PD11}a and Figure \ref{fig:PD11}b. First, the required critical values of $|V_0|$ are much lower in Fig. \ref{fig:PD11}b and second, there is an AFM phase for large $|V_0|$ in Figure \ref{fig:PD11}b. 

To explain the first difference we note that we have changed the values of $B$ and $M$ in moving from Fig.~\ref{fig:PD11}a to Fig.~\ref{fig:PD11}b. To understand why this leads to a decrease in the critical value of $|V_0|$ required to develop superconductivity we note that in these calculations we have fixed $\mu$ and so the number of particles in the system is permitted to fluctuate. As a result of this the number of particles in the system will depend on the band structure of the system and therefore will depend on $B$ and $M$. If we look at the band structure we see that the system in Fig.~\ref{fig:PD11}b is closer to half filling than the system in Fig.~\ref{fig:PD11}a. We argue that being closer to half filling means the tendency towards both antiferromagnetism and $d$-wave superconductivity will increase.  In the bare Hubbard model at half-filling the ground state is known to be an antiferromagnet and therefore this tendency is not surprising.  A technical understanding of this enhanced tendency can be achieved by looking at a fermionic susceptibility in the system. Closer to half filling the system has a larger fermionic susceptibility at $(\pi,\pi)$ \cite{Scalapino,Onari} and therefore the pairing vertex function in the nearest neighbour channel is enhanced. This enhanced pairing then translates to the lower phase boundary lines in panel b of Fig.~\ref{fig:PD11}. This is not surprising in view of our discussion of the role $V_0$ plays above. 

Let us now discuss the lower two panels of Fig.~\ref{fig:PD11}. These panels show slices of the phase diagram at the same values of $A,B$ and $M$ but different values of $\mu$. The same general behaviour as in panels a and b is again seen. Namely, for small $|V_0|$ both systems start in the N phase and as $|V_0|$ is increased transition to the d phase and then the d+id phase. The difference between the two diagrams is the critical values of $|V_0|$ and (as discussed in the previous section) the topology of the $d+id$-wave phase. First, the transition values in panel c are much lower than those in panel d. This is again explained by considering the filling of the system. In moving from panel c and panel d we have increased $\mu$ and thus moved further away from half filling. Second, the $d+id$-wave phase in Fig.~\ref{fig:PD11}c is topologically trivial while that of Fig.~\ref{fig:PD11}d is topologically non-trivial. We will discuss topology more in the next section of this chapter. 

We close this section with a discussion of Fig.~\ref{fig:PD2}. We see that this figure illustrates relative insensitivity of any of the order parameters to changes in the value of the in plane spin-orbit coupling $A$. All three curves show relatively small changes in the behaviour of the order parameters over the range of $A$ values shown. Next, we see that the plots of the superconducting order parameters as a function of $B$ consist of a peak around $B=0$ and as $|B|$ is increased these peaks drop to zero and superconductivity disappears. Eventually as $|B|$ is increased further we see two peaks in $\Delta^{(1)}$ and $\Delta^{(2)}$ almost evenly distributed about $B=0$. Finally, continuing away from $B=0$ the superconducting order parameters suddenly drop to zero and the system undergoes a phase transition to an AFM phase. This transition is signalled by a non-zero value of $S$ in the rightmost panel of Fig.~\ref{fig:PD2}. This crossover into an AFM phase exactly coincides with the value of $B$ for which the number of electrons in the system begins to decrease as a function of $B$.

\section{Topology}

In Ref. \cite{Alicea} Alicea made a connection between the heterostructure model of Sau and coworkers\cite{Sau} and $p_x+ip_y$ order parameter symmetry by performing a projection of the $s$-wave superconducting order parameter of the system onto the spin-orbit coupled bands of the quantum well. In the projected basis of Sau {\it et al}'s model, the effective pairing has $p_x+ip_y$ symmetry and so, if topological, it should be analogous to the Majorana fermion supporting state of Reference \cite{Read}. A similar projection can be done on the $d+id$-wave phase we have found in this chapter in order to show that in the projected subspace the pairing symmetry has odd parity. Thus what is left for us to show in arguing that this state should support a Majorana fermion is that this phase has non-trivial topology. To this end, we set out to classify the topology of the of the $d+id$-wave mean field state obtained above. It should be noted that topological superconductivity on its own isn't a sufficient condition for finding a Majorana fermion, but is necessary in order to obtain a state which is analogous to a {\em topological} $p_x+ip_y$ superconductor.

In order to classify the topology of the $d+id$-wave phase we must calculate an object known as a topological invariant. Much like the non-zero value of an order parameter tells us about the existence of a particular physical phase in a system, a non-zero value of a topological invariant tells us that our system is topological in nature. As will be discussed further (albeit in a different context) at the end of Chapter 5 of this thesis, calculating the topological invariant of an interacting ({\it i.e.} non-quadratic) Hamiltonian is not a well established problem as of yet and little is known about how one might do this. To avoid such complications our focus in calculating a topological invariant will not be on the full interacting Hamiltonian, but rather solely on the $d+id$ mean field state. 

{
Our argument for only focusing on this $d+id$ state is as follows. We know, based on the previous section of this chapter, that for some ranges of parameters the ground state of our model is a $d+id$-wave superconductor. It is this mean-field state that we are interested in comparing to the $p_x+ip_y$ state of Ref. \cite{Read}. Thus we are interested in the topology {\em of the resulting} mean field Hamiltonian. The question then is not whether the full Hamiltonian has non-trivial topology, but rather whether the resulting effective mean field Hamiltonian does. One can  argue that this would be some sort of approximation of the topology of the full system. The reason for this is that topology is a property of the {\em ground state} of a system and not the Hamiltonian. Our approximation for the ground state of the system based on the theory developed here is the (optimized) ground state of $H_{Aux}$. Thus the topology of this ground state is also an approximation to the topology of the actual ground state wave function of the full system. 

 }

With all of this in mind, the invariant we will calculate is called the TKNN number\cite{TKNN}(equivalent to the first Chern number). This calculation is done by using our optimized mean-field wave function. Specifically, we first choose a region of parameter space with $d+id$ superconductivity, then, using the BdG matrix Hamiltonian $\Lambda_{\kv}$ defined earlier in this chapter, we calculate\cite{Sato1}
 \begin{equation}
 I = \frac{1}{2\pi} \int d^2k \mathcal{F}({\bf k})
 \end{equation}
 where the so-called Berry curvature, $\mathcal{F}$, is defined using the eigenstates $\Lambda_{\kv}|\phi_n({\bf k})\rangle = E_n(\kv)|\phi_n({\bf k})\rangle$  in the following way\cite{Berry}
 \begin{equation}
  \mathcal{F}({\bf k}) = i\sum_{n}' \sum_{m\ne n} \epsilon^{ij}\left[\frac{\langle \phi_n({\bf k})|\frac{\partial \Lambda_{\kv}}{\partial k_i}|\phi_m({\bf k})\rangle\langle
  \phi_m({\bf k})|\frac{\partial \Lambda_{\kv}}{\partial k_j}|\phi_n({\bf k})\rangle}{(E_n({\bf k})-E_m({\bf k}))^2}\right],
 \end{equation}
where  the primed sum is a sum over filled bands ({that is to say quasiparticle modes with $E<0$}), $\epsilon^{1,2}=-\epsilon^{2,1}=1$ and $\epsilon^{i,i}=0$ and summation over the repeated indices $i$ and $j$ is implied. 

{The reader will recall the heuristic description of a topological superconductor given in the introduction of this thesis as a superconductor whose order parameter phase ``winds"   in momentum space. The rather technical object $I$ defined above can be thought of as counting the number of times the geometric phase\cite{Berry} of the filled quasiparticle states of the system winds as we cover the Brillouin zone. Thus $I$ quantifies a twisting of the phase of the system and, although further rigorous verification is needed, one can see how it could be related to the phase winding of the order parameter.     }

\begin{figure}[tb]
 \begin{center}
  \setlength{\unitlength}{1mm}

   \includegraphics[scale=.75]{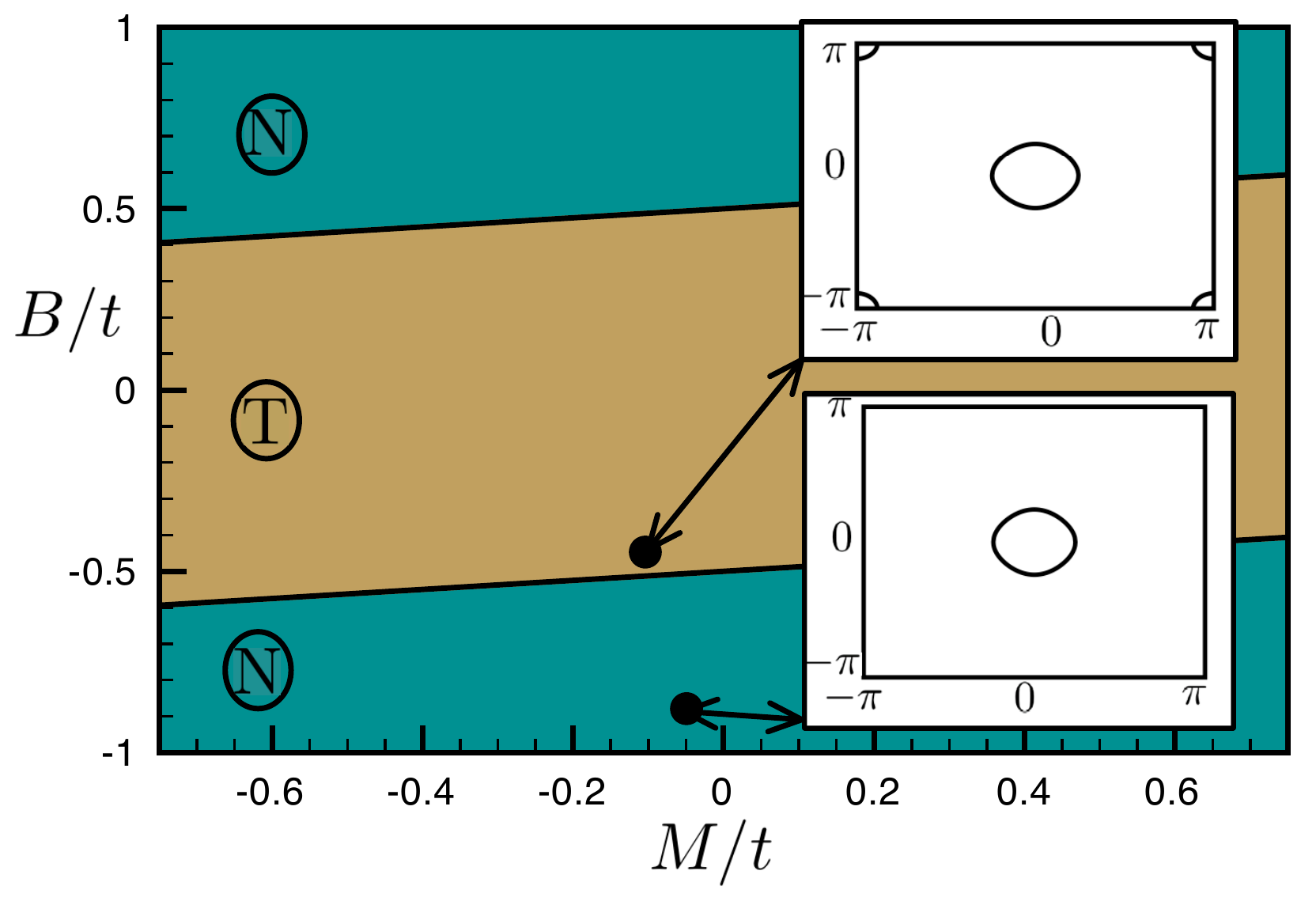}
   \end{center}
\caption{{\small
Topology of the $d+id$ ground state phase. In this figure we have set $A=0.25t$ and $\mu=0$. The topologically trivial phase is labeled T (beige) while the
non-trivial phase is labeled N (teal). The insets show an example of the Fermi surface in the first Brillouin zone for each phase. The lower plot of the Fermi surface, corresponding to the $d+id$ state of Fig. 2b, shows a single Fermi surface for the topologically non-trivial
phase while the upper one (the $d+id$ state in Fig. 2a) shows two Fermi surfaces in the topologically trivial region.
     }}\label{fig:top}
\end{figure}

The invariant $I$ allows us to classify the topology of the $d+id$ region of the model's phase diagram as either trivial ($I=0$) or non-trivial
 ($I=1$). Figure \ref{fig:top} shows the results of this calculation. To show that the topology of the system is related to the number of Fermi surfaces before interactions are turned on\cite{FuBerg}, we show a sample of the Fermi surface in the topologically trivial and the topologically nontrivial phases as insets in Figure \ref{fig:top}.

Figure~\ref{fig:top} shows that when treated at a mean field level our model has regions of $d+id$ topological superconductivity. For example, Fig.~\ref{fig:top}
shows that the $d+id$  state in Fig. \ref{fig:PD11}a is topologically trivial while the like in Fig. \ref{fig:PD11}b is topological. Further, Fig.  \ref{fig:top} suggests that some devices might exist in the topologically non-trivial region at some value of $M$. It may then be possible to move the system into a topological phase by
changing $M$ via either an applied field or proximity to a magnetic layer. Thus properly applying a Zeeman field ({\it
i.e.} tuning $M$) in a spin-orbit coupled superconductor may result in the transition of an ordinary superconductor to a topological one.

We can also find the position of Fig \ref{fig:PD11}c in Figure~\ref{fig:top}. We see that, as discussed above, Fig \ref{fig:PD11}c is found to lie in the trivial topological region. The reason that Fig \ref{fig:PD11}d is considered to have non-trivial topology is because when we tune $\mu$ the regions pictured in Fig.~\ref{fig:top} are altered. In particular, as we increase (decrease) $\mu$ from 0 the region of trivial topology in the center of this figure shrinks (grows). 

This concludes our mean field study of the model presented in Chapter 2. We have outlined the method used and highlighted the results of this method. Most importantly we have shown that mean field theory predicts regions of non-trivial topology in this system. This constitutes the ``weak coupling" study of this thesis. The remainder of this thesis will focus on the limit opposite to this, that of strong coupling. To this end the next chapter develops an effective Hamiltonian valid at strong coupling. The subsequent two chapters will discuss two different methods of studying this effective model.


\chapter{Strong Coupling Expansion}

\section{Introduction to the Strong Coupling Expansion}

\begin{figure}[tb]
 \begin{center}
  \setlength{\unitlength}{1mm}

   \includegraphics[scale=.65]{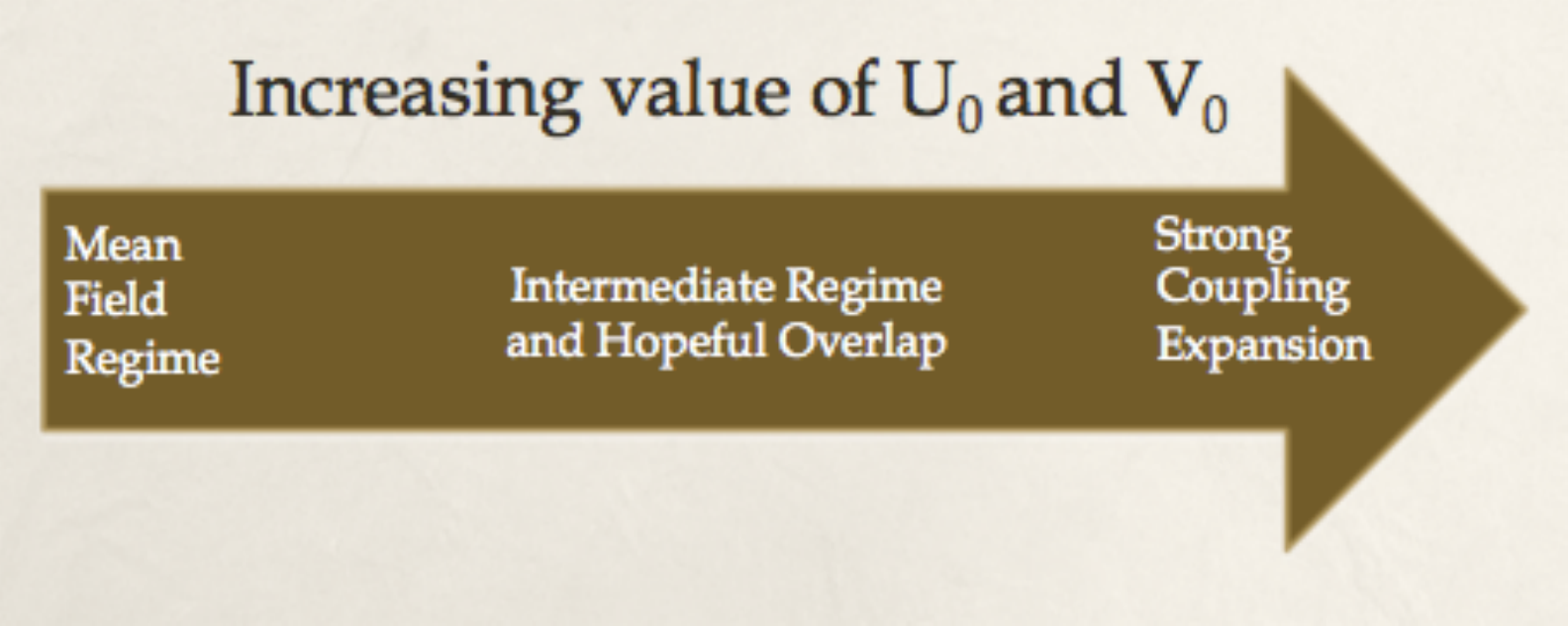}
   \end{center}
\caption{{\small
Illustration of the regions of applicability for the methods developed in this thesis. 
     }}\label{fig:app}
\end{figure}

The mean field theory of the previous section gives a good starting point for the study of our model but its applicability may be limited. In particular, its use at large values of $|V_0|$ and $U_0$ may be called into question. Owing to this, we now look at methods which should be considered applicable for very large values of  $|V_0|$ and $U_0$. Our hope is that some agreement between the two methods can be found ``in the middle", that is to say for intermediate values of  $V_0$ and $U_0$. This idea is illustrated in Figure \ref{fig:app}

The first step in this process of studying the regime of large interaction strengths is a calculation typically called the strong coupling expansion. In the context of the Hubbard model, $H_{Hub}=T+H_U$, this expansion has a rich history. Starting in 1978 the group of Chao {\it et al} \cite{chao} showed that a perturbative expansion of the Hubbard model near half-filling could be used to obtain the so-called t-J model. The t-J model is given by
\begin{equation}
H_{tJ} = -t\sum_{\langle i,j\rangle, \sigma} \left(P_Gc^\dagger_{i,\sigma} c_{j,\sigma}P_G +P_Gc^\dagger_{j,\sigma} c_{i,\sigma}P_G\right)+ J\sum_{\langle i,j\rangle} \Sv_{i}\cdot \Sv_{j}
\end{equation} 
where $P_G$ is a projection operator that allows no lattice site to have two electrons on it and 
\begin{equation}\label{sitespin}
\Sv_{i} = \frac{1}{2} \sum_{\alpha, \beta} c_{i, \alpha}^\dagger {\bf \sigma}_{\alpha,\beta}c_{i, \beta}
\end{equation}
 is an electronic spin operator at lattice site $i$. In the expansion of Chao and coworkers, the exchange constant $J$ is given by $J=4t^2/U$. The first term in the t-J model (the $t$ part) describes (restricted) hopping of electrons on nearest neighbours. Meanwhile the second term (the $J$ part), which can be recognized as the Heisenberg Hamiltonian, describes the interaction between spins on nearest neighbour lattice sites. The t-J model and the more general strong coupling expansion of the Hubbard model (valid even away from half filling) have been used to a high degree of success in order to describe the high temperature superconductors\cite{param, anderson2, zhang}.

The method of Chao and coworkers has been further developed and extended by many \cite{macdonald, dongen}. In the next section we will develop the main formalism used to perform the strong coupling expansion by using the Hubbard model as an example. The sections following this generalize the method to the full Hamiltonian presented in Chapter 2 while abdicating the more technical details of the calculation. The final section will present our analogy to the t-J model.

\section{The Hubbard Model at Strong Coupling}

Consider the Hubbard model given by the following Hamiltonian
\begin{equation}
H_{Hub} = T+H_U.
\end{equation}
Thinking about this model physically, $T$ moves an electron from one site to another while $H_U$ counts the number of sites on the lattice that contain two electrons and assigns an energy cost of $U_0$ to each instance of this. Let  $D = \sum_i n_{i,\uparrow} n_{i,\downarrow}$ count the number of doubly occupied lattice sites and consider the limit $U/t\to \infty$. In this limit having electrons on the same lattice site is extremely costly. The ground state of the model in this limit should then lie in the subspace of many-body states with the minimum number of double occupancies possible, $\{ |D^i_{min}\rangle\}$. We can imagine finding the matrix representation of $H_{Hub}$ in this subspace, $H_{i,j} = \langle D_{min}^i| H_{Hub} | D_{min}^j\rangle$. Inspecting this object we see that in the limit of very strong $U_0$ any parts of $H_{Hub}$ which connect states with a different eigenvalue of $D$ fall out of our many-body description of the system. 

The idea of the strong coupling expansion is then to begin with $H_{Hub}$ and perform a transformation that, to a certain order in $t/U$, eliminates processes connecting many-body states with different numbers of doubly occupied lattice sites. A more convenient, but entirely equivalent, way to look at this task turns out\cite{dongen} to be transforming $H_{Hub}$ in such a way that $H_U$ is a constant of motion to a certain order in $t/U$. This latter interpretation is more systematic and therefore more suited to our relatively complicated model Hamiltonian. 

To this end we apply a unitary transformation to the electron operators $c^\dagger_{i,\sigma} = e^{iS} \bar{c}_{i,\sigma}^\dagger e^{-iS}$ in order to obtain
\begin{equation}
\tilde {H} = e^{iS} (T+H_U) e^{-iS} \equiv H_U+T'
\end{equation}
In order for $H_U$ to be a constant of motion under the transformed Hamiltonian we require that
\begin{equation}\label{constant1}
[H_U, \tilde {H}] = [H_U, T']=0.
\end{equation} 
To satisfy this condition we expand both $T'$ and the transformation $S$ in a power series of $U_0$ as follows
\begin{equation}
S=-i\sum_{n=1}^\infty \frac{S_n}{U_0^n}, \ \ \ \ \ \ \ T' = \sum_{n=1}^{\infty} \frac{T'_{n}}{U_0^{n-1}}
\end{equation}
We now insist that Eq. (\ref{constant1}) is satisfied to a given order in $1/U_0$, {\it i.e.} at order $\tilde{n}$ in our expansion we have $[H_0, T'_{n}]=0$ for all $\tilde{n}\ge n\ge1$.

Our task is then to solve for the transformation $S$ which satisfies $[H_0, T'_{n}]=0$ to a given order $n$ and then plug this transformation back into $\tilde{H}$ to find the properly expanded Hamiltonian. To do this we first note that for any operator $X$ 
\begin{equation}
e^{iS} X e^{-iS}= X+[iS,X]+\frac{1}{2} [iS,[iS,X]]+...
\end{equation}
and so we have
\begin{equation}
\tilde {H} = H_U +(T+[iS, H_U]) +([iS,T]+\frac{1}{2}[iS, [iS, H_U+T]])+...
\end{equation}
Inserting the expansions for $S$ into the above and defining $H_U=U_0\tilde {H}_U$ we can read off
\begin{eqnarray}
&&T'_{1} = T + [S_1, \tilde {H}_U]  \\ \nonumber 
&&T'_{2} = [S_1, T] + [S_2, \tilde {H}_U] +\frac{1}{2}[S_1,[S_1, \tilde{H}_U]]
\end{eqnarray}
The first condition we must satisfy is then
\begin{equation}\label{conds1}
0 = [\ {H}_U, T + [S_1, \tilde {H}_U]]
\end{equation} 

\begin{figure}[tb]
 \begin{center}
  \setlength{\unitlength}{1mm}

   \includegraphics[scale=.65]{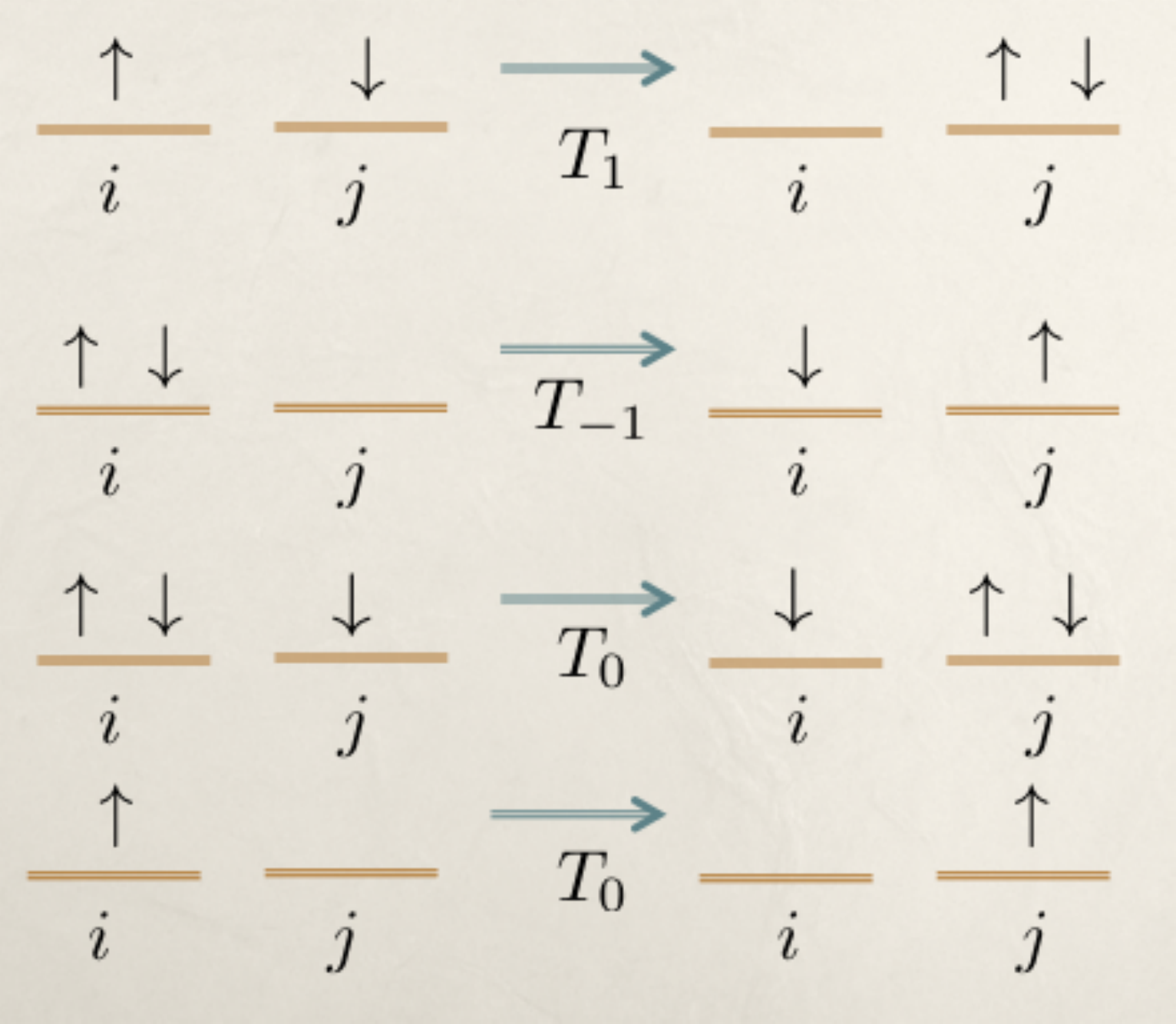}
   \end{center}
\caption{{\small
Possible hopping processes for a spin up electron under action of $T_{-1}$, $T_{1}$ and $T_{0}$. There are four analogous processes for a spin down electron. 
     }}\label{fig:hop}
\end{figure}

Solving the conditions above is made easier if we decompose $T$ into channels that change $H_U$ by a constant amount. $H_U$ counts the number of doubly occupied sites on the lattice and therefore the pieces of $T$ that change $H_U$ by a set amount must change the number of doubly occupied sites by some fixed amount. There are three possible hopping processes in $T$ that can do so: ones that increase the number of doubly occupied sites by 1, ones that don't change it and ones that decrease the number by 1. In order to find expressions for these terms we define the hole occupancy $h_{i,\sigma}=1-n_{i,\sigma}$ and then use the identity $1=n_{i,\sigma}+h_{i,\sigma}$ on either side of the summand in the definition of $T$. Doing so allows us to write the decomposition $T=T_0+T_1+T_{-1}$ where 
 \begin{eqnarray}
  \ {T}_{-1} &=& -t\sum_{i,\sigma,\delta} h_{i,\bar{\sigma}} c^\dagger_{i,\sigma}c_{i+\delta,\sigma'} n_{i+\delta,\bar{\sigma}} \\ \nonumber 
 \ {T}_{1} &=& -t\sum_{i,\sigma,\delta} n_{i,\bar{\sigma}} c^\dagger_{i,\sigma}c_{i+\delta,\sigma'} h_{i+\delta,\bar{\sigma}} \\ \nonumber 
     \ {T}_{0} &=&-t \sum_{i,\sigma,\delta} (n_{i,\bar{\sigma}} c^\dagger_{i,\sigma} c_{i+\delta,\sigma'}  n_{i+\delta,\bar{\sigma}} \\ \nonumber &&+ h_{i,\bar{\sigma}} c^\dagger_{i,\sigma}c_{i+\delta,\sigma}  h_{i+\delta,\bar{\sigma}}) 
 \end{eqnarray}
In the above $T_m$ will increase the value of $H_U$ by an amount $mU_0$. Reading the various projections above this is done as follows. $T_{1}$ moves an electron from a singly occupied site to an already singly occupied site, $T_{-1}$ moves an electron from a doubly occupied site to an empty site and $T_0$ either moves an electron from a doubly occupied site to an already singly occupied site or moves an electron from a singly occupied site to an empty site. These four processes are illustrated in Figure \ref{fig:hop}.

One can easily show that $[ H_U, T_m]=mU_0T_m$. This can be done rigorously by writing out the definition of the operators $T_m$ and $H_U$ or motivated on heuristic grounds. We present the heuristic argument here and leave the rigorous calculation to the appendix. Consider some many-body state that is an eigenstate of the double occupancy operator $D$, we will call this state $|D\rangle$. These states satisfy the following expressions $H_U|D\rangle = U_0 \tilde{D}|D\rangle$ and $T_m|D\rangle=|D+m\rangle$ where $\tilde{D}$ is the number of double occupancies, {\it i.e.} the eigenvalue of the operator $D$. Now consider the action of the commutator $[T_m, H_U]$ on this state
\begin{eqnarray}
[H_U, T_m]|D\rangle &=& H_UT_m|D\rangle-T_mH_U|D\rangle = H_U|D+m\rangle-T_mU_0\tilde{D}|D\rangle \\ \nonumber &=&  U_0(\tilde{D}+m)|D+m\rangle- U_0\tilde{D} T_m|D\rangle = U_0(\tilde{D}+m)T_m|D\rangle- U_0\tilde{D} T_m|D\rangle\\ \nonumber &=& U_0mT_m|D\rangle
\end{eqnarray}
which shows that $[ H_U, T_m]=mU_0T_m$. 

Note that there is a potentially confusing point in the notation we have used. $T_m$ (with no prime) denotes the pieces of $T$ that change the number of double occupancies by $m$, meanwhile $T'_{n}$ denotes terms in the expansion of the transformed $T$ which we have called $T'$. The two are in no way equivalent and should not be confused.

Inserting the channel decomposition $T = \sum_{m}{T}_{m}$ and using the fact that $[\tilde {H}_U, T] =  \frac{1}{U}\sum_{m}mU\ {T}_{m}$ the condition in Eq. (\ref{conds1}) is solved by
\begin{equation}
S_1 =T_1-T_{-1}
\end{equation}
Using this expression in the first order Hamiltonian we have
\begin{eqnarray}
&&\nonumber T'_{1} =T + [S_1, \tilde {H}_U] = T_0+T_{-1}+T_{1} + [T_1-T{-1}, \tilde {H}_U] = T_0+T_{-1}+T_{1}-T_{-1}-T_1 \\ 
&& T'_{1}=T_0
\end{eqnarray}
We see, not surprisingly, that the part of the expansion valid to $\mathcal{O}(1)$ involves the part of $T$ which does not change the number of doubly occupied lattice sites. 

We now move on to the second order calculation. Our first task in this calculation will be solving the condition 
\begin{equation}
0=[H_U, T_2'] = [H_U, [S_1, T] + [S_2, \tilde {H}_U] +\frac{1}{2}[S_1,[S_1, \tilde{H}_U]]]. 
\end{equation}
Solving the above must be accomplished by properly choosing $S_2$. In a task that we leave to the appendix one can show that $S_2$ is given by
\begin{equation}
S_2 = [T_1,T_0]+ [T_{-1}, T_0].
\end{equation}
Inserting this back into $T_2'$ we then obtain the second order term in the expansion of $T'$
\begin{equation}
T'_2 = \frac{1}{U}\left(T_1T_{-1}-T_{-1}T_1\right). 
\end{equation}
Heuristically both of these terms make sense. $T_1T_{-1}$ decreases the number of doubly occupancies by 1 and then increases it by 1 (the reverse is true of   $T_{-1}T_{1}$). The net result of this is no change in the number of doubly occupied sites. 

In principle one could continue this process to whatever order of $t/U$ is desired. The prescription remains the same: find an expression for $T'_n$, fix $S_n$ by insisting $[H_U, T'_n]=0$ and then plug this back into $T'_n$ to get the $(t/U_0)^{n-1}$ contribution to $\tilde{H}$. The pattern seen above continues order by order; $T'_n$ will involve a sum of all possible combinations of $n$ of the $T_m$ operators such that the net result of acting with these operators does not change the number of double occupancies. Symbolically these terms will look like $T_{m_1}T_{m_2}...T_{m_n}$ with $\sum_{i} m_{i}=0$. For practical purposes we end our discussion at $T'_2$. This gives an effective Hamiltonian for the Hubbard model which is valid to order $(t/U_0)^2$
\begin{equation}\label{stronghub}
\tilde{H} = H_U+ T_0 + \frac{1}{U_0}\left(T_1T_{-1}-T_{-1}T_1\right).
\end{equation}
If one considers the projection of the above Hamiltonian onto a subspace of states at half-filling and with no double occupancies, writes out the expressions for $T_1T_{-1}-T_{-1}T_1$ in terms of electron operators and performs the associated sums over spin indices then $J\sum_{\langle i,j\rangle} \Sv_i\cdot \Sv_j$ with $J=4t^2/U$ is obtained. This projection will be discussed later on in this chapter. 

The reader should note that the expansion model in Eq. (\ref{stronghub}) finds application in a broad range of systems. For example, the so-called $t-J$ model (which one obtains following the projection discussed in the previous paragraph) has seen success for decades in the field of high temperature superconductors \cite{param, zhang}. Depending on doping, the mean field ground state of the model in Eq. (\ref{stronghub}) has been shown to range from a fermi liquid, to a $d$-wave superconductor or a Mott insulator\cite{param}. Another field where the application of results presented here has been fruitful is in the physics of spin liquids\cite{spin1,spin2,spin3}.  In these systems, an expansion identical to the one outlined above is performed and a projection to half filling is also performed. The difference in these spin liquid studies is that they keep terms in the expansion up to fourth order in the $T_m$ operators. These fourth order terms lead to the phenomenon of ring exchange\cite{spin2,spin3}, an important process in spin liquid physics. Lastly, we would like to point out that an expansion similar to the one in Eq. (\ref{stronghub}) can be done to obtain the so-called ``Dyzaloshinskii-Moriya" interaction\cite{dzy1}, which has important applications in various fields of magnetism\cite{dzy2,dzy3} 


With the mechanics of our strong coupling method clearly illustrated we move on to the move complicated expansion of $H= T+H_{SO}+H_U+H_V$. We devote next two subsections to describing how the two additional terms in $H$ (namely $H_V$ and $H_{SO}$) alter the process we have outlined above.  

\section{Strong Coupling Expansion of the Full Hamiltonian}

\subsection{Spin-Orbit Coupling as Hopping}
In the previous section our expansion relied on a real space interpretation of $T$ and in particular how it changed the number of doubly occupied sites on a given lattice. Therefore it will also be important to have a real space interpretation of $H_{SO}$. In order to obtain this real space description we Fourier transform $H_{SO}$ in the standard way by letting $c_{\kv,\sigma} = \frac{1}{\sqrt{N}} \sum_{i} e^{i\kv\cdot \rv_i} c_{i,\sigma}$. This leads to the following real-space operator
\begin{eqnarray}\label{SOreal}
H_{SO} &=& \sum_{\sigma, i} \sigma (M-4B)c^\dagger _{i,\sigma} c _{i,\sigma} +\sum_{\sigma, i, \delta} B\sigma c^\dagger _{i,\sigma} c _{i+\delta,\sigma}\nonumber \\  &-&\frac{iA}{2}\sum_{\sigma, i, \delta} \vec{\delta}\cdot(\hat{x}-i\sigma\hat{y}) c^\dagger _{i,\sigma} c _{i+\delta,\bar{\sigma}}
\end{eqnarray}
where $\sigma=+1$ for spin up and $\sigma=-1$ for spin down and we have written $H_{SO}$ as three different pieces because each has a different physical meaning.

By making the substitution $(S_z)_i = \frac{1}{2}\left(c^\dagger _{i,\uparrow} c _{i,\uparrow} -c^\dagger _{i,\downarrow} c _{i,\downarrow} \right)$ the first term above can be written as $2(M-4B)\sum_{i} (S_z)_i \equiv H_Z$ and therefore this piece of $H_{SO}$ functions exactly as one would expect a Zeeman like term to function. Furthermore, because $H_Z$ does not change the real space configuration of electrons on the lattice, it does not change the interaction energy $H_U+H_V$. The next term in Eq. ({\ref{SOreal}) describes electrons hopping on nearest neighbours with the spin dependent hopping amplitude $\sigma B$. The third and final term describes electrons hopping to a nearest neighbour site while having its spin flipped. The effective hopping amplitude in this interpretation depends on both the spin of the initial electron as well as the direction in which the hop occurs. 

With the above physical interpretation as our motive, we define a generalized hopping matrix as follows
\begin{equation}\label{tgen}
\hat{t}_{\sigma,\sigma'}(\vec{\delta}) = (\sigma B-t)\delta_{\sigma,\sigma'} -\frac{Ai}{2}\vec{\delta}\cdot(\hat{x}-i\sigma\hat{y}) \delta_{\sigma',\bar{\sigma}}
\end{equation}  
then the tight binding kinetic energy and the spin-orbit coupling term in the original Hamiltonian can be written as follows
\begin{equation}
T+H_{SO} = \sum_{i,\delta,\sigma,\sigma'} c^\dagger_{i,\sigma} \hat{t}_{\sigma,\sigma'}(\vec{\delta})c_{i+\delta,\sigma'}+H_Z.
\end{equation}
We will call the first part of the above $H_1$, that is $H_1\equiv T+H_{SO}-H_Z$, or
\begin{equation}\label{h1}
H_1 = \sum_{i,\delta,\sigma,\sigma'} c^\dagger_{i,\sigma} \hat{t}_{\sigma,\sigma'}(\vec{\delta})c_{i+\delta,\sigma'}.
\end{equation}
$H_1$ as defined above will play the same role as $T$ in the expansion outlined in the previous section of this chapter. $H_1$ incorporates spin-orbit coupling into the model but at the same time still only describes electrons hopping on nearest neighbours (albeit with spin and directional dependence to the hopping amplitude). As with $T$ in the case of the Hubbard model, it will be useful to have a description of $H_1$ in terms of channels that change the interaction energy by a set amount.  

\subsection{Channel Decomposition of $H_1$}

Just as a key step in the expansion of the Hubbard model was to write $T$ in pieces that changed $H_U$ by a set amount so too will be writing $H_1$ in pieces that change the interaction $H_U+H_V$ by a set amount. We begin by considering pieces of $H_1$ that change $H_U$ by a set amount. In complete analogy to the Hubbard model we define   

 \begin{eqnarray}
  \ {T}_{-1} &=& \sum_{i,\sigma,\delta, \sigma'} h_{i,\bar{\sigma}} c^\dagger_{i,\sigma} \hat{t}_{\sigma,\sigma'}(\vec{\delta})c_{i+\delta,\sigma'} n_{i+\delta,\bar{\sigma}'} \\ \nonumber 
 \ {T}_{1} &=& \sum_{i,\sigma,\delta, \sigma'} n_{i,\bar{\sigma}} c^\dagger_{i,\sigma} \hat{t}_{\sigma,\sigma'}(\vec{\delta})c_{i+\delta,\sigma'} h_{i+\delta,\bar{\sigma}'} \\ \nonumber 
     \ {T}_{0} &=& \sum_{i,\sigma,\delta, \sigma'} (n_{i,\bar{\sigma}} c^\dagger_{i,\sigma} \hat{t}_{\sigma,\sigma'}(\vec{\delta})c_{i+\delta,\sigma'}  n_{i+\delta,\bar{\sigma}'} \\ \nonumber &&+ h_{i,\bar{\sigma}} c^\dagger_{i,\sigma} \hat{t}_{\sigma,\sigma'}(\vec{\delta})c_{i+\delta,\sigma'}  h_{i+\delta,\bar{\sigma}'}) 
 \end{eqnarray}
In the above $T_m$ will increase the value of $H_U$ by an amount $mU$ and one can show $[H_U,T_m]=mU_0T_m$. 

The next term to deal with is the nearest neighbour interaction term $H_V$.  In analogy with what we did above for $H_U$, the channels of $H_1$ we wish to find must change $H_V$ by a fixed amount. As such we look for the parts of $H_1$ that change the number of nearest neighbours on the lattice from $N_1$ to $N_2$. This is formally done by defining the nearest neighbour projection operator 
 \begin{eqnarray}
 {O}_i(\tilde{n}_{x,\uparrow}, \tilde{n}_{x,\downarrow},...\tilde{n}_{-y,\downarrow}) &=& \prod_{\delta,\alpha} (\tilde{n}_{\delta,\alpha}n_{i+\delta, \alpha}+(1-\tilde{n}_{\delta,\alpha})h_{i+\delta,\alpha}) \nonumber \\  &\equiv& {O}_i[n]
 \end{eqnarray}
which projects out all possible orientations of the nearest neighbours of atom site $i$ except the one labelled by $n$ where the arguments $\tilde{n}_{\delta,\alpha}$ take a value of 1 (a value of 0) when site $i+\delta$ is occupied (unoccupied). Inserting the identity $1=\prod_{\delta, \alpha} (n_{i+\delta,\alpha}+h_{i+\delta,\alpha})$ on either side of the summand in the channels ${T}_m$ (which we will denote $({T}_m)_{i,\delta,\sigma,\sigma'}$) we obtain the full decomposition operator
\begin{equation}\label{Tmnndef}
\ {T}_{m,N_2,N_1} =\small{ \sum_{i,\delta,\sigma, \sigma'} \sum_{S[n_1]=N_1} \sum_{S[n_2]=N_2} O_i[n_2] ({T}_m)_{i,\delta,\sigma, \sigma'}O_{i+\delta}[n_1] }
\end{equation}
where $S[n] \equiv \sum_{\delta, \alpha} n_{\delta,\alpha}$. The operator above now corresponds to a process where an electron hops from a site with $N_1$ occupied nearest neighbours to one with $N_2$ occupied nearest neighbours while changing the number of doubly occupied sites by $m$. An example of one possible process for the specific operator $T_{1,3,4}$ is illustrated in Figure \ref{fig:hop1}. A very tedious but straightforward calculation (left to the appendix) can be done to show that $ [\ {T}_{m,N_2,N_1}, H_U+H_V] = (mU_0+(N_2-N_1)V_0)\ {T}_{m,N_2,N_1}$. With these channels properly defined the strong coupling expansion can now be performed. 

Before moving on to the next section we feel it is worthwhile to clarify a potentially confusing point in the definition of the $T_{m,N_2, N_1}$ operators. The reader may be concerned of our claim that the $T_{m,N_2, N_1}$ operators change the value of $H_{V}$ by an amount $(N_2-N_1)V_0$ which only depends on the change in nearest neighbours of the electron we move from site $i$. One may argue that the electrons on sites neighbouring site $i$ have also lost nearest neighbours and hence should contribute to the change in the value of $H_{V}$. To remedy this confusion, we would like to recall that in our definition of $H_V$ the nearest neighbour sum $ \sum_{\langle i,j\rangle}$ counts a particular bond $i-j$ {\em only once}. If we were to take both the change in nearest neighbours of the electron at site $i$ and the change in nearest neighbours of the nearest neighbours to $i$ into account we would in fact be double counting the contribution of each nearest neighbour term. Therefore it is sufficient to concern ourselves with only the change in nearest neighbours of the electron that we are moving. 


\begin{figure}[tb]
 \begin{center}
  \setlength{\unitlength}{1mm}

   \includegraphics[scale=.65]{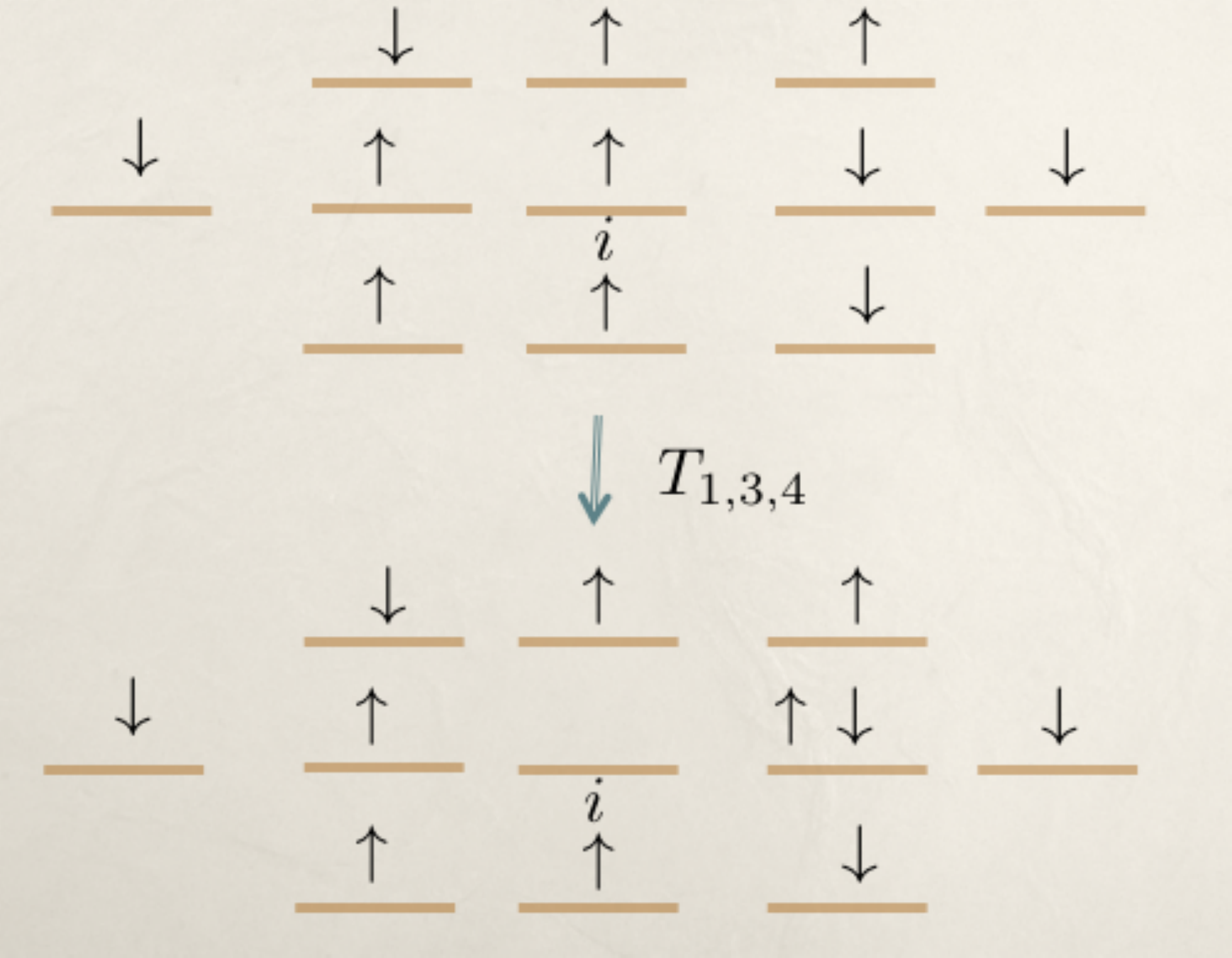}
   \end{center}
\caption{{\small
One possible process that could contribute to $T_{1,3,4}$. This image should be looked at as a bird's eye view of a square lattice. We have drawn site $i$, its 4 nearest neighbours, its 4 next nearest neighbours and two of its next, next nearest neighbours in order to fully illustrate this process.  
     }}\label{fig:hop1}
\end{figure}

\subsection{The expansion}
In order to begin the expansion we first define $H_0=H_U+H_V$ so that the full Hamiltonian now reads
\begin{equation}
H = H_1+H_0+H_Z.
\end{equation}
We again apply a unitary transformation to the electron operators $c^\dagger_{i,\sigma} = e^{iS} \bar{c}_{i,\sigma}^\dagger e^{-iS}$ but this time we wish to make $H_0$ a constant of motion. Applying this transformation gives us the following
\begin{equation}
\tilde {H} = e^{iS} (H_1+H_0+H_Z) e^{-iS} \equiv H_0+H_1'+H_Z
\end{equation}
where $H_Z$ does not change under this transformation because $[H_U+H_V, H_Z]=0$ and therefore $H_Z$ does not change $H_0$. In order for $H_0$ to be a constant of motion we require that
\begin{equation}\label{constant}
[H_0, \tilde {H}] = [H_0, H'_1]=0.
\end{equation} 
This time we expand both $H'_1$ and the transformation $S$ in a power series of $U_0$ as follows
\begin{equation}
S=-i\sum_{n=1}^\infty \frac{S_n}{U_0^n}, \ \ \ \ \ \ \ H_1 = \sum_{n=1}^{\infty} \frac{H'_{1,n}}{U_0^{n-1}}.
\end{equation}

Inserting the expansions for $S$ and $H_1'$ into the above and defining $H_0=U_0\tilde {H}_0$ we find
\begin{eqnarray}\label{conds}
&&H'_{1,1} = H_1 + [S_1, \tilde {H}_0]  \\ \nonumber 
&&H'_{2,1} = [S_1, {H}_1] + [S_2, \tilde {H}_0] +\frac{1}{2}[S_1,[S_1, \ {H}_0]]
\end{eqnarray}
The first condition we must satisfy is then
\begin{equation}
0 = [\ {H}_0, H_1 + [S_1, \tilde {H}_0]]
\end{equation} 
Inserting the channel decomposition $H_1 = \sum_{m,N_2,N_1}{T}_{m, N_2,N_1}$ and using the fact that $[\tilde {H}_0, H_1] =  \frac{1}{U}\sum_{m,N_2,N_1}(mU+(N_2-N_1)V)\ {T}_{m, N_2,N_1}$ the above is then solved by
\begin{equation}
S_1 = \sum'_{m,M,N} \frac{U_0\ {T}_{m, N_2,N_1}}{mU_0-(N_2-N_1)V_0}
\end{equation}
where the primed sum excludes terms for which $mU_0-(N_2-N_1)V_0=0$. Using this expression in the first order Hamiltonian we have
\begin{equation}
H'_{1,1} = \sum_{m,N_2,N_1}\ {T}_{m, N_2,N_1} -  \sum'_{m,N_2,N_1}\ {T}_{m, N_2,N_1} = \sum_N \ {T}_{0,N,N}
\end{equation}
This tells us to $\mathcal{O}(1)$ the piece of the full Hamiltonian that does not change $H_U+H_V$ involves the sum of all channels with hops that do not change the number of double occupancies and begin and end with $M$ occupied nearest neighbour bonds. We see that none of these operations change $H_U$ or $H_V$ justifying the above expression for $H'_{1,1}$.

Next we have to solve $[H_0, H'_{2,1}]=0$ by properly choosing $S_2$. The result of this is the following expression for $H'_{1,2}$
\begin{equation}
H'_{1,2} = \sum_{m,M_1, N_1, M_2}' \frac{U_0 [\ {T}_{m,M_1, N_1}, \ {T}_{-m, M_2, M_1+M_2-N_1}]}{2(mU_0+(M_1-N_1)V_0)}
\end{equation}
where again the sum is over all indices except ones for which $mU+(M_1-N_1)V=0$. This term is quadratic in the $T_{m,N_2,N_1}$ and the specific combination of the $T_{m,N_2,N_1}$ operators does not change the number of doubly occupied sites nor the number of occupied nearest neighbour bounds. We can see this by noting that ${T}_{-m, M_2, M_1+M_2-N_1}$ decreases the number of double occupied sites by $m$ and changes the number of occupied nearest neighbour bounds by $N_1-M_1$ and then ${T}_{m,M_1, N_1}$ increases the number of doubly occupied sites by $m$ and changes the number of occupied nearest neighbours by $M_1-N_1$. 

Using the two results above we can now write an effective Hamiltonian which is valid to $\mathcal{O}(1/U_0^2)$
\begin{eqnarray}\label{strongH}
\tilde {H} &=& H_U+H_V + H_Z + \sum_{M} \ {T}_{0,M,M} \\ \nonumber &+& \sum_{m,M_1,N_1,M_2}' \frac{ {T}_{m,M_1, N_1}  {T}_{-m, M_2, M_1+M_2-N_1}}{(mU_0+(M_1-N_1)V_0)}+\mathcal{O}(1/U_0^2),
\end{eqnarray}
where we made use of index relabeling symmetry to get rid of the commutator and the factor of a half in the second term of the expansion. In principal we could keep going to higher orders in $1/U_0$. The next order term would involve cubic combinations of the $T_{m,N_2,N_1}$ operators that collectively do not change the number of doubly occupied sites nor the number of occupied nearest neighbour bonds. We stop here for practicality as these higher order terms become very complicated to write down and are not needed to accomplish our current goal. 

Depending on both the type of lattice we work with as well as the doping of the system, certain symmetry arguments may be used to limit the powers of the $T_{m,N_2,N_1}$ operators at higher orders in this expansion. In particular, for the current problem ({\it i.e.} a square lattice) if we consider only systems at half filling then an argument can be made to limit the expansion to only even powers of  $T_{m,N_2,N_1}$. To see this consider taking matrix elements of the $T_{m,N_2, N_1}$ in a subspace of half-filled, singly occupied states. In this case, the net result of any power in the expansion on $\tilde{H}$ acting on one of these states must end up putting an electron back on the original lattice site it removed one from. For a square lattice it is impossible for this to happen using an odd number of the $T_{m,N_2,N_1}$ operators and therefore only even powers of these operators contribute to the expansion of $\tilde{H}$. To understand why this is the case it is useful to consider the terms in $\tilde{H}$ with two powers of $T_{m,N_2,N_1}$ as well as those with three powers of $T_{m,N_2,N_1}$. With two powers of the $T_{m,N_2,N_1}$ operators we start by moving an electron to a nearest neighbour site thereby making this site doubly occupied and the original site empty. The second  $T_{m,N_2,N_1}$ can then move one of the two electrons on this doubly occupied site back to the original lattice site thereby retaining the half-filled, singly occupied nature of the many-body configuration. On the other hand, three powers of the $T_{m,N_2,N_1}$ cannot accomplish this. Like the quadratic case, the first power must create a doubly occupied site. After the first power has acted there are numerous things the second power of  $T_{m,N_2,N_1}$ can do to the electron configuration, however none of these puts the state in a position for the third power to move us back in a half-filled, singly occupied state. For example, the second power could move one of the electrons on the doubly occupied site back to the original site, but then after this the third power will create a doubly occupied site again. 

We note that another way to view this strong coupling expansion is as a kind of perturbation theory. In this sense Eq. (\ref{strongH}) would correspond to second order perturbation theory. Taking this point of view for the time being, the energies of the ``virtual" states in this perturbation theory would then be $E_v = mU_0+(N_1-N_2)V_0$.  As these energies depend on a combination of both $U_0$ and $V_0$ one may be concerned with how ``small" these energies are and thus how appropriate the expansion is. Put another way, we might worry that although both $U_0$ and $|V_0|$ might independently be large, perhaps the combination in $E_v$ is small owing to some cancelling. To put these concerns to rest we use the fact that this thesis is most interested in systems at, or near, half-filling. Near half filling hopping processes where $M_1>N_1$ and $m=-1$ are most common. Inspecting the second order term in Eq. (\ref{strongH}), these correspond to two hop processes that first create a double occupancy (first hop) and then destroy it (second hop) in the meantime increasing the number of nearest neighbours on the second hop. These processes give $\tilde{E}_v = -(U_0-(M_1-N_1)V_0)$. For $V_0<0$ and $M_1>N_1$ we see that these most common hops lead to an {\em enhancement} of $E_v$ making $1/E_v$ smaller compared to the energies of the bare Hubbard model ({\it i.e. }$V_0=0$). Processes where $M_1<N_1$ are rare and despite the fact that they may lead to smaller values of $E_v$ they are statistically suppressed when acting on states near half-filling and are thus still ``small" 


\section{Projection to Half-Filling: An analogy to the $t-J$ model}
This section will give a projection of Eq. (\ref{strongH}) valid near half filling.  For illustrative purposes we begin by completing our derivation of the $t-J$ model from the Hubbard model. With this example in place we then outline the steps needed to derive an analogous effective model for our Hamiltonian. The projection we will outline is useful for obtaining both a more physical picture of the system and an effective model for the system that can be further probed. Physically, the projection will show us that some sort of magnetic order is present in the $\mathcal{O}(t/U)$ term in $\tilde{H}$. Meanwhile, from a calculation standpoint, this projection will isolate the parts of $\tilde{H}$ that are relevant near half filling. The next chapter of this thesis will use a renormalized version of the projection outlined here in order to study various phases of $\tilde{H}$.  

Our starting point in this discussion is Eq. (\ref{stronghub}). In the limit of very strong $H_U$ a system at half-filing or hole doped away from half filling, {\it i.e.} $\langle n \rangle = 1-x$ for $x\ge0$, will arrange itself in such a way that no doubly occupied lattice sites are present. If no double occupancies exist on the lattice then the number of double occupancies cannot be decreased and therefore any terms in Eq. (\ref{stronghub}) with $T_{-1}$ acting first immediately project to zero. Writing out what is left in the order $1/U$ term in (\ref{stronghub}) we have
\begin{equation}
H_{J} = -\frac{t^2}{U} \sum_{i_1, \delta_1, i_2, \delta_2, \sigma, \sigma'}h_{i_2,\bar{\sigma}} c^\dagger_{i_2,\sigma}c_{i_2+\delta_2,\sigma} n_{i_2+\delta_2,\bar{\sigma}}  n_{i_1,\bar{\sigma}'} c^\dagger_{i_1,\sigma'}c_{i_1+\delta_1,\sigma'} h_{i_1+\delta_1,\bar{\sigma}'}
\end{equation}

Let us now consider each term in the above (from right to left) as it acts on a state that is in the subspace of singly occupied, half-filled states. First $ h_{i_1+\delta_1,\bar{\sigma}'}$ acts and on the subspace of states we are interested in this term can be dropped. The rational behind this is that if state $ (i_1+\delta_1,\bar{\sigma}')$ is not vacant state $i_1+\delta_1,{\sigma}'$ will be and the next term in the expression, namely $c_{i_1+\delta_1,\sigma'}$, will give zero anyways. Next,  $c^\dagger_{i_1,\sigma'}c_{i_1+\delta_1,\sigma'} $ moves the electron to site $i_1$. The projections $n_{i_2+\delta_2,\bar{\sigma}}  n_{i_1,\bar{\sigma}'}$ can also be dropped for a similar reason as to why we could drop $ h_{i_1+\delta_1,\bar{\sigma}'}$.  After this, the operator $c^\dagger_{i_2,\sigma}c_{i_2+\delta_2,\sigma}$ acts, moving an electron from $i_2+\delta_2$ to $i_2$. We now note that if we wish to begin and end in the singly occupied, half-filled subspace then we require  $i_2=i_1+\delta_1$ and $i_1=i_2+\delta_2$. The reason why is because the action of $ c^\dagger_{i_1,\sigma'}c_{i_1+\delta_1,\sigma'} $ creates a double occupancy and therefore $c^\dagger_{i_2,\sigma}c_{i_2+\delta_2,\sigma}$ must destroy this double occupancy in order to maintain $\tilde{D}=0$. We therefore have (after reorganizing the electronic operators and ignoring a term that will be constant in the half-filled subspace) 
\begin{equation}
H_{J} = \frac{t^2}{U} \sum_{i, \delta, \sigma, \sigma'}c^\dagger_{i,\sigma}c_{i,\sigma'}c^\dagger_{i+\delta,\sigma'}c_{i+\delta,\sigma}  
\end{equation}
Writing out the terms in the two spin sums above and comparing to the electron spin operators $\Sv_i$ defined at the beginning of this chapter one can easily verify that (again dropping a constant term)
\begin{equation}
H_{J} = \frac{4t^2}{U} \sum_{\langle i, j\rangle }\Sv_i\cdot \Sv_j
\end{equation}

We  now obtain the $t-J$ model as follows. First, we write out the $\mathcal{O}(1)$ term in the expansion ({\it i.e.} $T_0$). Second, we replace the two order $1/U_0$ terms in (\ref{stronghub}) with $H_{J}$ defined above. Lastly, we neglect $H_U$ as it will be zero in the subspace of singly occupied, many-body states. Following this prescription we obtain
\begin{equation}
H_{tJ} = -t\sum_{\langle i,j\rangle, \sigma} P_G c_{i, \sigma}^\dagger c_{j,\sigma} P_G +J\sum_{\langle i, j\rangle }\Sv_i\cdot \Sv_j,
\end{equation}
where $J=4t^2/U$ and we have introduced the Gutzwiller projection operator $P_G$ which insists a state have no double occupancies. The above model should be viewed as an effective theory for Eq. (\ref{stronghub}) valid {\em near} half filling. Notice that for systems at half filling the first term above projects to zero and we are simply left with the Heisenberg model $H_J$.  With this example concretely in place let us proceed to the analogous projection of our present model.

We again begin by considering some state $|\psi\rangle$ at half-filling. For strong $U_0$ the electrons will avoid configurations with double occupancies and the state $|\psi\rangle$ will consist of a lattice of singly occupied sites. Let us consider the action of the following term on this subspace of possible states $|\psi\rangle$
\begin{equation}
 \sum_{m,M_1,N_1,M_2}' \frac{ {T}_{m,M_1, N_1}  {T}_{-m, M_2, M_1+M_2-N_1}}{(mU_0+(M_1-N_1)V_0)}.
\end{equation} 
To begin with $|\psi\rangle$ has only singly occupied sites and so we can neither maintain the number of doubly occupied sites ($-m=0$), nor decrease it ($-m=1$). Therefore the only possible value of $-m$ is $-m=1$. Next, all sites in $|\psi\rangle$ have $4$ occupied nearest neighbours and so $M_1+M_2-N_1=4$. After a hop has occurred the electron will be on a site with $3$ nearest neighbours and so $M_2=3$. If we want to begin and end in a subspace of all possible half-filled states with no double occupancies the second part of the term above, namely $ {T}_{m,M_1, N_1}$, must now destroy the doubly occupied site ${T}_{-m, M_2, M_1+M_2-N_1}$ has created. For this to be the case we must set $m=-1$ and $N_1=3$ from which it follows that $M_1=4$. Plugging in all of these results we get the following projection to half-filled, singly occupied states
\begin{equation}
 \sum_{m,M_1,N_1,M_2}' \frac{ {T}_{m,M_1, N_1}  {T}_{-m, M_2, M_1+M_2-N_1}}{(mU+_0(M_1-N_1)V_0)}\to -\frac{ {T}_{-1,4, 3}  {T}_{1, 3,4}}{U_0-V_0} 
\end{equation}
Taking all of our projections into account and inserting the definition of $T_{m,N_2,N_1}$ into the above we then find
\begin{equation}\label{spin2}
 \frac{ {T}_{-1,4, 3}  {T}_{1, 3,4}}{U_0-V_0} =\small{\sum_{i,\delta,\sigma, \sigma', \alpha,\alpha'}\frac{c^\dagger_{i+\delta,\alpha} \hat{t}_{\alpha,\alpha'}(-\vec{\delta})c_{i,\alpha'} c^\dagger_{i,\sigma} \hat{t}_{\sigma,\sigma'}(\vec{\delta})c_{i+\delta,\sigma'}}{U_0-V_0}  }
\end{equation}
where we have dropped all projection operators for reasons analogous to why they were dropped in our presentation of the Hubbard model. 

The spin sums in the above term can again be eliminated if we write things in terms of the lattice spin operators $(\Sv)_i$. Recalling the structure of the $\hat{t}_{\sigma,\sigma'}(\vec{\delta})$ hopping element we see that there are four terms that can contribute to the spin sums in Eq. (\ref{spin2}). These contributions correspond to $\sigma=\sigma'$ with $\alpha=\alpha'$, $\sigma=\bar{\sigma}'$ with $\alpha=\alpha'$, $\sigma=\sigma'$ with $ \alpha=\bar{\alpha}'$ and $\sigma=\bar{\sigma}'$ with $ \alpha=\bar{\alpha}'$. Performing all of these spin sums is tedious but straightforward. Here we will sketch the process for the $\sigma=\bar{\sigma}'$ with $ \alpha=\bar{\alpha}'$ sum and then give the result of doing the rest of the sums.

Setting $\sigma=\bar{\sigma}'$ and $ \alpha=\bar{\alpha}'$ in the summand of Eq. (\ref{spin2})  gives us the following
\begin{equation}
 \sum_{i,\delta,\sigma, \alpha,}\frac{A^2\vec{\delta}\cdot(\hat{x}-i\sigma\hat{y}) \vec{\delta}\cdot(\hat{x}-i\alpha\hat{y})}{4(U_0-V_0)}c^\dagger_{i+\delta,\alpha}c_{i,\bar{\alpha}} c^\dagger_{i,\sigma}c_{i+\delta,\bar{\sigma}}
\end{equation}
We first consider the case where $\vec{\delta}=\pm\hat{x}$. After making one commutation we have
\begin{equation}
 n_i-\sum_{\sigma, \alpha} c^\dagger_{i+\delta,\alpha}c_{i+\delta,\bar{\sigma}} c^\dagger_{i,\sigma}c_{i,\bar{\alpha}}
\end{equation}
where we have only written the sum over spins and for brevity we have dropped the prefactor $A^2/(4(U_0-V_0))$ which will be added back later. Writing out all four terms in the sum above, using the definition in Eq. (\ref{sitespin}) to identify spin operators, recalling that we are interested in the action of our Hamiltonian on the subspace of half-filled singly occupied states and doing a few straightforward manipulations  we arrive at
 \begin{equation}
 2\left(-(S_y)_{i+\delta}(S_y)_{i}+(S_x)_{i+\delta}(S_x)_{i}+(S_z)_{i+\delta}(S_z)_{i}+\frac{n_in_{i+\delta}}{4}\right).
\end{equation}

The rest of the calculation is completed in a manner similar to the one outlined above. In the end we arrive at the following result
\begin{equation}
 \frac{ {T}_{-1,4, 3}  {T}_{1, 3,4}}{U_0-V_0}  =E_0 +\sum_{i,\delta} \Sv_{i+\delta} J_\delta \Sv^T_{i}
\end{equation}
where we have defined the constant 
\begin{equation}
E_0 =-\frac{2Nt^2}{U_0-V_0} - \frac{NA^2}{2(U_0-V_0)}-\frac{2NB^2}{U_0-V_0}
\end{equation}
and the $3\times3$ matrix 
\begin{eqnarray}
 J_{\delta}= \bar{U}\left(\begin{matrix} 
      4t^2+\tilde{A}^2-4B^2 & 0& -4At\hat{y}\cdot \vec{\delta} \\
      0 & 4t^2-\tilde{A}^2-4B^2&4At\hat{x}\cdot \vec{\delta} \\
      4At\hat{y}\cdot \vec{\delta}&-4At\hat{x}\cdot \vec{\delta}&4t^2-A^2+4B^2\\
   \end{matrix}\right)
\end{eqnarray}
where $\tilde{A} =A(\vec{\delta}\cdot(\hat{x}+i\hat{y}))$ and $\bar{U} = \frac{1}{2(U_0-V_0)}$. Note that in the limit $V_0,A,B\to0$ the above result readily simplifies to the Heisenberg model for spins on a lattice as $J_\delta$ becomes the identity matrix times $2t^2/U$ and an additional factor of $2$ comes from writing the sum over $i$ and $\delta$ as a nearest neighbour sum. 

To complete our analogy to the $t-J$ model we drop the overall constant $E_0$ from the above result. Next, we note that near half filling and at strong coupling the terms $H_U$ and $H_V$ will be approximately constant and equal to $0$ and $2N$ respectively. This is because at strong coupling there will be no doubly occupied sites (hence $H_U=0$) and near half filling the number of nearest neighbours will not strongly fluctuate (hence $H_V$ constant). For the purposes of obtaining an effective model we ignore these constants. Taking all of this into account we obtain
\begin{equation}\label{tjd}
H_{t,J_\delta} = H_Z + \sum_{N} T_{0,N,N} +\sum_{i,\delta} \Sv_{i+\delta} J_\delta \Sv^T_{i}.
\end{equation}
Compared to the $t-J$ model we have several obvious differences. First, with spin-orbit coupling the exchange constant $J$ has now become a matrix. Second, our term $\sum_{M} T_{0,M,M}$ restricts both the number of double occupancies {\em and} the number of occupied nearest neighbours. Lastly, because of the Zeeman/mass term in the original model we have the magnetic like contribution $H_Z$.  

\section{Studying the Strongly Coupled Hamiltonian} 
The remainder of this thesis is dedicated to studying the strong coupling Hamiltonian given in Equation (\ref{strongH}). There are numerous methods for studying strong coupling Hamiltonians; here we will outline two distinct methods that have enjoyed success (in the context of the Hubbard model) in the past. Both of these paths involve methods that are based on forms of mean field theory and the motivation for both is as follows. We may wish to look for a certain type of order in the effective Hamiltonian (\ref{strongH}). As such we begin by considering a mean field ground state with this order built into it via some tunable order parameters, for illustrative purposes let us call this $|\psi_{MF}\rangle$. We now wish to let this wave function know that electron-electron interactions are the dominant energy scale in the problem. In order to encode strong correlations into this mean field ground state one then applies the so-called Gutzwiller projection operator $P_G$ to project out any part of the mean field ground state that contains doubly occupied sites. To illustrate this consider expanding $|\psi_{MF}\rangle$ in a basis of real space electron configurations $\{ |x\rangle\}$ where $x$ labels the possible position and spin of all electrons in the system. Then we have
\beq
|\psi_{MF}\rangle = \sum_x \langle x|\psi_{MF}\rangle |x\rangle 
\eeq
The action of the projection operator is then
\beq\label{trailstrong}
P_G|\psi_{MF}\rangle = \sum'_x \langle x|\psi_{MF}\rangle |x\rangle 
\eeq
where the primed sum now excludes real space configurations, $x$, with more than one electron on any lattice site. Equation (\ref{trailstrong}) now constitutes the ``trial" ground state we wish to use in a variational study. 

It is when one tries to actually evaluate to required variational energy $E_t=\langle \psi_{MF}|P_G HP_G|\psi_{MF}\rangle$ in order to minimize with respect to the order parameters built into $|\psi_{MF}\rangle$ that a choice of method must be made. The reason for this is that the Gutzwiller projection in this expectation value makes the variational energy impossible to evaluate by hand and so one must resort to other methods in order to complete this task. The first of our two methods evaluates $E_t$ numerically using Variational Monte Carlo techniques. For example, this is the method of choice in Reference \cite{param}. Although Monte Carlo methods have been refined to the point that they can practically be considered exact, the necessity to resort to numerical techniques to calculate various quantities of interest provides reduced intuition when compared to analytic approaches. As a result, our second method looks at approximately applying the Gutzwiller projection operator. In this method the Gutzwiller operator is exchanged for a density dependent ``renormalization" parameter. With this exchange of an operator for a number one can calculate the variational energy by hand in order to perform the required minimization. This method has become known as the ``Gutzwiller Approximation" and was developed several decades ago for the Hubbard model\cite{zhang}.

The proceeding two chapters look at generalizing and applying these two methods to the strong coupling Hamiltonian derived in this chapter. The next chapter develops a Gutzwiller approximation and derives renormalized mean field theory results. The penultimate chapter of this thesis then presents a variational Monte Carlo study of the strong coupling model Hamiltonian derived here.

\chapter{Gutzwiller Approximation}

\section{Gutzwiller Approximation for the Extended Hubbard Model}

We begin our discussion of the Gutzwiller approximation by studying the limit $A,B,M=0$ of the strong coupling model derived in the previous section. This will not only serve as a good illustration of how the method should be applied, but will also provide results we can compare with the traditional application of the Gutzwiller approximation to the t-J model. 

Our effective Hamiltonian of interest is then
\begin{eqnarray}\label{HtJV}
H_{U,V} &=&  H_U+H_V + \sum_{M} \ {T}_{0,M,M} \\ \nonumber &+& \sum_{m,M_1,N_1,M_2}' \frac{ {T}_{m,M_1, N_1}  {T}_{-m, M_2, M_1+M_2-N_1}}{(mU+(M_1-N_1)V)}.
\end{eqnarray}
As mentioned earlier in this thesis, we are concerned with systems that are at half filling or hole doped away from half filling. The methods developed here can in principle be extended to electron doped systems by performing a particle-hole transformation on the Hamiltonian\cite{weber}.  

For systems hole doped away from half filling we take $\langle n \rangle =1-x$ where $x\ge0$ is the concentration of holes in the system. We further assume that the system is completely unpolarized, {\it i.e.} that $\langle n_{\uparrow}\rangle =\langle n_{\downarrow} \rangle = \frac{1}{2}\langle n\rangle$.  We are interested in looking for superconductivity in this strong coupling Hamiltonian. A good starting point will then be to consider as our mean-field ground state the standard $BCS$ ground state
\begin{equation}
|\psi_{BCS}\rangle = \prod_{\kv} (u_\kv+v_{\kv}c^\dagger_{\kv,\uparrow} c^\dagger_{-\kv,\downarrow})|0\rangle 
\end{equation}
where $u_\kv = \sqrt{\frac{E_\kv+\xi_\kv}{2E_\kv}}$ and $v_\kv = e^{i\phi_\kv}\sqrt{\frac{E_\kv-\xi_\kv}{2E_\kv}}$ where $E_\kv = \sqrt{\xi_\kv^2 +|\Delta_{\kv}|^2}$ and $\phi_\kv$ is the phase of $\Delta_{\kv}$. The functions $\Delta_\kv$ and $\xi_\kv$ are the variational parameters we are interested in optimizing. To encode strong interactions on this ground state we apply the Gutzwiller projection operator $P_G$ to get the variational ground state $|\psi_{Var}\rangle = P_G |\psi_{BCS}\rangle$. The trail energy we must optimize is then
\begin{equation}
E_{var}= \frac{\langle \psi_{BCS} | P_G H_{U,V} P_G|\psi_{BCS}\rangle}{\langle \psi_{BCS}  |P_GP_G|\psi_{BCS}\rangle}
\end{equation}
      
 The above object is impossible to calculate analytically because of the presence of the $P_G$ operators. The goal of the Gutzwiller projection is to approximately apply these operators by replacing them with an appropriately chosen constant. When this is done the variational energy simply becomes an expectation value in the BCS ground state with renormalized parameters $t\to g_t t, J\to g_J J$; something that is tractable analytically.       
       
To begin this process we consider the expectation value of the hopping term given by the following
 \begin{eqnarray}\label{kg}
 K_G &=& \sum_N \frac{\langle \psi_{BCS} | P_G  T_{0,N,N}P_G|\psi_{BCS}\rangle}{\langle \psi_{BCS}  |P_GP_G|\psi_{BCS}\rangle}
\\ \nonumber &=&\sum_{N,i,\delta,\sigma} \sum_{S[n_1], S[n_2]=N} \frac{\langle \psi_{BCS} | P_G O_{i+\delta}[n_2] c^\dagger_{i+\delta, \sigma}c_{i, \sigma}O_{i}[n_1]P_G|\psi_{BCS}\rangle}{\langle \psi_{BCS}  |P_GP_G|\psi_{BCS}\rangle}
 \end{eqnarray}      
Let us consider the action of $O_{i+\delta}[n_2] c^\dagger_{i+\delta, \sigma}c_{i, \sigma}O_{i}[n_1]P_G$ on the BCS ground state. First $P_G$ removes all possible doubly occupancies. For a system at half filling (or slightly hole doped away from half filling) this projection means that the only contributions left in $|\psi_{BCS}\rangle$ are many body states with each lattice site either singly occupied or unoccupied. $T_{0,N,N}$ then moves an electron from an occupied site with $N$ nearest neighbours to an empty site with $N$ nearest neighbours. 

Now we wish to compare this highly restricted hopping process to the action of the normal kinetic energy operator, $c^\dagger_{i+\delta, \sigma}c_{i, \sigma}$, acting on just the BCS ground state. When this term acts, it finds a lattice site that is occupied by an electron of some spin and then moves it to a neighbour site that is not occupied by this particular spin. This process has considerably more degrees of freedom than the one described in the previous paragraph and because of this we expect the value of $K_G$ defined in Eq. (\ref{kg}) to be reduced compared to the expectation value of the normal kinetic energy operator in the (unprojected) BCS ground state.

In order to reflect this reduction in $K_G$ and at the same time remove the projection operator $P_G$ we must make an approximation. First, we let $K_{BCS}$ be the kinetic energy of the BCS ground state $K_{BCS}=\langle \psi_{BCS}|c^\dagger_{i+\delta, \sigma}c_{i, \sigma}|\psi_{BCS}\rangle/\langle \psi_{BCS}|\psi_{BCS}\rangle$. Second, we let $p(G)$ be the probability any particular hopping process described above for $K_G$ will be a success and $p(BCS)$ be the probability the hop $c^\dagger_{i+\delta, \sigma}c_{i, \sigma}$ occurs successfully in the BCS ground state. We now make the following approximate claim\cite{zhang}

\beq
\frac{K_G}{K_{BCS}}\simeq \frac{p(G)}{p(BCS)}
\eeq
The final statement of the Gutzwiller approximation for the kinetic energy $K_G$ is then the following
\beq
{K_G}\simeq \frac{p(G)}{p(BCS)}K_{BCS} \equiv g_t K_{BCS}
\eeq
We can think of the above renormalization process in a more physical way as follows. For a strongly interacting system having two electrons on the same lattice site is so energetically costly that the electrons will avoid many-body configurations where this happens. This avoidance of doubly occupied states greatly reduces the phase space of processes contributing to electron's kinetic energy. Therefore when we wish to study the model in mean field theory we should reflect this reduction by replacing $t\to g_tt$ in our expression for the kinetic energy. 

In order to calculate $g_t$ we must assign relative probabilities to the hopping processes occurring in $K_G$ and $K_{BCS}$. For the latter case (plain kinetic energy operator acting on a BCS ground state) the probability that the process occurs is just the probability that the lattice site the electron moves to is void of the particular spin we are moving. To assign a probability to this we assume that the probability a particular lattice site is occupied by a particular spin is independent of any other lattice site being occupied by a particular spin. The probability of the particular move we just described is then $p(BCS)=1-\langle n_{i,\sigma}\rangle= (1+x)/2$. 

For the case of the projected expectation value there are several concurrent requirements we must meet. First, we require the lattice site where the electron is moved to be empty, a probability of $x$ under Gutzwiller projection. Second, the site we move the electron from must have exactly $N_n$ occupied nearest neighbours sites. The site we are moving the electron to is a nearest neighbour to the site we are moving the electron from, because of this the number of occupied nearest neighbours must be 3 or less. The probability of having $N_n\le3$ occupied nearest neighbour sites is the binomial probability of having exactly $N_n$ successes in 3 trials with success probability $1-x$, in other words $\binom{3}{N_n}(1-x)^{N_n}x^{3-N_n}$. Finally, the site we move the electron to must also have exactly $N_n$ occupied nearest neighbours, this again occurs with probability $\binom{3}{N_n}(1-x)^{N_n}x^{3-N_n}$. Combining these three observations we find that the probability of the entire series of events occurring is $p(G)=x\sum_{N_n=0}^3 (\binom{3}{N_n}(1-x)^{N_n}x^{3-N_n})^2$. 

Appealing to the above discussion we can now write the following
\beq
g_t = \frac{p(G)}{p(BCS)} = \frac{2x}{1+x}\sum_{N_n=0}^3\left(\binom{3}{N_n}(1-x)^{N_n}x^{3-N_n}\right)^2
\eeq
We then have the following Gutzwiller approximation for the kinetic energy 
\begin{equation}\label{Kgut}
K_G \simeq 2g_t \sum_{\kv} \epsilon_{\kv} |v_\kv|^2
\end{equation}
where $\epsilon_\kv = -2t(\cos{k_x}+\cos{k_y})$ is the tight binding spectrum and we have used the standard BCS result $K_{BCS}=2\sum_{\kv} \epsilon_\kv |v_{\kv}|^2$. To get an idea of how well this approximation of the kinetic energy does at different doping levels and for different order parameter symmetries we have carried out Monte Carlo calculations in order to compare to the approximate result given by Equation (\ref{Kgut}). The results of these Monte Carlo calculations as well as the value of the kinetic energy $K_G$ are both plotted in Fig. \ref{fig:GA_compare} at doping values of $x\simeq 0.4878$ and $x\simeq 0.09756$ for $d$-wave and $s$-wave order parameter symmetries. Following the case made in Ref. \cite{zhang}, we contend that this approximation does a fair job at reproducing the quantitative values of $K_G$ and an excellent job at reproducing the qualitative behaviour of $K_G$ as it gets the general trend of the curve correct. These results are very satisfactory when we consider that we have reduced the complicated expectation value given in Eq. (\ref{kg}), which can only be calculated using sophisticated numerical methods, to the relatively simple result in Equation (\ref{Kgut}). The reader should note that the errors involved in the approximation for $K_G$ made here for the extended Hubbard model are comparable to those for the Hubbard model in Ref. \cite{zhang}.   

 
 \begin{figure}[h]
  \setlength{\unitlength}{1mm}

   \includegraphics[scale=.5]{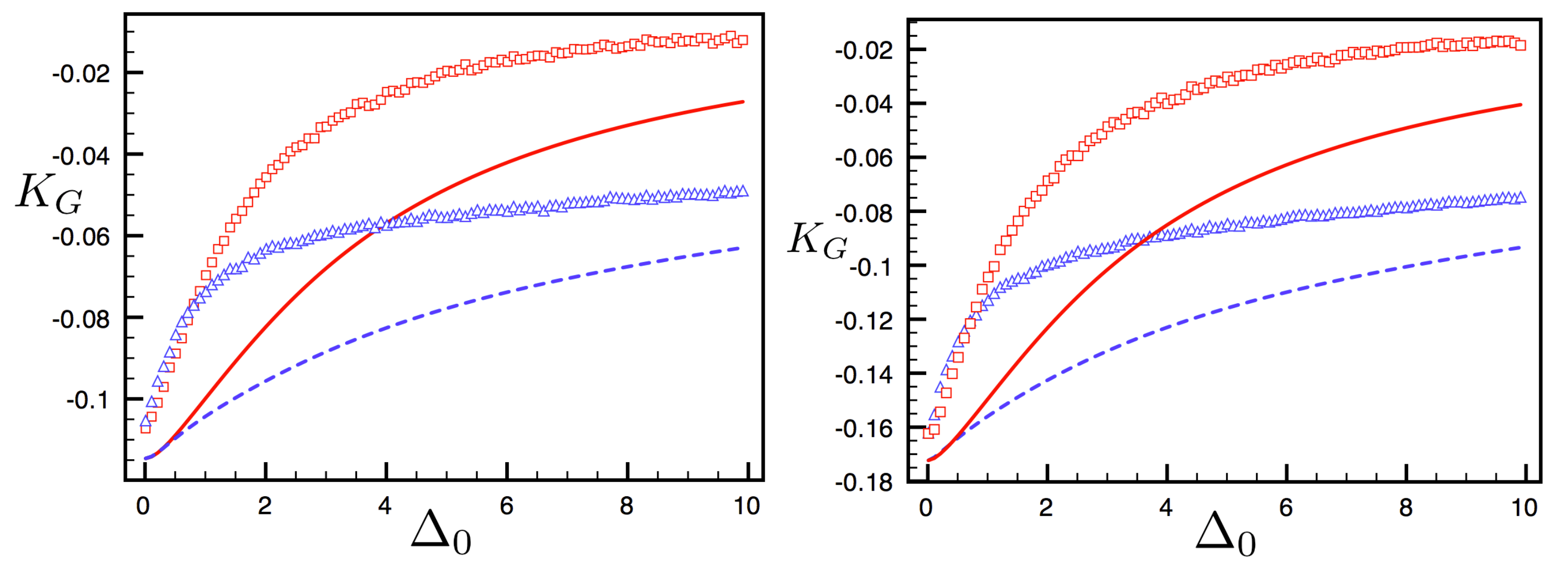}
\caption{{\small
Comparison of Monte Carlo and Gutzwiller approximation results for the kinetic energy $K_G$. Both of these plots are for $N=82$ lattice sites, the left plot is for a system with $N_H=4$ holes while the right plot is $N_H=8$ holes. At each doping level we have tried two different order parameter symmetries, $d$-wave and $s$-wave. For $d$-wave symmetry with $\Delta_\kv = \Delta_0(\cos(k_x)-\cos(k_y))$ we have plotted the Gutzwiller approximation given in Eq. (\ref{Kgut})  using a blue dashed line and numerical Monte Carlo results using blue triangles. For $s$-wave symmetry we have $\Delta_\kv = \Delta_0$ and have plotted the Gutzwiller approximation using a red solid line and numerical Monte Carlo results using red squares. In both figures we have set $t=1$ and $\xi_\kv = \epsilon_\kv -\mu$ where $\epsilon_\kv$ is defined in the text and $\mu$ is the chemical potential.  We have also scaled the kinetic energy by the number of lattice sites, $N$.
     }
     }\label{fig:GA_compare}
\end{figure}

Now we move on the to Gutzwiller approximation for the second order term in $H_{U,V}$. We are faced with the following
\begin{equation}\label{defVg}
V_G= \sum_{m,M_1,N_1,M_2}' \frac{ 1}{(mU_0+(M_1-N_1)V_0)}  \frac{\langle \psi_{BCS} | P_G {T}_{m,M_1, N_1}  {T}_{-m, M_2, M_1+M_2-N_1}P_G|\psi_{BCS}\rangle}{\langle \psi_{BCS}  |P_GP_G|\psi_{BCS}\rangle} 
\end{equation} 
Let us first consider the specific values that $m$ can take. We are interested in systems with hole doping $x\ge0$ and so after $P_G$ is applied to $|\psi_{BCS}\rangle$ there are only singly occupied or empty sites left on the lattice. This means that $-m$ can never be $-1$; for a hole doped system it is impossible to reduce the number of doubly occupied sites. This leaves the possibility for $m$ to be either $-1$ or $0$. Next, in order to keep things simple we will ignore the $m=0$ terms above as these terms require the movement of two electrons to two empty sites and so are a factor of $x^2$ less probable than their $m=-1$ counterparts. 

Taking the above considerations into account leaves only the $m=-1$ terms in Eq. (\ref{defVg}). Let us consider what it is the term ${T}_{-1,M_1, N_1}  {T}_{1, M_2, M_1+M_2-N_1}$ does. First, it creates a doubly occupied site and then it destroys one. In order to remain in the subspace of no doubly occupied sites it must actually destroy the same doubly occupied site it creates. This can happen in two ways: (1) an electron is hopped to an occupied site and then one of the two electrons on this site is moved back to the original site or, (2) an electron is moved to an occupied nearest neighbour and then one of the two electrons on this site are moved to an empty next-nearest neighbour site. Keeping with our motive of simplicity, we will ignore the latter ``three site" move as it is less probable near half filling. Taking all of these concerns into account leaves the following
\begin{equation}\label{defVg3}
V_G= \sum_{M_1,N_1} \frac{\langle \psi_{BCS} | P_G {T}_{-1,M_1, N_1}  {T}_{1 N_1, M_1}P_G|\psi_{BCS}\rangle}{((M_1-N_1)V_0-U_0)\langle \psi_{BCS}  |P_GP_G|\psi_{BCS}\rangle} 
\end{equation} 

In order to develop a Gutzwiller approximation for the above quantity we again replace the average in the state $P_G|\psi_{BCS}\rangle$ with an average in the state $|\psi_{BCS}\rangle$ times the ratio of the probability for the process to occur in each state. To derive expressions for these probabilities we begin by considering the case where the $T$ operators do nothing but move an electron (indiscriminate of the number of nearest neighbours) and $P_G$ is ignored. This BCS limit involves an electron being moved to an occupied site and then one of the two electrons at the site being moved back. In order for this to occur both sites must only be occupied by a specific flavour of spin, let us say spin up and spin down for concreteness. The probability of having a site only occupied by a spin up electron and a site only occupied by a spin down electron is then $p_V(BCS)=\left(\frac{1-x}{2}\right)\left(\frac{1+x}{2}\right)\left(\frac{1-x}{2}\right) \left(\frac{1+x}{2}\right)$. Now we consider the probability with all of the appropriate projections. Consider the net result of $P_G {T}_{-1,M_1, N_1}  {T}_{1 N_1, M_1}P_G$ . First we hop an electron, let's say spin up, from a site with $M_1$ nearest neighbours to a site occupied by a spin down electron with $N_1$ nearest neighbours. Given the specifics of this, $M_1$ can range from $1$ to $4$ while $N_1$ can take values from $0$ to $3$. The probability of this happening is then $\left(\frac{1-x}{2}\right)^2 \binom{3}{N_1}(1-x)^{N_1} x^{3-N_1}\binom{4}{M_1}(1-x)^{M_1}x^{4-M_1}$. 

Weighting these two probabilities against each other, our approximation amounts to the following expression for $V_G$ 

\begin{equation}\label{defVg2}
V_G\simeq -\frac{t^2}{\tilde{U}}\sum_{i,\delta,\sigma,\sigma'} {\langle \psi_{BCS} | c_{i+\delta,\sigma}^\dagger c_{i,\sigma} c_{i,\sigma'}^\dagger c_{i+\delta, \sigma'}|\psi_{BCS}\rangle}
\end{equation} 
 where to make connection with the $t-J$ model we have defined the renormalized coupling $\tilde{U}$ given by

 \begin{equation}\label{exJ}
 \tilde{U}^{-1} = \frac{4}{U_0(1+x)^2} \sum_{N_1=0}^3 \sum_{ M_1=1}^4 \frac{(1-x)^{N_1} x^{3-N_1}(1-x)^{M_1}x^{4-M_1}}{1-(M_1-N_1)v}  \binom{3}{N_1}\binom{4}{M_1}. 
 \end{equation}
where $v\equiv V_0/U_0$. 

In the approximation for $V_G$ above we must isolate the parts of  $c_{i+\delta,\sigma}^\dagger c_{i,\sigma} c_{i,\sigma'}^\dagger c_{i+\delta, \sigma'}$ that are relevant at or near half filling. To do this we rewrite this object in terms of spin operators $\Sv_i$ and a nearest neighbour contribution $n_in_{i+\delta}$. We drop the nearest neighbour contribution because near half filling and with the constraint that all sites must be singly occupied the number of occupied nearest neighbour bounds will not strongly depend on any variational parameter. This is consistent with dropping the $-\sum_{\langle i,j\rangle} n_in_j$ term in the traditional treatment of the $t-J$ model. The result of this rewriting is

\begin{equation}\label{defVg1}
V_G\simeq -\tilde{J}\sum_{\langle i,j\rangle} {\langle \psi_{BCS} | \Sv_i \cdot \Sv_j|\psi_{BCS}\rangle}
\end{equation} 
where we have defined the renormalized exchange constant $\tilde{J}=4t^2/\tilde{U}$.  The value of $V_G$ can now be evaluated in terms of the BCS ground state and is given by
\begin{equation}\label{vg}
V_G \simeq -\frac{3\tilde{J}}{4N} \sum_{\kv, \kv'} W_{\kv,\kv'}\left(|v_\kv|^2|v_{\kv'}|^2+u_\kv v_{\kv}^*u_{\kv'}^*v_{\kv'}\right)
\end{equation}
where $W_{\kv,\kv'} = 2\left(\cos(k_x-k_x')+\cos(k_y-k_y')\right)$. The plots in Fig. \ref{fig:GA_compare2} compare this approximation to Monte Carlo calculations in the same way Fig. \ref{fig:GA_compare} does for the kinetic energy $K_G$. We again see that the qualitative and quantitative accuracy of this approximation are satisfactory considering the great amount of simplification involved in moving from Eq. (\ref{defVg}) to Equation (\ref{vg}).

 \begin{figure}[h]
  \setlength{\unitlength}{1mm}

   \includegraphics[scale=.65]{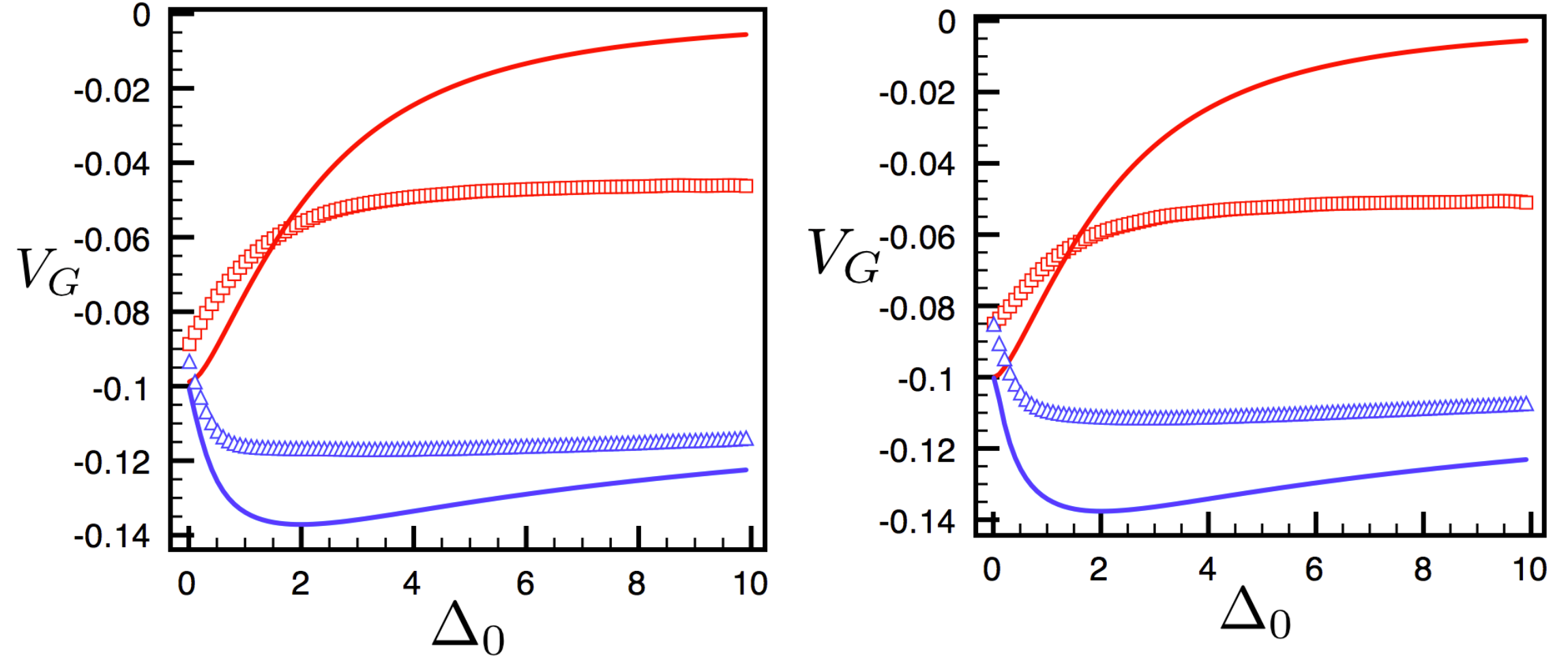}
\caption{{\small
Comparison of Monte Carlo and Gutzwiller approximation results for the magnetic energy $V_G$. Both of these plots are for $N=82$ lattice sites, the left plot is for a system with $N_H=4$ holes while the right plot is $N_H=8$ holes. Following Fig. \ref{fig:GA_compare}, at each doping level we have tried two different order parameter symmetries, $d$-wave and $s$-wave. For $d$-wave symmetry with $\Delta_\kv = \Delta_0(\cos(k_x)-\cos(k_y))$ we have plotted the Gutzwiller approximation given in Eq. (\ref{vg})  using a blue dashed line and numerical Monte Carlo results using blue triangles. For $s$-wave symmetry we have $\Delta_\kv = \Delta_0$ and have plotted the Gutzwiller approximation using a red solid line and numerical Monte Carlo results using red squares. In both figures we have set $t=1$ and $\xi_\kv = \epsilon_\kv -\mu$ where $\epsilon_\kv$ is defined in the text and $\mu$ is the chemical potential.  
     }
     }\label{fig:GA_compare2}
\end{figure}

 \begin{figure}[h]
  \setlength{\unitlength}{1mm}

   \includegraphics[scale=.6]{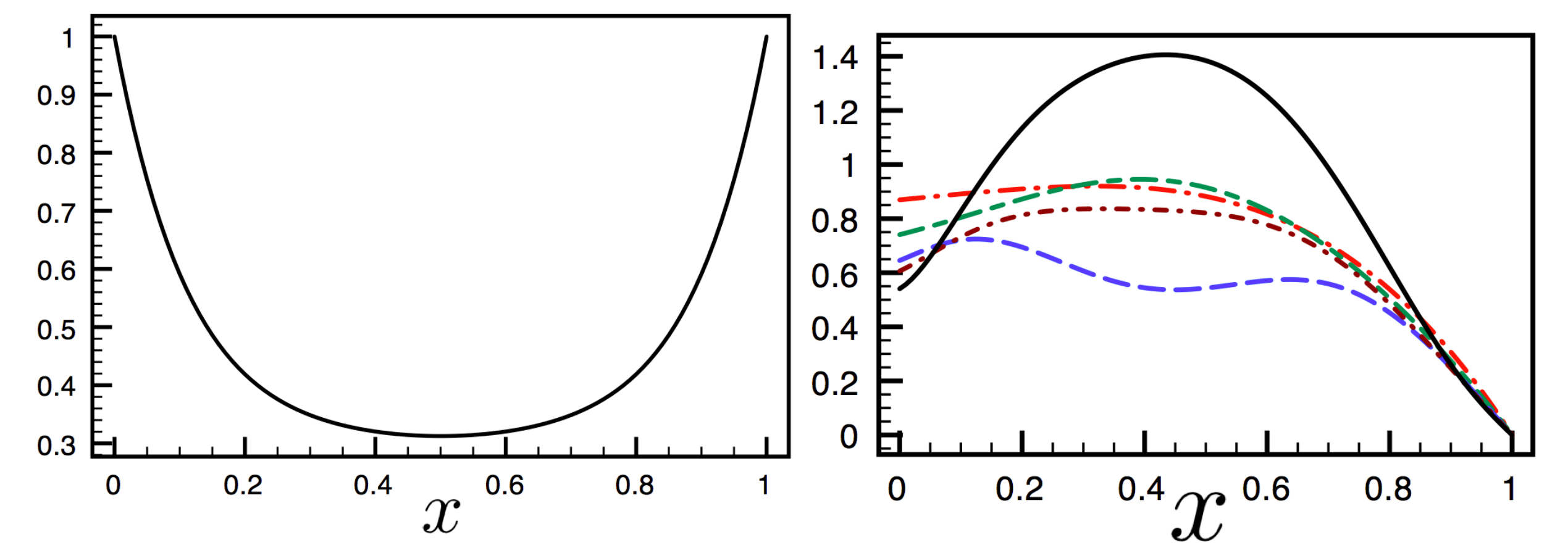}
\caption{{\small
Comparison between the renormalized parameters of the work in this thesis and the renormalized parameters of standard $t-J$ model. The horizontal axis in both figures plots the hole doping $x$. The left plot is the ratio of the renormalization $g_t$ as defined in the text to the $t-J$ model's $2x/(1+x)$ while the right plot shows the ratio between the renormalized $J$ of the $t-J$ model and our $\tilde{J}=4t^2/\tilde{U}$. In the bottom plot the dashed-dot line (red) is for $v=-.15$, the dash-dash line (online) for $v-.35$, the short-dashed and dotted line (online) is for $v=-.55$, the dashed line (online) for $v=-.65$ and the solid line (black) is for $v=-.85$
     }
     }\label{fig:doping}
\end{figure}


\begin{figure}[h]
  \setlength{\unitlength}{1mm}

   \includegraphics[scale=.6]{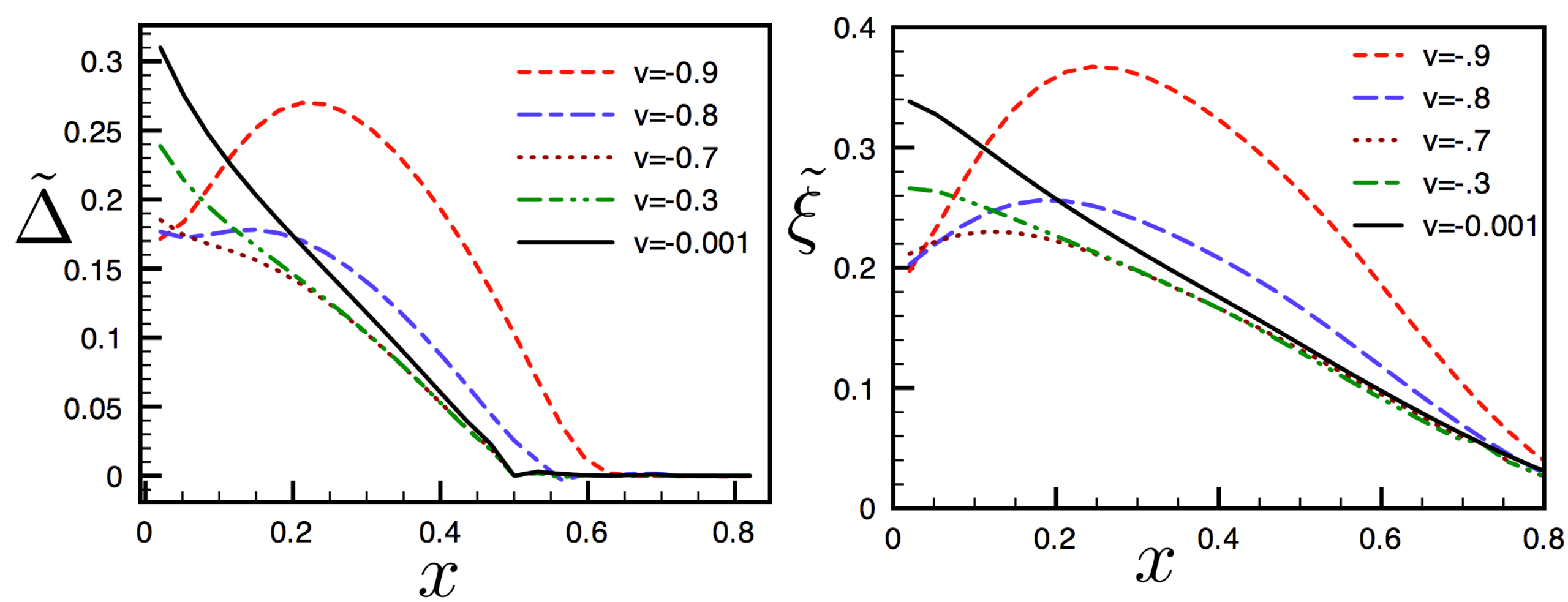}
\caption{{\small
Results of self-consistent calculations for varying values of $v$. The left figure shows the results for $\tilde{\Delta}$ while the right one shows $\tilde{\xi}$. For this calculation we have fixed $U_0=12t$. 
     }
     }\label{fig:OP_vary_v}
\end{figure}

\begin{figure}[h]
  \setlength{\unitlength}{1mm}
\begin{center}
   \includegraphics[scale=.7]{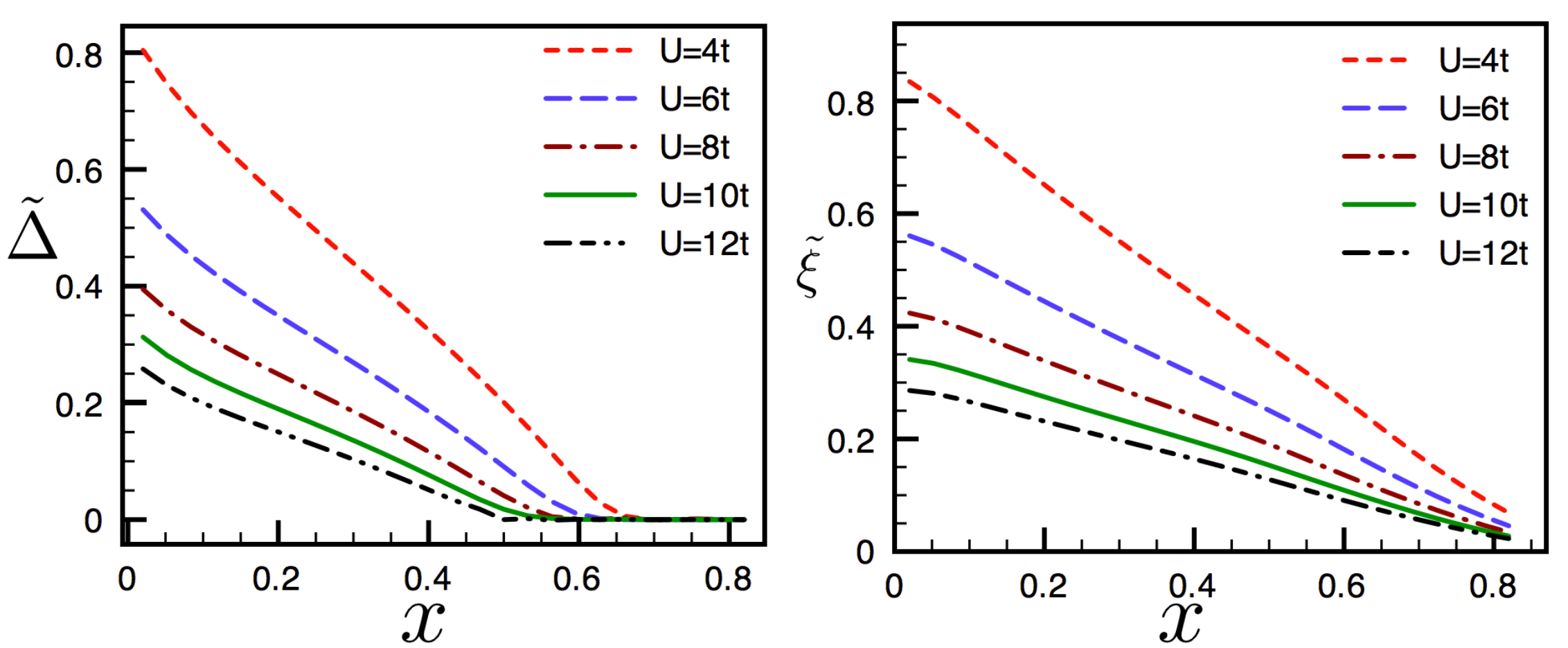}
   \end{center}
\caption{{\small
Results of self-consistent calculations for varying values of $U_0$. The left figure shows the results for $\tilde{\Delta}$ while the right one shows $\tilde{\xi}$. For this calculation we have fixed $v=-0.2$. 
     }
     }\label{fig:OP_vary_U}
\end{figure}

Using the results in Eqs. (\ref{Kgut}) and (\ref{vg}) we may write the variational energy of this problem as follows
\begin{equation}
E(\Delta_\kv,\xi_\kv) = -\sum_{\kv} (g_t\epsilon_\kv-\mu) \left(\frac{\xi_\kv}{E_\kv}\right) -\frac{3\tilde{J}}{8N} (d_x^2+d_y^2+2e_x^2+2e_y^2)
\end{equation}
where we have defined $d_{i} = \sum_{\kv} \left(\frac{\Delta_\kv \cos(k_i)}{E_\kv}\right)$, $e_{i} = \sum_{\kv} \left(\frac{\xi_\kv \cos(k_i)}{E_\kv}\right)$ and $\mu$ is determined self-consistently through $x = \frac{1}{N}\sum_\kv\left(\frac{\xi_\kv }{E_\kv}\right)$.  Minimizing the above functional with respect to $\Delta_\kv$ and $\xi_\kv$ leads to the following results
\begin{eqnarray}
&&\Delta_\kv = \Delta_x \cos(k_x)+\Delta_y \cos(k_y) \\ \nonumber 
&&\xi_\kv =g_t\epsilon_\kv -\mu + \xi_x \cos(k_x)+\xi_y \cos(k_y) \\ \nonumber 
\end{eqnarray}
Where the parameters $\Delta_i$ and $\xi_i$ must be determined self-consistently by solving the following equations  
\begin{eqnarray}\label{SelfCons}
&&\Delta_i = \frac{3\tilde{J}}{4N}\sum_{\kv} \left(\frac{\Delta_\kv \cos(k_i)}{E_\kv}\right)  \\ \nonumber 
&&\xi_i=\frac{3\tilde{J}}{4N}\sum_{\kv} \left(\frac{\xi_\kv \cos(k_i)}{E_\kv}\right)  \\ \nonumber 
\end{eqnarray}

These self-consistent conditions are well known in context of Gutzwiller approximations to the $t-J$ model, however in their above context the parameters are altered in comparison to the $t-J$ model. The difference occurs in the kinetic energy renormalization factor $g_t$ and the exchange coupling $\tilde{J}$. The $g_t$ term in our model is different from the work in Ref. \cite{zhang} by the doping dependent factor $\sum_{N_n}\left(\binom{3}{N_n}(1-x)^{N_n}x^{3-N_n}\right)^2$, which we have plotted in Fig.  \ref{fig:doping}. Inspecting the figure, we see that away from half filling this term leads to a lower kinetic energy than in analogous studies of the $t-J$ model. The second difference between our self-consistent work and that done for the $t-J$ model is the exchange coupling $\tilde{J}$ given in Eq. (\ref{exJ}) which compares to $16t^2/(U_0(1+x)^2)$ in the $t-J$ model. We see that the exchange constant in this work is renormalized by a complicated $x$ dependent factor. The right panel of Fig. \ref{fig:doping} further explores the differences between the two for different values of $v$. The figure shows that for moderate to large values of $v$ the exchange constant is lowered by the existence of the nearest neighbour interaction. Generally speaking this suppression is non-monotonic in $x$ and becomes greatest as $x\to1$. For larger values of $v$ such as the $v=-.85$ curve, we see that the renormalized exchange coupling is enhanced for moderate $x$ values but suppressed as $x\to0$ and $x\to1$.   

 We have solved the self consistent equations in Eq. (\ref{SelfCons}) over a space of $x$, $U_0$ and $V_0$ values. We have found that the solution is that of a $d$-wave superconductor, $\Delta_x=-\Delta_y$ with $|\Delta_x|=|\Delta_y|=\tilde{\Delta}$, and $\xi_x=\xi_y=-|\tilde{\xi}|$. The results of our calculations are presented in Figs. \ref{fig:OP_vary_v} and \ref{fig:OP_vary_U}. In Fig. \ref{fig:OP_vary_v} we have fixed the value of $U_0=12t$\cite{zhang, param} and explored how changing $v$ affects the order parameters. For small to moderate values of $v$ (up to about -0.7) both $\tilde{\Delta}$ and $\tilde{\xi}$ are monotonically decreasing functions of the hole doping $x$. We see that $\tilde{\Delta}$ goes sharply, but continuously, to zero at some critical value $x_c$. This critical value of $x$ is independent of $v$ up until large values of $v$. Past $v=-0.7$ we see that the monotonicity of the parameters breaks down and $x_c$ begins to increase with $v$.    In Fig. \ref{fig:OP_vary_U} we fix $v=-.2$ and vary $U_0$. In the figure, we see that the value of $U_0$ has little to no affect on the shape of the curves in Fig. \ref{fig:OP_vary_U}, it merely changes the magnitude of the parameters and increases $x_c$ in $\tilde{\Delta}$. 
 
 
To close this section we will present calculations of the superconducting order parameter. The parameter $\tilde{\Delta}$ is not exactly the ``order parameter" as we must take into account the Gutzwiller projection operator. We define the superconducting order parameter using the following function
\begin{equation}
\Delta_{SC}(i,j) = \frac{\langle \psi_{BCS} |P_G\left( c_{i,\uparrow}^\dagger c_{j,\downarrow}^\dagger-c_{i,\downarrow}^\dagger c_{j,\uparrow}^\dagger\right)P_G|\psi_{BCS}\rangle}{\langle \psi_{BCS} |P_GP_G|\psi_{BCS}\rangle}
\end{equation}
Using the same weighting of relative probabilities as was done for $K_G$ and $V_G$ above we can write a Gutzwiller approximation for this quantity as follows
 \begin{equation}
\Delta_{SC}(i,j) \simeq g_{\Delta} \frac{\langle \psi_{BCS} |\left( c_{i,\uparrow}^\dagger c_{j,\downarrow}^\dagger-c_{i,\downarrow}^\dagger c_{j,\uparrow}^\dagger\right)|\psi_{BCS}\rangle}{\langle \psi_{BCS} |\psi_{BCS}\rangle}
\end{equation}
 where $g_{\Delta}=2x/(1+x)$ comes from a proper weighting using relative probabilities. Upon writing the above in momentum space one can easily show the following result for nearest neighbour $(i,j)$. 
   \begin{equation}\label{OPP}
\Delta_{SC}(i) =\frac{ 4g_{\Delta}}{3 \tilde{J}} \Delta_{i}
\end{equation}
We have plotted our results for this quantity at the same value of $U_0$ as in Fig.  \ref{fig:OP_vary_v} and have compared our results to those of the bare $t-J$ model. The results are in Figure  \ref{fig:SCOP}. We see that the order parameter $\Delta_{SC}$ is a non-monotonic function of the doping $x$ which grows from zero to a maximum and then returns to zero at $x_c$. Both the maximum value of $\Delta_{SC}$ and the $x$ value at which it is achieved vary only slightly with $v$ and for large $v$ both tend to increase with $v$. For the sake of comparison we have also plotted the results obtained for this order parameter from the traditional $t-J$ model. We see that our analysis here leads to an enhancement of $\Delta_{SC}$. This makes physical sense as the difference between the model considered here and the $t-J$ model is an attractive nearest neighbour interaction, something that should favour $d$-wave superconductivity.

\begin{figure}[h]
  \setlength{\unitlength}{1mm}

   \includegraphics[scale=.8]{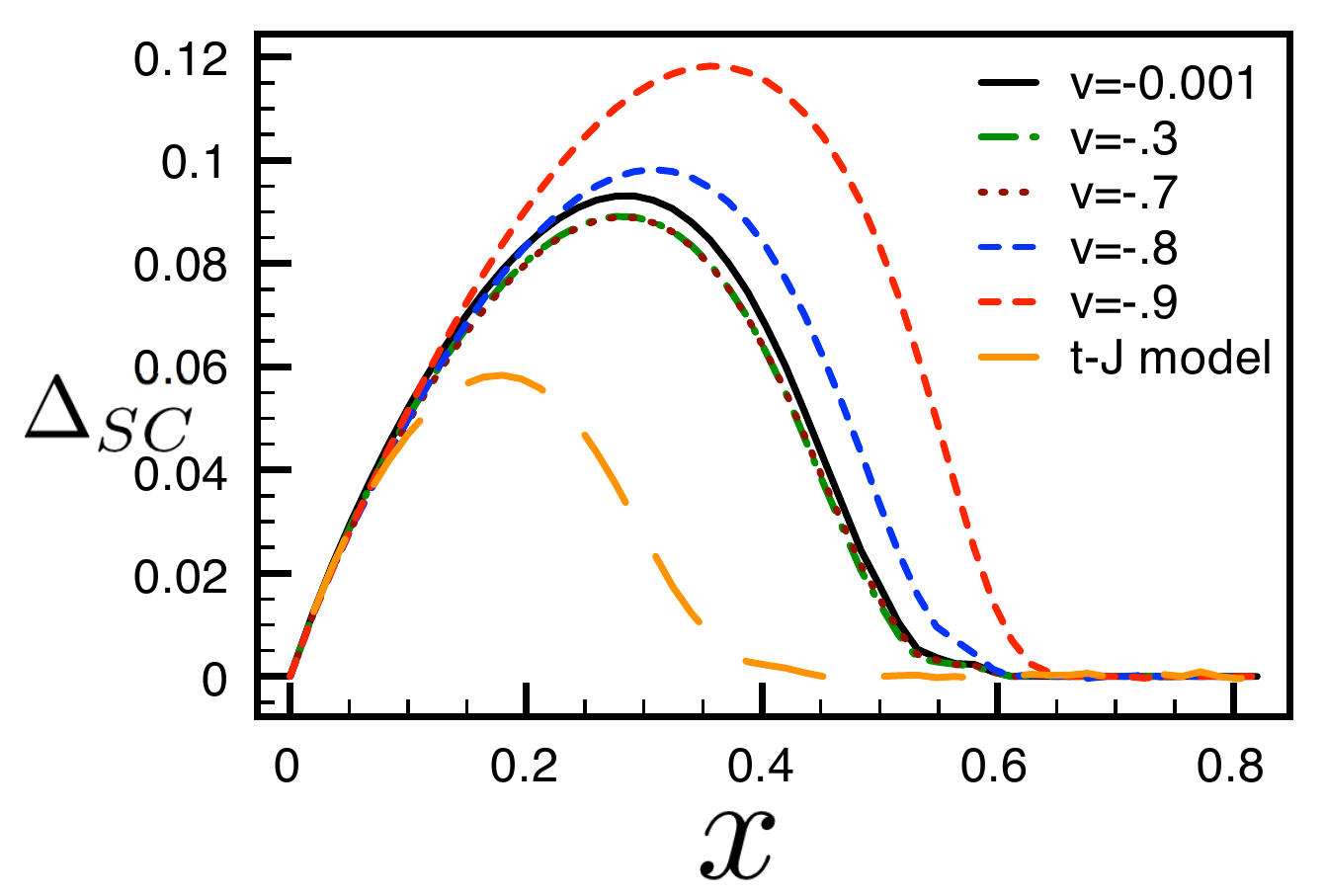}
\caption{{\small
Full order parameter as defined in Eq. (\ref{OPP}) of the text. We have determined this order parameter for several different values of $v$ and fixed $U_0=12t$. 
     }
     }\label{fig:SCOP}
\end{figure}

This concludes our discussion of the Gutzwiller approximation method at zero spin-orbit coupling. This section has outlined the method with the transparent example of a system with zero spin-orbit coupling. Furthermore, to our knowledge, a strong coupling treatment of the extended Hubbard model has not been presented in the literature thus far, so this section also represents a new contribution to the field. We now move on to treat the archetypal system of interest in this thesis, interacting electrons with finite spin-orbit coupling. 

\section{Renormalized Mean Field Theory at Finite Spin-Orbit Coupling}

 In this section we follow the same basic steps as in the last section except this time we keep the parameters $A,B$ and $M$ finite. We are still interested in the model in Equation (\ref{strongH}}).  Keeping either $M$ or $B$ finite leads to spin-polarization of the electrons in the system as in general the Zeeman term $H_Z$ leads to an energetic preference for one spin flavour over the other. This inequality in the number of spin-up and spin-down electrons complicates the nature of the Gutzwiller approximation as the tacit assumption we made in the last section that $\langle n_{\uparrow}\rangle   =\langle n_{\downarrow}\rangle   =\langle n \rangle/2$ now no longer holds. For an inhomogenous system such as this we must renormalize processes involving different spin flavours separately. 
 
 It should be noted that treating this problem in a general fashion is not at all straightforward\cite{Ede1,Ede2}. Another tacit assumption we made in our Gutzwiller approximation in the previous section was that $\langle \psi|n_{\sigma}|\psi\rangle=  \langle \psi|P_Gn_{\sigma}P_G|\psi\rangle  $, that is the expectation value of number operators is the same in the projected and non-projected mean field ground state. In a system with spin homogeneity ({\it i.e.} $\langle n_{\uparrow}\rangle   =\langle n_{\downarrow}\rangle   =\langle n \rangle/2$) this is indeed the case\cite{Ede2}, however in moving away from the homogeneous case this assumption is no longer true. Solving this problem in a rigorous fashion is far from trivial and we will avoid such a task in this thesis. Thus we will work under the approximation $\langle \psi|n_{\sigma}|\psi\rangle \simeq  \langle \psi|P_Gn_{\sigma}P_G|\psi\rangle  $. 

A second complication arising when attempting to apply the methods of the previous section to systems with spin-orbit coupling is that in the BCS wave function the average $\langle c_{\kv,\sigma} ^\dagger c_{\kv,\sigma'}\rangle_{BCS} $ is zero for $\sigma\ne \sigma'$. This fact will lead to several important terms in the Hamiltonian in (\ref{strongH}) having zero expectation value. This loss of information is avoided if we abandon the BCS wave function and resort to a more general self-consistent study. 

To this end we define the following auxiliary Hamiltonian 
\begin{eqnarray}\label{Haux}
H_{Aux}&=& \sum_{\kv,\sigma} \xi_{\kv, \sigma} c^\dagger_{\kv,\sigma} c_{\kv,\sigma} +\sum_{\kv,\sigma} \alpha_{\kv, \sigma} c^\dagger_{\kv,\sigma} c_{\kv,\bar{\sigma}} \\ \nonumber  &-& \frac{1}{2}\sum_{\kv,\sigma, \sigma'} \left(\Delta_{\kv, \sigma, \sigma'} c^\dagger_{\kv,\sigma} c^\dagger_{-\kv,\sigma'} +\text{h.c.}\right)
\end{eqnarray}
where $\xi_{\kv, \sigma} $, $\alpha_{\kv, \sigma} $ and $ \Delta_{\kv, \sigma, \sigma'}$ are all functions that must be determined. Our goal will be to perform a mean field decoupling of our full quartic Hamiltonian to bring it into the above form. From this decoupled Hamiltonian we will be able to read off the expressions for $\xi_{\kv, \sigma} $, $\alpha_{\kv, \sigma} $ and $ \Delta_{\kv, \sigma, \sigma'}$. 

The mean field decoupling we will perform is completed in two steps. First, we renormalize all of the terms in the full Hamiltonian to take strong correlations into account. This part is our Gutzwiller approximation. Second, we perform a standard mean field decoupling of the quartic terms in this renormalized full Hamiltonian.

We begin with the hopping term $T_{0,M,M}$ in Eq. (\ref{strongH}). Again our goal is to replace the projection operators with a more convenient renormalization constant.  Let us first consider the projection that restricts the number of nearest neighbours permitted before and after an electron is moved from one site to the other. This part of the projection does not care about spin but instead only cares about whether a site is occupied or not. Owing to this, we can take care of it using the same relative probabilities as in the previous section. That is we must include the following factor
\beq 
g_{NN} = \sum_{N_n=0}^3\left(\binom{3}{N_n}(1-x)^{N_n}x^{3-N_n}\right)^2
\eeq
in our projection operator. 

Next we treat the different parts of $T_{0,M,M}$ separately. Recall that we defined the generalized hopping matrix, $\hat{t}$, in Eq. (\ref{tgen}) as having two distinct parts. One part moves an electron from one site to another whilst maintaining that electron's spin, while a second part moves an electron to a nearest neighbour site and flips its spin. We renormalize each of these parts separately. By considering all possible ways this process can occur in some ground state $|\psi\rangle$ and all possible ways they can occur in a Gutzwiller projected state $ P_G |\psi\rangle$ we obtain the following factors\cite{zhang, Ko, Chou}.
\begin{eqnarray}
&&g_t(\sigma) = \frac{x}{1-\langle n_{\sigma} \rangle } \\ \nonumber
&&g_A= \frac{x}{\sqrt{(1-\langle n_{\uparrow} \rangle)(1-\langle n_{\downarrow} \rangle)} } 
\end{eqnarray}
where these factors are used in defining the renormalized hopping matrix
\beq
\hat{t}^{g}(\delta)   = (\sigma B-t)g_t(\sigma) g_{NN}\delta_{\sigma,\sigma'} -g_A g_{NN}\frac{Ai}{2}\vec{\delta}\cdot(\hat{x}-i\sigma\hat{y}) \delta_{\sigma',\bar{\sigma}}
\eeq
which leads to the following renormalized kinetic energy operator
\beq 
H_{T,SO} = \sum_{i,\delta,\sigma, \sigma'}    \hat{t}^{g}(\delta)  c_{i,\sigma}^\dagger c_{j,\sigma'}.
\eeq

Inspecting $g_t(\sigma)$ we see that it is simply the relative probability of a hop of a spin $\sigma$ in the two states; in the state  $ P_G |\psi\rangle$ we require an empty site to make the hop while in $|\psi\rangle$ we simply need the destination site to not already have an electron of spin $\sigma$. We can understand $g_A$ in almost exactly the same way. The process in $ P_G |\psi\rangle$ requires an empty site and hence the factor of $x$ in the numerator. Meanwhile we need the nearest neighbour site to be void of an electron with a spin opposite to that of the electron we are moving. This fact accounts for a factor $1- \langle n_{\sigma} \rangle$.  In defining these renormalization constants we must equally weight the probability for the process to happen in the forward direction with the probability for it to happen in the reverse direction\cite{Ko}. This practice ensures that our renormalized Hamiltonian is Hermitian. With this in mind we must also include in $g_A$ the probability for the process to happen in reverse, this is where the factor of $1- \langle n_{\bar{\sigma}} \rangle$ and the square root come from. 

We can now renormalize the second order term in equation (\ref{strongH}). We will accomplish this task in two steps. First, we account for the restrictions having to do with nearest neighbours as they are not spin dependent and should thus be renormalized in the same way they were in the first section of this chapter. Second, we renormalize the spin dependent part. 

The first step above proceeds as follows. We again neglect three site hopping and processes where the $m$ in the $T_{m,N_1,N_2}$ operators is zero. This leaves us with only one site terms. Neglecting contributions that will be only very slowly varying functions of any order parameters (terms involving things like $n_i n_j$) we can write this in the form
\beq 
H_{\tilde{J}} = \sum_{i,\delta} \Sv_{i+\delta} \tilde{J}_{\delta} \Sv^T_{i}.
\eeq

with the coupling matrix 
\begin{eqnarray}
 \tilde{J}_{\delta}=\frac{1}{2} \left(\begin{matrix} 
      J_1+J_2a(\delta)-J_4 & 0& -J_3\hat{y}\cdot \vec{\delta} \\
      0 & J_1-J_2a(\delta)-J_4&J_3\hat{x}\cdot \vec{\delta} \\
      J_3\hat{y}\cdot \vec{\delta}&-J_3\hat{x}\cdot \vec{\delta}&J_1-J_2+J_4\\
         \end{matrix}\right)
\end{eqnarray}
where we have defined the three couplings $J_1=4t^2/\tilde{u}$,  $J_2=A^2/\tilde{u}$,  $J_3=4At/\tilde{u}$ and $J_4 = 4B^2/\tilde{u}$ and the function $a(\delta)$ is 1 if $\delta=\pm \hat{x}$ and -1 if $\delta=\pm\hat{y}$.

In the above the terms we have kept are exactly the terms that contribute to the analogy to the t-J model we made at the end of Chapter 4. The rationale behind this is these terms are the relevant parts of the second order term in (\ref{strongH}) near half filling.  The difference between the above and our projected Hamiltonian from Chapter 4 is that we have allowed for systems that are hole doped away from half-filling by approximately taking into account the restrictions on the number of nearest neighbours permitted by a given hop. This is contrary to the treatment in Chapter 4 where we only kept hops that can happen at half-filling. This renormalization is reflected in the renormalized $\tilde{u}$ as follows
\beq
\frac{1}{\tilde{u}}= \frac{1}{U} \sum_{N_1=0}^3 \sum_{ M_1=1}^4 \frac{(1-x)^{N_1} x^{3-N_1}(1-x)^{M_1}x^{4-M_1}}{1-(M_1-N_1)v}  \binom{3}{N_1}\binom{4}{M_1}
\eeq
This is done in much the same way we defined $\tilde{U}$ in the previous section with the only difference being that this time we have not yet taken the on site projection requirements into account. This is why the factor $4/(1+x)^2$ is absent from the above definition. 

We can now split this piece of the Hamiltonian into different terms and renormalize each piece by comparing amplitudes in the non-projected Gutzwiller state to those in the Gutzwiller state. We then write $H_{\tilde{J}}$ in terms of the following four terms
\begin{eqnarray} \label{exterms}
H_{J,1}&=&\frac{J_1-J_4}{4} \sum_{i,\delta} \left(S_{i+\delta}^{+} S_{i}^{-} +S_{i+\delta}^{-} S_{i}^{+} \right)\\ \nonumber 
H_{J,2}&=&\frac{J_1-J_2+J_4}{2} \sum_{i,\delta}  S_{i+\delta}^{z} S_{i}^{z} \\ \nonumber
H_{J,3}&=&\frac{J_2}{4} \sum_{i,\delta}a(\delta) ( S_{i+\delta}^{+} S_{i}^{+} +S_{i+\delta}^{-} S_{i}^{-})  \\ \nonumber
H_{J,4}&=&\frac{J_3}{2} \sum_{i,\delta}\left(  \delta \cdot (\hat{y}+i \hat{x})S_{i+\delta}^{z} S_{i}^{+}+ \delta \cdot (\hat{y}-i \hat{x})S_{i+\delta}^{z} S_{i}^{-}\right)
\end{eqnarray}
where we have defined the spin raising and lowering operators $S^+_{i} = c_{i,\uparrow}^\dagger c_{i,\downarrow}$ and $S^-_{i} = c_{i,\downarrow}^\dagger c_{i,\uparrow}$. Let us discuss the renormalization of each term above in the order listed. 

$H_{J,1}$ takes two occupied nearest neighbour sites and flips their spins. We discuss $S_{i+\delta}^{+} S_{i}^{-}$ for concreteness while keeping in mind that this discussion immediately extends to the second term. In a Gutzwiller projected state we only require that sites $i$ and $i+\delta$ are occupied  by an up spin and a down spin respectively, while in a non-projected state we not only require states $(i,\uparrow)$ and $(i+\delta, \downarrow)$ to be occupied but also need $(i,\downarrow)$ and $(i+\delta, \uparrow)$ to be empty. These considerations lead to the renormalization factor 
  \beq
  g_{J,1} = \frac{1}{(1-\langle n_{\uparrow} \rangle)(1-\langle n_{\downarrow}\rangle)}
  \eeq
  and we make the approximation $H_{J,1} \to   g_{J,1} H_{J,1}$. The other three contributions are then renormalized by the following factors
  \begin{eqnarray}
  g_{J,2} &=& 1 \\ \nonumber 
    g_{J,3} &=&  \frac{1}{(1-\langle n_{\uparrow}\rangle)(1-\langle n_{\downarrow}\rangle)} \\ \nonumber 
      g_{J,4} &=&  \frac{1}{\sqrt{(1-\langle n_{\uparrow}\rangle)(1-\langle n_{\downarrow}\rangle)}}
 \end{eqnarray}
  where our approximation amounts to $H_{J,\alpha} \to g_{J,\alpha}H_{J,\alpha}$ for each of $\alpha = 2,3,4$. With these approximations in place our fully renormalized $H_{\tilde{J}}$ is as follows
  
\beq 
H_{\tilde{J}} = \sum_{i,\delta} \Sv_{i+\delta} \mathcal{J}_{\delta} \Sv^T_{i}.
\eeq
 Where the new coupling matrix reads 
  \begin{eqnarray}
 \tilde{J}_{\delta}= \left(\begin{matrix} 
      \mathcal{J}_2+\mathcal{J}_3a(\delta) & 0& -\mathcal{J}_4\hat{y}\cdot \vec{\delta} \\
      0 & \mathcal{J}_2-\mathcal{J}_3a(\delta) &\mathcal{J}_4\hat{x}\cdot \vec{\delta} \\
      \mathcal{J}_4\hat{y}\cdot \vec{\delta}&-\mathcal{J}_4\hat{x}\cdot \vec{\delta}&\mathcal{J}_1\\
         \end{matrix}\right)
\end{eqnarray}
with $\mathcal{J}_1= \frac{J_1-J_2+J_4}{2}$, $\mathcal{J}_2 = \frac{g_{J,1} (J_1-J_4)}{2}$,  $\mathcal{J}_3=\frac{g_{J,3}J_2}{2}$ and finally $\mathcal{J}_4=\frac{g_{J,4}J_3}{2}$. This concludes the first step in defining the mean field parameters $\xi_{\kv,\alpha}, \alpha_{\kv,\alpha}$ and $\Delta_{\kv\sigma, \sigma'}$. We proceed to study the mean field theory of the renormalized Hamiltonian $H_{GA} = H_{T,SO}+H_{\tilde{J}}$. 

To perform a mean field decoupling of $H_{GA}$ we replace all quartic combinations of the electron operators in $H_{\tilde{J}}$ by the following 
\begin{eqnarray}
&&c^{\dagger}_{\kv_1,\alpha_1}c^{\dagger}_{\kv_2,\alpha_2}c_{\kv_3,\alpha_3}c_{\kv_4,\alpha_4}\\ \nonumber  &\to& \langle c^{\dagger}_{\kv_1,\alpha_1}c^{\dagger}_{\kv_2,\alpha_2} \rangle c_{\kv_3,\alpha_3}c_{\kv_4,\alpha_4} + c^{\dagger}_{\kv_1,\alpha_1}c^{\dagger}_{\kv_2,\alpha_2}\langle c_{\kv_3,\alpha_3}c_{\kv_4,\alpha_4} \rangle -\langle c^{\dagger}_{\kv_1,\alpha_1}c^{\dagger}_{\kv_2,\alpha_2}\rangle \langle c_{\kv_3,\alpha_3}c_{\kv_4,\alpha_4} \rangle \\ \nonumber &+&\langle c^{\dagger}_{\kv_1,\alpha_1} c_{\kv_4,\alpha_4} \rangle  c^{\dagger}_{\kv_2,\alpha_2}c_{\kv_3,\alpha_3} + c^{\dagger}_{\kv_1,\alpha_1}c_{\kv_4,\alpha_4} \langle  c^{\dagger}_{\kv_2,\alpha_2}c_{\kv_3,\alpha_3}\rangle -\langle c^{\dagger}_{\kv_1,\alpha_1}c_{\kv_4,\alpha_4}\rangle \langle  c^{\dagger}_{\kv_2,\alpha_2}c_{\kv_3,\alpha_3} \rangle \\ \nonumber &-&\langle c^{\dagger}_{\kv_1,\alpha_1} c_{\kv_3,\alpha_3} \rangle  c^{\dagger}_{\kv_2,\alpha_2}c_{\kv_4,\alpha_4} - c^{\dagger}_{\kv_1,\alpha_1}c_{\kv_3,\alpha_3} \langle  c^{\dagger}_{\kv_2,\alpha_2}c_{\kv_4,\alpha_4}\rangle +\langle c^{\dagger}_{\kv_1,\alpha_1}c_{\kv_3,\alpha_3}\rangle \langle  c^{\dagger}_{\kv_2,\alpha_2}c_{\kv_4,\alpha_4} \rangle  .
\end{eqnarray}
We then assume that $\langle c^{\dagger}_{\kv_1,\alpha_1}c^{\dagger}_{\kv_2,\alpha_2} \rangle \propto \delta_{\kv_1,-\kv_2}$ and $\langle c^{\dagger}_{\kv_1,\alpha_1} c_{\kv_2,\alpha_2} \rangle \propto \delta_{\kv_1,\kv_2}$. Ignoring the overall constant that comes from this process and comparing the result to Equation (\ref{Haux}) we find that the free parameters built into this auxiliary model must have the following form
\begin{eqnarray} \label{Scparam}
\xi_{\kv,\sigma} &=& \left(1-\sigma \frac{B}{t}\right)g_t(\sigma) g_{NN}\epsilon_{\kv}-\mu \\ \nonumber &+& \hat{\xi}_{x, \sigma} \cos{k_x} + \hat{\xi}_{ y,\sigma}\cos{k_y}+ \sigma(M-4B+M_{\text{eff}}) \\ \nonumber 
\alpha_{\kv,\sigma} &=& g_A g_{NN}A(\sin{k_x}-i\sigma \sin{k_y}) +\hat{A}_x\sin{k_x}-\hat{A}_yi\sigma \sin{k_y} \\ \nonumber
\Delta_{\kv,\uparrow,\downarrow} &=&-\Delta_{\kv,\downarrow, \uparrow}= \hat{\Delta}_x \cos(k_x)+\hat{\Delta}_y \cos(k_y) \\ \nonumber 
\Delta_{\kv,\uparrow,\uparrow} &=&-\hat{\Delta}^{\uparrow, \uparrow}_x \sin(k_x)+i\hat{\Delta}^{\uparrow, \uparrow}_y \sin(k_y) \\ \nonumber 
\Delta_{\kv,\downarrow,\downarrow} &=&\hat{\Delta}^{\downarrow, \downarrow}_x \sin(k_x)+i\hat{\Delta}^{\downarrow, \downarrow}_y \sin(k_y) \\ \nonumber 
\end{eqnarray}
where we have defined the effective field $M_{\text{eff}} = 2\mathcal{J}_1(\langle n_{\uparrow} \rangle -\langle n_{\downarrow} \rangle)$ and the following parameters
\begin{eqnarray}\label{Selfconsist}
\hat{\xi}_{i,\sigma} &=& -2\mathcal{J}_2  \tilde{\xi}_{i,\bar{\sigma}}- {\mathcal{J}_1} \tilde{\xi}_{i,\sigma}+{\mathcal{J}_4}( \tilde{A}_i+ \tilde{A}_i^*) \\ \nonumber
\hat{A}_i &=& -{2\mathcal{J}_3} \tilde{A}_{i}+{\mathcal{J}_1} \tilde{A}_{i}^*+{\mathcal{J}_4}( \tilde{\xi}_{i,\uparrow}+ \tilde{\xi}_{i,\downarrow})  \\ \nonumber 
\hat{\Delta}_i &=&{2\mathcal{J}_2}\tilde{\Delta}_i+{\mathcal{J}_1}\tilde{\Delta}_i-{\mathcal{J}_4}\left(\tilde{\Delta}_i^{\uparrow,\uparrow}+\tilde{\Delta}_i^{\downarrow,\downarrow}\right) \\ \nonumber 
  \hat{\Delta}_i^{\sigma, \sigma} &=&{2\mathcal{J}_3} \tilde{\Delta}_i^{\bar{\sigma},\bar{\sigma}}-{\mathcal{J}_1} \tilde{\Delta}_i^{\sigma,\sigma} -{2\mathcal{J}_4} \tilde{\Delta}_i
\end{eqnarray}
In the above there are twelve free parameters that must be fixed through ensuring that the theory is self-consistent.  The self-consistency conditions we must satisfy are
\begin{eqnarray}\label{SCeqn}
  \tilde{\xi}_{j, \sigma}= \frac{1}{N}\sum_{\kv} \cos{k_j} \langle c_{\kv,\sigma}^\dagger c_{\kv,\sigma}\rangle \\ \nonumber 
   \tilde{A}_j= \frac{1}{N}\sum_{\kv} q_{j} \sin{k_j} \langle c_{\kv,\uparrow}^\dagger c_{\kv, \downarrow}\rangle \\ \nonumber 
  \tilde{\Delta}_j=  \frac{1}{N}\sum_{\kv} \cos{k_j} \langle c_{\kv,\uparrow}^\dagger c^\dagger_{-\kv, \downarrow}\rangle^* \\ \nonumber 
    \tilde{\Delta}^{\sigma, \sigma}_j= \frac{1}{N} \sum_{\kv} \tilde{q}_{j, \sigma}\sin{k_j} \langle c_{\kv,\sigma}^\dagger c^\dagger_{-\kv, \sigma}\rangle^* \\ \nonumber 
\end{eqnarray}
where to make the equations compact we have used $q_x=1$ and $q_y=-i$ and $\tilde{q}_{y, \sigma}=-i$ while $\tilde{q}_{x,\sigma} = -\sigma$. These twelve equations must be solved simultaneously with the particle number constraints $\langle n_{\sigma}\rangle = \frac{1}{N} \sum_{\kv} \langle  c_{\kv,\sigma}^\dagger c_{\kv,\sigma}\rangle $ and $x=1-\langle n_{\uparrow}\rangle - \langle n_{\downarrow}\rangle$. The theory above then comes down to solving a fifteen by fifteen system of non-linear (integral) equations. 

To ensure that the notation we have defined above is clear, let us make explicit mention of it here. In moving from Eq. (\ref{Scparam}) to Eq. (\ref{Selfconsist}) and finally to Equation (\ref{SCeqn}) we have used the same symbols but introduced a hat and tilde in the latter two steps. We have used bare symbols, {\it e.g.} $\xi_{\kv,\sigma}$, to denote the $\kv$ dependent values in the auxiliary Hamiltonian. The parameters with the hat symbol, {\it e.g.} $\hat{\xi}_{i,\sigma}$, are constants that enter the definition of $\kv$ dependent functions used to define the auxiliary Hamiltonian. Finally, we have used the tilde symbol, {\it e.g.} $\tilde{\xi}_{i,\sigma}$, for self-consistent parameters that are heretofore unknown until we solve the equations in  (\ref{SCeqn}).  

A quick check can be done to show that the above theory reduces to the theory developed in the previous section of this chapter. In the limit $A,B,M\to0$ we have $\langle n_{\uparrow}\rangle=\langle n_{\downarrow}\rangle=(1-x)/2$, $\mathcal{J}_3, \mathcal{J}_4\to0$ we also have $\mathcal{J}_1\to \tilde{J}/2$. In this limit we must also send $\mathcal{J}_2\to \tilde{J}/2$ as is discussed in Reference \cite{Ko}. In this limit the equations in (\ref{Selfconsist}) reduce to the system we solved in the previous section provided we set $\hat{\Delta}_i^{\sigma,\sigma}=\hat{A}_i=0$, both something that the previous section tacitly assumed and a valid solution to the self consistency equations in (\ref{SCeqn}).  

With finite spin-orbit coupling the conditions in Eq. (\ref{SCeqn}) cannot simply be solved by setting terms like the triplet superconductivity $\tilde{\Delta}_j^{\sigma,\sigma}$ to zero. This stems from the non-trivial coupling between $\hat{\xi}_{i,\sigma}$ and $\tilde{A}_i$,   $\hat{A}_{i}$ and $\tilde{\xi}_{i,\sigma}$, $\hat{\Delta}_{i}$ and $\tilde{\Delta}^{\sigma,\sigma}_i$ and $\hat{\Delta}^{\sigma,\sigma}_i$  and $\tilde{\Delta}_{i}$ that is in place as a results of the coupling $\mathcal{J}_4$. This can be seen by inspecting the last term in each of the definitions in Equation (\ref{Selfconsist}). 

To see why this coupling refuses solutions where some of the parameters are identically zero while others are not, consider looking for a solution where $\tilde{\Delta}_i^{\sigma,\sigma}=0$. The last two equations in Eq. (\ref{Selfconsist}) then become
\begin{eqnarray}\label{demo1}
\hat{\Delta}_i &=&-{2\mathcal{J}_2}\tilde{\Delta}_i-{\mathcal{J}_1}\tilde{\Delta}_i\\ \nonumber 
  \hat{\Delta}_i^{\sigma, \sigma} &=&{2\mathcal{J}_4} \tilde{\Delta}_i
\end{eqnarray}
and the relevant equations we must solve are 
\begin{eqnarray}\label{nonzero}
  \tilde{\Delta}_j=  \frac{1}{N}\sum_{\kv} \cos{k_j} \langle c_{\kv,\uparrow}^\dagger c^\dagger_{-\kv, \downarrow}\rangle^* \\ \nonumber 
0= \frac{1}{N} \sum_{\kv} \tilde{q}_{j, \sigma}\sin{k_j} \langle c_{\kv,\sigma}^\dagger c^\dagger_{-\kv, \sigma}\rangle^* \\ \nonumber 
\end{eqnarray}
It is impractical to find analytic expressions for averages such as $\langle c_{\kv,\sigma}^\dagger c^\dagger_{-\kv, \sigma}\rangle$ (they depend on eigenvectors of a complicated four by four matrix) however we can find these numerically. Looking at numerics and using some intuition, we argue that roughly $\langle c_{\kv,\sigma}^\dagger c^\dagger_{-\kv, \sigma}\rangle \propto \hat{\Delta}_i^{\sigma, \sigma}$ while Eq. (\ref{demo1}) tells us that $\hat{\Delta}_i^{\sigma, \sigma}\sim \tilde{\Delta}_i$. So in general either $\tilde{\Delta}_i=0$ or the sum on the right of of the second equation in (\ref{nonzero}) is nonzero and therefore $\tilde{\Delta}_i^{\sigma,\sigma}$ cannot be zero. 

With this point made we proceed to solve the self-consistency equations numerically. Where these parameters are non-zero the general trend we see in the solutions is that $|\hat{\xi}_{x,\sigma}| = |\hat{\xi}_{y,\sigma}| = \hat{\xi}_{\sigma}$, $\hat{A}_x=\hat{A}_y = \hat{A}$, $\hat{\Delta}_x = - \hat{\Delta}_y$ and $\hat{\Delta}^{\sigma,\sigma}_x = - \hat{\Delta}_y^{\sigma,\sigma}$. This equality of parameters with $x$ and $y$ directionality is not surprising as nowhere in our model is there anything that prefers the $x$ direction over the $y$.   

 \begin{figure}[tb]
  \setlength{\unitlength}{1mm}

   \includegraphics[scale=.6]{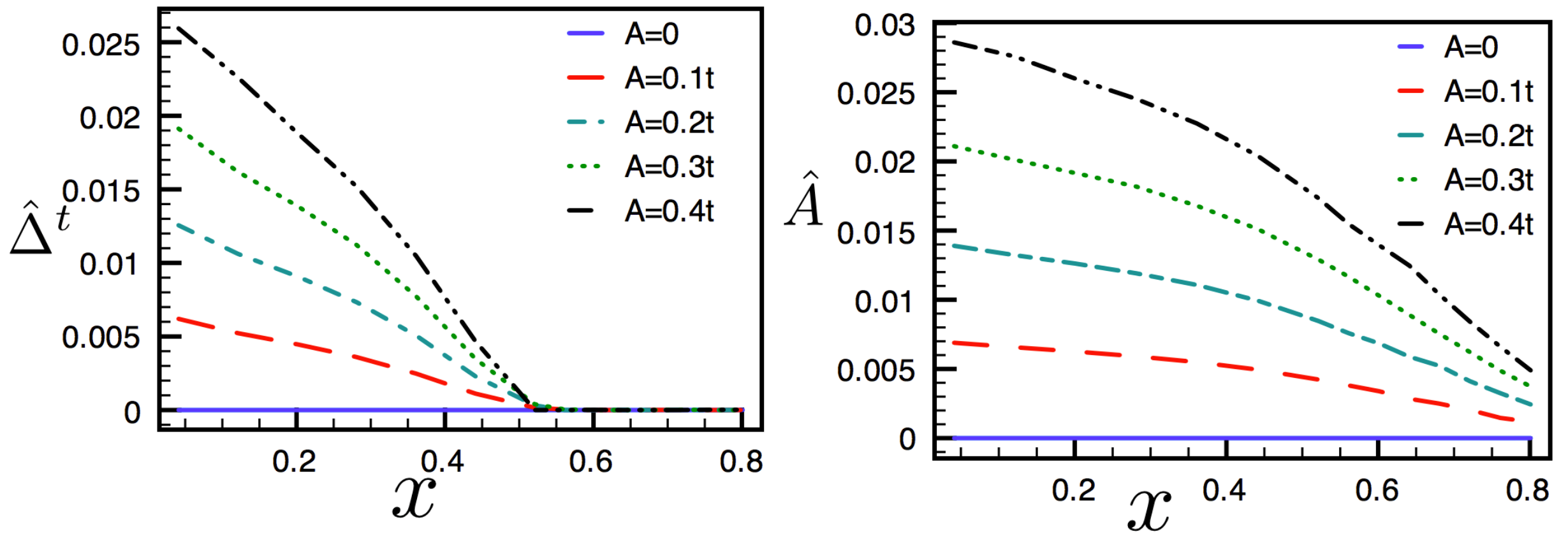}
\caption{{\small
Plot of the parameters $\hat{\Delta}^{t}$ and $\hat{A}$ for various values of the bare spin-orbit coupling parameter $A$. We have kept $M=B=0$ in this plot and the order parameters are plotted in energies of $t$. We have set $U_0=12t$ and $v=-0.2$ in this figure. 
     }
     }\label{fig:SC_f_A}
\end{figure}

For simplicity we have chosen to consider  $M,B=0$ and $A\ne0$. In this special case of the spin-orbit coupling parameters there is not a preferred spin direction and we have $\hat{\xi}_{\sigma}= \hat{\xi}$, $\langle n \rangle _{\sigma} = (1-x)/2$ and $\hat{\Delta}_i^{\uparrow,\uparrow}= \hat{\Delta}_i^{\downarrow,\downarrow}= \hat{\Delta}^t_i$. Further, when we solve the self-consistency conditions we find that the values of $\hat{\xi}$ and $\hat{\Delta}$ are effected very little by letting $A$ become finite and their behaviour remains as described in the previous section. With this in mind Fig.~ \ref{fig:SC_f_A} plots the doping dependence of $\hat{\Delta}^{t}$ and $\hat{A}$ for various values of $A$.  

Looking at the figure we see not surprisingly that for $A=0$  $\hat{\Delta}^{t}$ and $\hat{A}$ are zero for all values of doping. As we allow $A$ to become finite both develop behaviour that is monotonically decreasing in the doping $x$. Like $\hat{\xi}$, $\hat{A}$ is finite over the whole range of $x$ values shown while like $\hat{\Delta}$ we see that $\hat{\Delta}^{t}$ goes to zero at some critical $x_c$ which both matches the $x_c$ of $\hat{\Delta}$ and is unaffected by $A$. Both  $\hat{\Delta}^{t}$ and $\hat{A}$ increase in magnitude with $A$ but their general shape remains unchanged. We also note that the values of these parameters is an order of magnitude less than their counterparts   $\hat{\xi}$ and $\hat{\Delta}$. Therefore the dominant phase seen in this treatment is again $d$-wave superconductivity.  

Work on completing the study of the phase diagram defined by the fifteen conditions in Eq. (\ref{SCeqn}) is still ongoing. In particular, we are still looking at the dependence of the various order parameters on the parameters $B$ and $M$. 

In Chapter 3 of this thesis we said that the most interesting phase we found from a topological standpoint was the $d+id$-wave superconductor yet this kind of phase is absent from the discussion of the current section. This is a topic for future study, in the analysis above we have neglected so-called ``three-site" hopping terms from our discussion. These three site hopping terms, if included, should lead to pairing on next-nearest neighbour bonds and thus $d+id$-wave superconductivity. Therefore, although the above discussion hasn't directly found $d+id$-wave superconductivity it would be, in principle, straightforward (but very tedious) to include it. 

\section{Topology of the Strongly Coupled System}

With our discussion of the renormalized mean-field theory for a strongly interacting system of spin-orbit coupled electrons completed, we move on to discuss the topology of the system. In the mean field study in Chapter 3 we were able to define a topological invariant for the mean field ground state in terms of the topological properties of the auxiliary Hamiltonian, $H_{Aux}$. We were free to do this because the ground state $|\psi_{Var}\rangle$ was, by construction, the exact ground state of $H_{Aux}$. On the contrary, the system treated in this chapter has the ground state $P_G|\psi_{ MF}\rangle$. Although the topological properties of $|\psi_{ MF}\rangle$ may again be classified by studying the auxiliary Hamiltonian $H_{Aux}$, a complication arises when we apply the Gutzwiller projection $P_G$.  This complicated comes from the fact that it is unclear whether the projection operator $P_G$ will change the topology of $|\psi_{Aux}\rangle$. 

The ease of defining a topological invariant for the auxiliary Hamiltonian $H_{Aux}$ comes from the fact that it can be written in terms of ``good" single particle states\cite{Berry}. For a general interacting many-body state such as  $P_G|\psi_{ MF}\rangle$ this is not possible. The literature on finding the topological properties of a complicated many-body state is not well developed and therefore classifying $P_G|\psi_{ MF}\rangle$ at the time of writing is a difficult and not well defined task. One possibility the author has considered lies in studying the entanglement entropy of this wave function\cite{alictop,bergtop}, but little progress in this direction has been made. Owing to this difficulty, we leave classifying the topology of the system to future study.

\newpage

\newpage
\chapter{Variational Monte Carlo}

\section{Outline of this Chapter} 

This chapter, like the previous one, is interested in studying the strong coupling Hamiltonian derived in Chapter 4. This time we opt to evaluate the relevant variational energy numerically using Monte Carlo techniques. The next section of this chapter describes the ground state wave function we will study, the one after outlines the technique to be used and the closing section discusses some preliminary results of this method. 

\section{Variational Ground State}

The variational wave function we will use is chosen in complete analogy with similar work in the context of the high-$T_c$ superconductors\cite{param}. In the context of the high-$T_c$'s, one begins with a standard BCS type wave function and then encodes strong correlations by applying the so-called Gutzwiller projection. In place of the BCS mean field wave function, we will use a more general ``auxiliary" wave function which is the ground state of a suitably chosen auxiliary Hamiltonian. Towards this end, the variational ground state we use is
\begin{equation}\label{var}
|\psi_{Var}\rangle =  \exp(-iS) P_GP_{S_z}P_N |\psi_{Aux}\rangle.
\end{equation}  
The remainder of this section will be spent describing each of the 5 terms above. 

To begin, the auxiliary state $ |\psi_{Aux}\rangle$ is our analog of $|\psi_{BCS}\rangle$. It is obtained by diagonalizing some auxiliary Hamiltonian, $H_{Aux}$, built out of variational parameters. The cost of additional variational parameters in variational Monte Carlo is very numerically expensive. Each additional parameter introduces a new dimension in which we must minimize and therefore increases the number of function evaluations our minimization routine must perform. Function evaluations in Variational Monte Carlo calculations are the most costly part of the algorithm and so each additional variational parameter slows down the routine by a considerable amount. One is then left with the job of weighing the numerical cost of each variational parameters against the benefit of the additional physical information it may provide. In our cost-benefit weighting we have chosen our auxiliary Hamiltonian to have the enough variational parameters to allow for a constructive comparison with the variational mean field theory and Gutzwiller approximation results from earlier in this thesis. 

To this end we take the following auxiliary Hamiltonian 
\begin{equation}
H_{Aux} = T+H_{SO} + H_{SC}
\end{equation}
where
\begin{equation}
H_{SC} = \sum_{\kv} (\Delta_\kv c_{\kv,\uparrow}^\dagger c_{-\kv, \downarrow}^\dagger +\text{h.c.})
\end{equation}
with $\Delta_\kv$ is built out of variational parameters and taken to have whatever symmetry we wish to study. This amounts to an $S=0$ version of the auxiliary Hamiltonian in Chapter 3. 

The method of diagonalizing $H_{Aux}$ closely parallels the method outlined in Chapter 3 with the exception that the form of the ground state we require for Variational Monte Carlo is different. We consider the process of diagonalizing the following auxiliary Hamiltonian
\begin{eqnarray}
H_{Aux} &=& \sum_{\kv,\sigma} (\xi_\kv c^\dagger _{\kv, \sigma} c_{\kv, \sigma} + (d_1-i\sigma d_2)c^\dagger _{\kv, \sigma} c_{\kv, \bar{\sigma}})  \\ \nonumber  &+&\sum_{\kv,\sigma}\sigma d_3 c^\dagger _{\kv, \sigma} c_{\kv, \sigma}+\sum_{\kv} (\Delta_\kv c_{\kv, \uparrow}^\dagger  c_{-\kv, \downarrow}^\dagger +h.c.)
\end{eqnarray}
where $\xi_\kv = \epsilon_\kv -\tilde{\mu}$ and $\epsilon_\kv = -2t(\cos{k_x}+\cos{k_y})$. The above result can equivalently be written as
\begin{equation}
H_{Aux} = \frac{1}{2} \sum_{\kv} \Psi_{\kv}^\dagger \Lambda_\kv \Psi_\kv
\end{equation}
where we have defined $\Psi_\kv = (c_{\kv,\uparrow}, c_{\kv, \downarrow} , c_{-\kv, \uparrow}^\dagger, c_{-\kv, \downarrow}^\dagger)^T$ and
\begin{eqnarray}
\Lambda_\kv &=& \left (  \begin{matrix} 
      \xi_\kv +d_3 & d_1-id_2 &0& \Delta_\kv \\
      d_1+id_2 & \xi_\kv -d_3 &-\Delta_\kv&0 \\
      0 & -\Delta_\kv& -\xi_\kv -d_3 & d_1+id_2\\
      \Delta_\kv & 0 & d_1-id_2 & -\xi_\kv +d_3\\
   \end{matrix}\right)
  \nonumber \\   &\equiv& \left(   \begin{matrix} 
         h_\kv & i\Delta_\kv \sigma_y \\
        -i\Delta_\kv \sigma_y & -h_{-\kv}^* \\
      \end{matrix}\right)
\end{eqnarray}
We now wish to bring this auxiliary Hamiltonian to the diagonal form
\begin{equation}
H_{Aux} = \frac{1}{2} \sum_{\kv} \Phi_{\kv}^\dagger \bar{\Lambda}_\kv \Phi_\kv
\end{equation}
where $\Phi_\kv = (\gamma_{\kv,\uparrow}, \gamma_{\kv, \downarrow} , \gamma_{-\kv, \uparrow}^\dagger, \gamma_{-\kv, \downarrow}^\dagger)^T$ and
\begin{equation}
\bar{\Lambda}_\kv = \left (  \begin{matrix} 
      E_{\kv, \uparrow} & 0&0&0 \\
      0&E_{\kv,\downarrow}&0&0 \\
      0&0&-E_{-\kv,\uparrow}&0\\
      0&0&0&-E_{-\kv,\downarrow}\\
   \end{matrix}\right)
\end{equation}
The fact that we can write the eigenvalues in terms of $E_{\kv,\sigma}$ and $-E_{-\kv,\sigma}$ is guaranteed by the particle hole symmetry of the system, namely the fact that $\Gamma \Lambda_\kv \Gamma = -\Lambda_{-\kv}^*$ where 
\begin{equation}
\Lambda = \left (  \begin{matrix} 
     0&I_{2\times2} \\
      I_{2\times2} &0 \\
   \end{matrix}\right).
 \end{equation}
Diagonalization of $\Lambda_\kv$ is of course done by making the unitary transformation $\bar{\Lambda}_\kv=W_\kv ^\dagger \Lambda_\kv W_\kv  $ and $\Psi_\kv = W_\kv \Phi_\kv$. We can write the unitary matrix $W_\kv$ in terms of $2\times2$ blocks as follows
\begin{equation}
W_\kv = \left(   \begin{matrix} 
         u_\kv &v^*_{-\kv} \\
        v_\kv & u_{-\kv}^* \\
      \end{matrix}\right)
\end{equation}
the form of which is again made possible by the particle-hole symmetry of the BdG Hamiltonian $\Lambda_\kv$. Our new annihilation operators are then given by
\begin{equation}\label{quasi}
\gamma_{\kv, \alpha } = \sum_{\beta} (u^*_{\kv, \beta,\alpha} c_{\kv, \beta} + v^*_{\kv, \beta, \alpha} c^\dagger _{-\kv, \beta})
\end{equation}
and the final form of our auxiliary Hamiltonian is
\begin{equation}
H_{Aux} = \sum_{\kv, \alpha} E_{\kv, \alpha} \gamma_{\kv, \alpha}^\dagger \gamma_{\kv, \alpha} +E_0
\end{equation}

For positive dispersion $E_{\kv, \alpha}>0$ the ground state of the above Hamiltonian is a state of zero $\gamma$ quasiparticles, that is to say we define $\gamma_{\kv, \alpha} |\psi_{Aux}\rangle =0$ $\forall (\kv, \alpha)$. In principle we can find a form of this ground state if we begin with the electron vacuum and then ``empty" it of any quasiparticles 
\begin{equation}
|\psi_{Aux}\rangle = \prod_{\kv, \alpha} (\gamma_{\kv, \alpha}) |0\rangle. 
\end{equation} 
The property $\gamma_{\kv, \alpha} |\psi_{Aux}\rangle =0$ is then guaranteed by the fact that the $\gamma$ particles obey anti-commutation relations (or, more specifically, because of the fact that the electron operators and $\gamma$ operators are related by a unitary transformation). The above definition, although compact, is not in a form convenient to our Monte Carlo integration and so we are forced to adopt a different approach to finding $|\psi_{Aux}\rangle$. 

To obtain a form of the auxiliary ground state this is suitable for Monte Carlo integration we begin with the following {\it ansatz} for the ground state
\begin{equation}
|\psi_{Aux}\rangle = \exp{\left(\sum_{\kv, \alpha,\beta} a_{\alpha,\beta}(\kv) c^\dagger_{\kv, \alpha} c^\dagger_{-\kv,\beta}\right)} |0\rangle 
\end{equation}
where the matrix $a_{\alpha,\beta}(\kv)$ is initially unknown. In order to fix $a_{\alpha,\beta}(\kv)$ we begin by insisting that this form of $|\psi_{Aux}\rangle$ is indeed a vacuum of $\gamma$ particles. In other words we must have the following property
\begin{equation}
\gamma_{\kv, \alpha} |\psi_{Aux}\rangle =0  \ \ \ \text{          or          }  \ \ \  \sum_{\beta} (u^*_{\kv, \beta,\alpha} c_{\kv, \beta} + v^*_{\kv, \beta, \alpha} c^\dagger _{-\kv, \beta})|\psi_{Aux}\rangle =0
\end{equation}
We now define the following vectors $C_{\kv} = (c_{\kv,\uparrow}, c_{\kv,\downarrow})^T$ and $C_{-\kv}^\dagger = (c^\dagger_{-\kv,\uparrow}, c^\dagger_{-\kv,\downarrow})^T$ and the condition on the ground state above reads
\begin{equation}\label{req1}
(u^\dagger_{\kv} C_{\kv} + v^\dagger_{\kv} C^\dagger _{-\kv})|\psi_{Aux}\rangle =0
\end{equation}

Let us now focus on the actual effect of $c_{\kv,\alpha}$ acting on $|\psi_{Aux}\rangle$. We first define $B \equiv  \sum_{\kv, \alpha,\beta} a_{\alpha,\beta}(\kv) c^\dagger_{\kv, \alpha} c^\dagger_{-\kv,\beta}$ and note that the canonical anticommutation relations for the electron operators and some commutator identities can be used to show the following
\begin{equation}
[c_{\kv,\alpha}, B ] = \sum_{ \alpha'} \left(a_{\alpha,\alpha'}(\kv)-a_{\alpha',\alpha}(-\kv) \right)c^\dagger_{-\kv,\alpha'} \equiv Y_{\kv,\alpha}
\end{equation}
Using the above and also the fact that $[B, Y_{\kv, \alpha}]=0$ one can easily show $[c_{\kv,\alpha}, e^{B}] = Y_{\kv, \alpha} e^{B}$. Using this essential commutation relation and noting that our ground state can be written $|\psi_{Aux}\rangle = e^B|0\rangle$ we have the following 
\beq
c_{\kv,\alpha} |\psi_{Aux}\rangle = c_{\kv,\alpha} e^B|0\rangle = Y_{\kv,\alpha} e^B|0\rangle+ e^Bc_{\kv,\alpha}|0\rangle = Y_{\kv,\alpha} |\psi_{Aux}\rangle
\eeq 
Straightforward manipulation of the above then yields 
\begin{eqnarray}
&&(c_{\kv,\alpha} - \sum_{\alpha'}\tilde{a}_{\kv, \alpha,\alpha'} c_{-\kv, \alpha'}^\dagger)|\psi_{Aux}\rangle =0\\ \nonumber &&  \ \text{or} \ (C_\kv-\tilde{a}C^\dagger_{-\kv})|\psi_{Aux}\rangle =0
\end{eqnarray}
where $\tilde{a}_{\kv, \alpha,\alpha'}  = \left(a_{\alpha,\alpha'}(\kv)-a_{\alpha',\alpha}(-\kv) \right)$. Through comparison with Eq. (\ref{req1}) we find
\begin{equation}
a_{\alpha,\alpha'}(\kv)-a_{\alpha',\alpha}(-\kv)  = -\sum_{\gamma} (u_\kv ^\dagger)^{-1} _{\alpha,\delta} (v_\kv^\dagger)_{\delta, \alpha'}
\end{equation}
We finally obtain the desired result for the coefficients $a_{\alpha,\alpha'}(\kv)$ by making note of a symmetry they must obey. The combination $c_{\kv,\alpha}^\dagger c_{-\kv,\beta}^\dagger$ in the exponent of $|\psi_{Aux}\rangle$ is antisymmetric under the label exchange $\kv\to-\kv$, $\alpha \leftrightarrow\beta$. Owing to this fact,  $a_{\alpha,\alpha'}(\kv)$ must also be antisymmetric under this exchange of labels: $a_{\alpha,\alpha'}(\kv)=-a_{\alpha',\alpha}(-\kv)$. We therefore have
\begin{equation}\label{aaak}
a_{\alpha,\alpha'}(\kv)  = -\frac{1}{2}\sum_{\delta} (u_\kv ^\dagger)^{-1} _{\alpha,\delta} (v_\kv^\dagger)_{\delta, \alpha'}
\end{equation}
We then write the Fourier transform $a_{\alpha,\beta}(i-j) = \frac{1}{N} \sum_{\kv} e^{-i\kv\cdot(\rv_i-\rv_j)} a_{\alpha,\beta }(\kv) $ or
\begin{equation}
a_{\alpha,\beta}(i-j)= - \frac{1}{2N} \sum_{\kv, \delta} e^{-i\kv\cdot(\rv_i-\rv_j)} (u_\kv ^\dagger)^{-1} _{\alpha,\delta} (v_\kv^\dagger)_{\delta, \beta}
\end{equation}
Fourier transforming both annihilation operators in the exponent of  $|\psi_{Aux}\rangle$ then results in 
\begin{equation}\label{aux}
|\psi_{Aux}\rangle = \exp{\left(\sum_{i,j, \alpha,\beta} a_{\alpha,\beta}(i-j) c^\dagger_{i, \alpha} c^\dagger_{j,\beta}\right)} |0\rangle 
\end{equation}
This and the expression for $a_{\alpha,\alpha'}(\kv)$ given in Eq. (\ref{aaak}) constitutes the final form of the auxiliary ground state necessary for a Monte Carlo study of the system. 

Readers familiar with implementation of Monte Carlo integration with the BCS wave function should note that the above result readily simplifies to the BCS result when the appropriate limit is taken. This is shown by noting that in the BCS limit $A,B,M\to0$ we have the matrices
\begin{equation}
u_{\kv, BCS} =   \left( \begin{matrix} 
      \tilde{u}_\kv^* & 0 \\
      0 & \tilde{u}_{-\kv}^* \\
   \end{matrix}\right)  \ \ \ \ \
   v_{\kv, BCS} =   \left( \begin{matrix} 
      0&  \tilde{v}_{-\kv}^*  \\
     -\tilde{v}_\kv^* & 0 \\
   \end{matrix}\right)
\end{equation}
where $\tilde{u}_\kv$ and $\tilde{v}_\kv$ are the familiar BCS coherence factors. We then have
\begin{equation}
(u^\dagger_{\kv, BCS})^{-1} =   \left( \begin{matrix} 
     \frac{1}{ \tilde{u}_\kv} & 0 \\
      0 & \frac{1}{\tilde{u}_{-\kv}} \\
   \end{matrix}\right)  \ \ \ \ \
   v^\dagger_{\kv, BCS} =   \left( \begin{matrix} 
      0&  -\tilde{v}_{\kv}  \\
     \tilde{v}_{-\kv} & 0 \\
   \end{matrix}\right)
\end{equation}
Then we immediately find
\begin{equation}
a_{\alpha,\beta}(\kv) =\frac{1}{2}\left( \begin{matrix} 
    0 & \frac{\tilde{v}_\kv}{\tilde{u}_\kv} \\
      -\frac{\tilde{v}_{-\kv}}{\tilde{u}_{-\kv}} & 0 \\
   \end{matrix}\right)
\end{equation}
leading to
\begin{equation}
|\psi_{Aux}\rangle_{BCS} = \exp{\left(\sum_\kv  \frac{\tilde{v}_\kv}{\tilde{u}_\kv} c^\dagger_{\kv,\uparrow} c_{-\kv, \downarrow}^\dagger\right)} |0\rangle 
\end{equation}
which is an alternative form of the standard BCS wave function and in particular the form that has been used in Monte Carlo calculations \cite{param}

The next two terms in $|\psi_{Var}\rangle$ are the projection operators $P_N$ and $P_{S_z}$. These operators are applied in order to make the wave function suitable for Monte Carlo integration. First, we note that, like the BCS ground state, $|\psi_{Aux}\rangle$ as written in Eq. (\ref{aux}) {\em does not} have a fixed number of particles. To take care of this $P_N$ projects the state onto a subspace with exactly $N$ pairs of electrons. Next, $|\psi_{Aux}\rangle$ does not have a fixed total $z$ component of the spin. To address this we follow Ref. \cite{weber} and apply the projection $P_{S_z}$, which projects us onto a space with $S_z=0$.   

So far we have not taken strong correlations into account, this role is filled by the final two terms in Equation (\ref{aux}). For large $U_0$ the system will avoid configurations where more than one electron occupies a single lattice site, {\it i.e.} in the language of Chapter 4 it will minimize $\tilde{D}$. As our current study is focused on a system which is either at half filling or hole doped away from half filling, the extreme limit $U_0\to \infty$ will result in $D=0$. We begin by taking this infinite $U_0$ limit and therefore apply the Gutzwiller projection, $P_G$, to the above wave function. The expression for the Gutzwiller projection operator is given by
\begin{equation}
P_G = \prod_{i} (1-n_{i,\uparrow}n_{i, \downarrow}).
\end{equation} 

In order to back off from the rather strict infinite $U$ requirement, a variational parameter $0< g <1$ is usually introduced to define the partial Gutzwiller projector
\begin{equation}
P_G = \prod_{i} (1-(1-g)n_{i,\uparrow}n_{i, \downarrow}).
\end{equation} 
To avoid this additional variational parameter, the authors of \cite{param} have instead suggested the introduction of a unitary operator $e^{-iS}$ which ``builds in the effects of double occupancy in perturbatively". This unitary operator is {\em the same} as the one defined in transforming $H$ to $\tilde{H}$ in Chapter 4. We follow the example set by Ref. \cite{param} and finally arrive at the trial wave function we started with in Equation (\ref{var}). 

With our variational wave function defined and fully motivated we move on to a description of Variational Monte Carlo methods. 

\section{Overview of Variational Monte Carlo Methods}
\subsection{Variational Energy and the Local Energy}
At the core of Variational Monte Carlo (VMC) is a trial wave function constructed out of a set of variational parameters $\{\lambda\}$ and denoted $|\psi(\lambda)\rangle$. We are again interested in the calculation of a trial energy
\begin{equation}\label{varE1}
E_t(\lambda)= \frac{\langle \psi(\lambda) |H|\psi(\lambda)\rangle}{\langle (\lambda) |\psi(\lambda)\rangle}.
\end{equation}
In principle VMC can {\em always} be used to calculate this variational energy, however it is most practical when $E_t$ cannot be calculated analytically. 

In deriving the results needed to perform VMC we begin by considering real space electronic states  $|\rv^{2N_e}\rangle = c_{\rv_1, \sigma_1}^\dagger c_{\rv_2, \sigma_2}^\dagger...c_{\rv_{2N}, \sigma_{2N_e}}^\dagger|0\rangle$ where $\rv^{2N_e}=   ((\rv_1, \sigma_1),(\rv_2, \sigma_2)...(\rv_{2N}, \sigma_{2N_e}))$ labels the real space position and spin of $2N_e$ electrons on our $N$ site lattice. We then form the identity $1=\sum_{\rv^{2N_e}}  |\rv^{2N_e}\rangle \langle\rv^{2N_e}|$ and use a copy of it in both the numerator and denominator of Eq. (\ref{varE1}) to write
\begin{equation}
E_{t}(\lambda) = \sum_{\rv^{2N_e}} \frac{\psi_\lambda(\rv^{2N_e}){H}\psi_\lambda(\rv^{2N_e})}{\sum_{\rv'^{2N_e}}|\psi_\lambda(\rv'^{2N_e})|^2}
\end{equation}
where we have defined $\psi_\lambda(\rv^{2N_e}) = \langle \rv^{2N_e}|\psi(\lambda)\rangle$ and ${H}\psi_\lambda(\rv^{2N_e}) = \langle \rv^{2N_e}|{H}|\psi(\lambda)\rangle$. It is now useful to define the local energy $E_\lambda(\rv^{2N_e}) =  \frac{{H}\psi_\lambda(\rv^{2N_e})}{\psi_\lambda(\rv^{2N_e})}$ and its ``probability" $P_\lambda(\rv^{2N_e}) = \frac{|\psi_\lambda(\rv^{2N_e})|^2}{\sum_{\rv'^{2N_e}}|\psi_\lambda(\rv'^{2N_e})|^2}\ge0$. In terms of these definitions $E_{t}$ is written in the suggestive form
\begin{equation}\label{evarmc}
E_{t}(\lambda) = \sum_{\rv^{2N_e}} P_\lambda(\rv^{2N_e}) E_\lambda(\rv^{2N_e}). 
\end{equation}
We now make the following key observation: as written above, $E_t(\lambda)$ looks like a statistical average of the random variable $E_\lambda(\rv^{2N_e}) $ which occurs with probability $ P_\lambda(\rv^{2N_e}) $. In principle the above average could be performed to give the exact energy $E_{t}(\lambda)$ as up until this point we have yet to make any approximations. This line of thinking becomes problematic when we recall that the sum in Eq. (\ref{evarmc}) is over {\em all} possible many-body configurations of the $2N$ electrons. The number of terms in this statistical average is thus enormous ({\it i.e.} incalculable) for even a relatively small lattice.

We must now think of intelligent ways to approximate the sum in Equation (\ref{evarmc}). One naive way might be to select a few highly probable terms in the sum and only take their contribution into account. This suggestion falls short when one recalls the definition $P_\lambda(\rv^{2N}) = \frac{|\psi_\lambda(\rv^{2N_e})|^2}{\sum_{\rv'^{2N_e}}|\psi_\lambda(\rv'^{2N_e})|^2}$; we see just to know the probability of a configuration $\rv^{2N_e}$ we still require a sum over all possible configurations. 

Up to this point we have not yet gotten to the Monte Carlo part of the technique, we have only outlined the Variational part of the theory. The Monte Carlo part comes in as a practical way to approximate (to very high precision) the sum involved in $E_t(\lambda)$. This approximation is done as follows. First, we use the Metropolis algorithm (to be defined shortly) to generate a sequence of many-body configurations, $\{\rv^{2N_e}\}_{MC}$. This sequence is generated as a ``walk" through configuration space where steps are given statistical weight based on {\em relative} probabilities. With this sequence generated the variational energy is approximated as a simple arithmetic average over the local energies of each configuration in the sequence. Thus,
\begin{equation}
E_{v}(\lambda) = \frac{1}{N_{mc}} \sum_{\{\rv^{2N_e}\}_{MC}} E_\lambda(\rv^{2N_e})
\end{equation}
where $N_{mc}$ is the number of configurations sampled. 

What's left to do is describe how the sequence of configurations is generated. The set of configurations $\{\rv^{2N_e}\}_{MC}$ is determined by beginning at some (random) configuration $\rv^{2N_e}_i$ and proposing a random move to a new configuration $\rv'^{2N_e}_i$. This move to a new configuration is in turn accepted with the Metropolis probability
\begin{equation}
P = \min\left\{1, \left|\frac{\langle \rv'^{2N_e}_i |{\psi}(\lambda)\rangle}{\langle \rv^{2N_e}_i |\tilde{\psi}(\lambda)\rangle}\right|^2\right\}.
\end{equation}
This process generates a string of configurations with a high {\em relative} probability. Note the very important fact that $\left|\frac{\langle \rv'^{2N_e}_i |{\psi}(\lambda)\rangle}{\langle \rv^{2N_e}_i |\tilde{\psi}(\lambda)\rangle}\right|^2$ does not contain a sum over all possible configurations, the problematic denominator $\sum_{\rv'^{2N_e}}|\psi_\lambda(\rv'^{2N_e})|^2$ is cancelled when considering relative probabilities.

\subsection{Fast Updates Using Pfaffians}

In the previous subsection we have shown that the task of finding $E_{v}$ is totally dependent on one's ability to find ratios of the form $\frac{\langle \rv'^{2N_e} |{\psi}(\lambda)\rangle}{\langle \rv^{2N_e} |{\psi}(\lambda)\rangle}$ as well as to calculate the local energy $E(\rv^{2N_e})$. Both of these objects rely on the ability to calculate the ratio of real space overlaps where the two configurations in question differ by only a few electron positions. It is then very important to have efficient ways of calculating these ratios. This subsection will discuss some efficient methods used in the past as well as an overview of how we have generalized the methods to be suitable for the current problem. 

At this point we specialize to discuss the variational wave function $|\psi(\lambda)\rangle=|\psi_{Var}\rangle $, where $|\psi_{Var}\rangle $ is as defined in Equation (\ref{var}). We then require the calculation of objects like $\langle \rv^{2N_e}|{\psi}_{Var}\rangle$. Let us assume for the sake of illustration that a specific configuration $|\rv^{2N_e}\rangle$ contains no double occupancies and has $S_z=0$. Then, ignoring the unitary transformation $e^{iS}$ for reasons that will be discussed in the next section, we are required to calculate 
\begin{equation}
\langle \rv^{2N_e} |{\psi}_{Var}\rangle = \langle \rv^{2N_e} |{\psi}_{Aux}\rangle
\end{equation}
Writing out the expression for $|\psi_{Aux}\rangle$ given in Eq. (\ref{aux}), expanding the exponential, keeping only the term with $2N$ electrons and then noting that $|\rv^{2N_e}\rangle = c_{\rv_1, \sigma_1}^\dagger c_{\rv_2, \sigma_2}^\dagger... c_{\rv_{2N_e}, \sigma_{2N_e}}^\dagger|0\rangle$ the above becomes

\begin{eqnarray}
 \langle \rv^{2N_e} |{\psi}_{Aux}\rangle &=&  \sum_{i_1...i_{N_e}, j_1...j_{N_e}}\sum_{ \alpha_1....\alpha_{N_e}, \beta_1...\beta_{N_e}} a_{\alpha_1,\beta_1}(i_1-j_1)  a_{\alpha_2,\beta_2}(i_2-j_2)...a_{\alpha_{N_e},\beta_{N_e}}(i_{N_e}-j_{N_e}) \nonumber  \\ &\times& \langle 0| c_{\rv_{2N_e}, \sigma_{2N_e}}... c_{\rv_2, \sigma_2} c_{\rv_1, \sigma_1} c^\dagger_{i_1, \alpha_1} c^\dagger_{j_1,\beta_1}...c^\dagger_{i_{N_e}, \alpha_{N_e}} c^\dagger_{j_{N_e},\beta_{N_e}}|0\rangle
 \end{eqnarray}
Inspection of the right hand side of the above object reveals that it is precisely the expression for the Pfaffian of an appropriately defined matrix\cite{weber}. Specifically, 
\begin{equation}
 \langle \rv^{2N_e} |{\psi}_{Aux}\rangle  = \Pf(Q)
\end{equation}
where $Q_{i,j} = a_{\sigma_i,\sigma_j} (\rv_i-\rv_j)-a_{\sigma_j,\sigma_i} (\rv_j-\rv_i)$. The Pfaffian is a property only defined for even dimensional, skew-symmetric matrices (note the important fact that $Q^T=-Q$).  It obeys the important relation $\Pf(Q)^2=\text{Det}(Q)$. An extensive discussion of the properties of Pfaffians is found in Reference \cite{bajdich}. In what follows, we will make use of the properties outlined in this reference in order to outline fast update procedures. 

Writing things in terms of Pfaffians, the central object required in determining whether or not a configuration is accepted by the Metropolis criterion is 
\begin{equation}
\frac{\langle \rv'^{2N_e} |{\psi}_{Aux}\rangle}{\langle \rv^{2N_e} |{\psi}_{Aux}\rangle} = \frac{\Pf(R)}{\Pf(Q)}
\end{equation}
where $R_{i,j} = a_{\sigma'_i,\sigma'_j} (\rv'_i-\rv'_j)-a_{\sigma'_j,\sigma'_i} (\rv'_j-\rv'_i)$ is the overlap matrix of the updated configuration. Calculating both Pfaffians would be computationally expensive as both require an order $N^3$ algorithm. Fortunately there are ways around calculating the full pfaffian.

Let us begin by considering a proposed update $\rv^{2N_e} \to \rv'^{2N_e}$ which only moves a single electron, say electron $\ell$. In this case the matrix $R$ differs from the matrix $Q$ by only the $\ell^{th}$ row {\em and} the $\ell^{th}$ column. For matrices obeying this property there is the following identity\cite{bajdich}
\begin{equation}
\text{Det}(\tilde{R}) = \Pf(Q)\Pf(R)
\end{equation} 
where $\tilde{R}$ is identical to $Q$ except for its $\ell^{th}$ row which is identical to the $\ell^{th}$ row of $R$. Using $\Pf(Q)^2=\text{Det}(Q)$ the above is easily manipulated to give
\begin{equation}
\frac{\Pf(R)}{\Pf(Q)} = \frac{\text{Det}(\tilde{R})}{\text{Det}(Q)}
\end{equation} 
The ratio of two determinants that differ by only a single row has been well known since very early in Variational Monte Carlo methods\cite{ceperley} and the following result immediately follows
\begin{equation}\label{update}
\frac{\Pf(R)}{\Pf(Q)} = \sum_{j} R_{\ell, j} Q^{-1}_{j,\ell}
\end{equation} 

According to the above, the cost of not needing of calculate a full Pfaffian is storing the inverse of $Q$. If the new configuration $\rv'^{2N_e}$ is accepted we must therefore update the inverse. Doing this involves use of the so-called Sherman-Morrison formula\cite{ceperley}. Two applications of the Sherman-Morrison formula and using the fact that a skew-symmetric matrix has a skew-symmetric inverse, {\it i.e.} $Q^{-1}_{i,i}=0$, gives\cite{tahara}
\begin{equation}\label{sm1}
R^{-1}_{i,j} = Q^{-1}_{i,j} -\frac{1}{r} \left(\beta_i Q^{-1}_{\ell,j} +Q^{-1}_{i,\ell}\beta_j \right) 
\end{equation}
where $\beta_i =   \sum_m R_{\ell,m} Q^{-1}_{m,i}-\delta_{i,\ell} $ and $r=\sum_m R_{\ell, m} Q^{-1}_{m,\ell}$. Equations (\ref{update}) and (\ref{sm1}) enable one to calculate any possible one-particle updates to the system in a computationally efficient way.    

We now move on to generalize the equation above to processes that move two electrons. This is a very important case for the strong coupling Hamiltonian we will study because, as outlined in the appendix, finding the local energy for this Hamiltonian involves updates where two electrons have been moved. 

The object we are interested in here is
\begin{equation}
\frac{\Pf(R)}{\Pf(Q)}
\end{equation}
where this time $R$ and $Q$ are identical except for the entries in their $\ell_1$ and $\ell_2$ rows and $\ell_1$ and $\ell_2$ columns. This difference corresponds to moving two electrons in the initial configuration. The above can be calculated by introducing the intermediate $S$ which is different from $Q$ by only one electron position and hence only column and row $\ell_1$. This gives
\begin{equation}
\frac{\Pf(R)}{\Pf(Q)} = \left(\frac{\Pf(R)}{\Pf(S)}\right)\left(\frac{\Pf(S)}{\Pf(Q)}\right)
\end{equation}
and it follows that we now have two, one electron updates to calculate. This corresponds to thinking of the update as first moving electron $\ell_1$ and then moving electron $\ell_2$. By using two versions of Eq. (\ref{update})  we have
\begin{equation}
\frac{\Pf(R)}{\Pf(Q)} = \left(\sum_{j} R_{\ell_2, j} S^{-1}_{j,\ell_2}\right)\left(\sum_{j} S_{\ell_1, j} Q^{-1}_{j,\ell_1}\right)
\end{equation}
The above would be efficient if we had information on $S^{-1}$, unfortunately we only know about $Q^{-1}$. As a result we need to write $S^{-1}$ in terms of $Q^{-1}$ using (\ref{sm1}). The result is
\begin{equation}
\frac{\Pf(R)}{\Pf(Q)} =s \left(\sum_{j} R_{\ell_2, j} (Q^{-1}_{j,\ell_2} -\frac{1}{s} \left(\beta_j Q^{-1}_{\ell_1,\ell_2} +Q^{-1}_{j,\ell_1}\beta_{\ell_2} \right)  ) \right)
\end{equation}
where $\beta_i = (-\delta_{i,\ell_1} + \sum_m S_{\ell_1,m} Q^{-1}_{m,i}) $ and $s=\sum_m S_{\ell_1, m} Q^{-1}_{m,\ell_1}$. Defining the extra shorthand notation
\begin{eqnarray}
&&\bar{s} = \sum_m S_{\ell_1, m} Q^{-1}_{m,\ell_2}  \\ \nonumber
&&r = \sum_m R_{\ell_2, m} Q^{-1}_{m,\ell_2} \\ \nonumber 
&&\bar{r} = \sum_m R_{\ell_2, m} Q^{-1}_{m,\ell_1}
\end{eqnarray}
the two electron update ratio can be written as
\begin{equation}
\frac{\Pf(R)}{\Pf(Q)} = Q^{-1}_{\ell_1,\ell_2}(R_{\ell_2,\ell_1} -(RQ^{-1}S)_{\ell_2, \ell_1}) +sr-\bar{s}\bar{r}+\bar{r}\delta_{\ell_1, \ell_2}
\end{equation}
In the trivial case $\ell_1=\ell_2$ we have $\bar{s},\bar{r},s\to r$ and the above reduces back to the one particle result $\frac{\Pf(R)}{\Pf(Q)} =r$. 

One could in principle keep going with this process for an arbitrary number of particle moves. After two particles the expressions grow increasingly complicated and so we stop here for practicality. The calculation in these higher order updates is identical to the  we have outlined above: we define intermediates and write the $n$ particle update as $n$ single particle updates, then calculate the Pfaffian ratio by application of Equations (\ref{update}) and (\ref{sm1}). 

The calculation of the local energy $E(\rv^{2N_e})$ for the strong coupling Hamiltonian we are interested in is very technical and as such we defer details of it to the appendix. We now proceed to discuss some preliminary results from our VMC study. 

\section{Preliminary Results}

 We are interested in studying the following variational energy of the full Hamiltonian $H$ defined in Chapter 2, (note that when we write $|\psi_{Aux}\rangle$ from this point forward we mean $P_{S_z}P_N|\psi_{Aux}\rangle$ )
\begin{eqnarray}
E_{Var} &=& \frac{\langle \psi_{Aux} | P_G  \exp(iS) H  \exp(-iS) P_G |\psi_{Aux}\rangle}{\langle \psi_{Aux} | P_G P_G |\psi_{Aux}\rangle} \nonumber  \\ &=& \frac{\langle \psi_{Aux} | P_G \tilde{H} P_G |\psi_{Aux}\rangle}{\langle \psi_{Aux} | P_G P_G |\psi_{Aux}\rangle}
\end{eqnarray}
where $\tilde{H} =  \exp(iS) H  \exp(-iS)$ is {\em exactly} the strong coupling transformed Hamiltonian from Chapter 4. Our problem is then equivalent to the variational study of the Hamiltonian $\tilde{H}$ in the ground state $P_G|\psi_{Aux}\rangle$. 

$E_{Var}$ is dependent on how we choose the function $\Delta_\kv$. Following what we found in Chapter 3, the most interesting phase of the model Hamiltonian $H$ from a topological standpoint is a $d+id$ superconductor. As such we choose $\Delta_\kv = \Delta^{(1)}(\cos{k_x}-\cos{k_y}) +i\Delta^{(2)}\sin{k_x}\sin{k_y}$ where $ \Delta^{(1)}$ and $ \Delta^{(2)}$ are variational constants. This choice leads to $E_{Var} = E_{Var}( \Delta^{(1)},  \Delta^{(2)})$ and our immediate task is  to minimize $E_{Var}$ over the 2D space $ ( \Delta^{(1)},  \Delta^{(2)})$. 

Work in this direction is still ongoing and to date only rough, preliminary results have been obtained. Consistent with what was found in Chapter's 3 and 5, these results support the claim that the variational ground state of the Hamiltonian $H$ (in some parameter regions) is $d+id$-wave superconductivity as we have found a minimum in $E_{Var}$ away from $( \Delta^{(1)},  \Delta^{(2)})=(0,0)$. Our goal with this work is to continue mapping a VMC phase diagram in order to form a firm comparison with the VMF results of Chapter 3. As additional future work we would also like to classify the topology of $P_G |\psi_{Aux}\rangle$ at points in the phase diagram where $d+id$-wave superconductivity is present. Similar to what was discussed in Chapter 5, such a study at the present time is made difficult by the fact that $P_G|\psi_{Aux}\rangle$ is not a ground state of some quadratic Hamiltonian.


\chapter{Conclusions}

\section{Summary of this Thesis}
Motivated by past work on semiconductor heterostructures \cite{Sau, Alicea}, this thesis has addressed the question of whether a system of interacting, spin-orbit coupled electrons in a Zeeman field can develop topological superconductivity. We began by outlining a model Hamiltonian which we feel appropriately encapsulates the physics of such a system. We then moved on to present an analysis of the system that was valid at weak electron-electron interaction strengths. This study found that some regions of our phases diagram do indeed support topological superconductivity. With these results presented we moved on to study the system of interest at large electron-electron interaction strengths. This saw us perform a strong coupling expansion in order to eliminate the parts of the Hamiltonian that are not relevant at strong coupling. We went on to develop two different methods for studying this effective strongly interacting Hamiltonian; the Gutzwiller approximation and Variational Monte Carlo. Additionally, we have shown and discussed preliminary results obtained using these two methods.   

The prevailing theme in the numerous methods for studying the model has been that $d+id$-wave superconductivity is the preferred ground state of our model in various regions of the phase diagram. At week coupling this $d+id$ wave phase can be shown to be topological in some parameter regimes. At strong coupling we have been unable to rigorously classify the topology of the $d+id$ superconductor, but are hopeful that it will be shown to be topological in the near future.

\section{Future Work}

The first category of future work we will discuss has to do with completing and extending some of the strong coupling work presented here. First, we would like to complete the numerical variational Monte Carlo study described in Chapter 6 in order to get a more complete view of the phase diagram for this model. Second, we would like to extend the Gutzwiller approximation in Chapter 5 to include next nearest neighbour terms. Although these terms will complicate the theory, they will ultimately lead to tendency towards $d+id$-wave pairing.  Having this tendancy towards $d+id$-wave pairing in the theory will be useful in making a more complete comparison between the Gutzwiller approximation and the results for both our VMF and VMC studies. Third, we would like to develop a method to classify the topology of the strongly interacting wave function $P_G|\psi_{Aux}\rangle$ in order to describe the topology of the system in the strong coupling limit. 

In addition to the extensions of the work presented here, there is also further practical work to be done. The results here provide a good proof-of-principle that a topological superconducting state may be found in a system of interacting, spin-orbit coupled electrons. Ultimately we would like to look at possible values for the parameters $t,A,B,M,\mu, U_0$ and $V_0$ in order to propose an actual device that may be studied in the laboratory. In principle this would involve {\it ab-initio} calculations and then comparison with known values of these parameters in real materials. 

An alternative to the above would be to look towards systems where our parameters could easily be tuned.  With recent advancements showing that spin-orbit coupling can be simulated in systems of cold-atoms\cite{Lin}, the model proposed and studied here may be of relevance to cold-atoms experiments. Cold-atomic systems have the advantage that the strengths of interaction and spin-orbit coupling parameters may be tuned in the  are not fixed as they are in materials, they can be tuned in the laboratory. With this in mind, our work suggests that if spin-orbit coupling and interactions can be tuned correctly, a topological superconductor may be simulated in a cold atoms system.    


\appendix

\chapter{Details of the Strong Coupling Expansion}

\section{Commutators for Channel Decomposition}

\subsection{The Hubbard Model}
Here we provide a rigorous derivation of the commutator identity $[H_U, T_m]=mU_0T_m$. To do this we begin by recalling the following commutator properties
\begin{equation}
[c_{i,\sigma}^\dagger c_{j,\sigma}, n_{\ell,\alpha}] = \delta_{\sigma,\alpha}(\delta_{j,\ell} - \delta_{i,\ell}) c_{i,\alpha}^\dagger c_{j,\alpha}
\end{equation}
\begin{equation}
[n_{\ell,\alpha}, n_{m,\beta} ] = - [n_{\ell,\alpha}, h_{m,\beta} ] =0
\end{equation}  
We will now use these to explicitly show that $[H_U, T_1]=U_0T_1$ and argue that results for $m=0$ and $m=-1$ follow in an identical fashion. We have
 \begin{eqnarray}
 [H_U, T_1] &=& -tU_0\sum_{\langle i,j\rangle ,\sigma,\ell}  [n_{\ell, \uparrow}n_{\ell,\downarrow}, n_{i,\bar{\sigma}}c_{i,\sigma}^\dagger c_{j,\sigma}h_{j,\bar{\sigma}}]
 \\ \nonumber &=& tU_0\sum_{\langle i,j\rangle ,\sigma,\ell} \left( n_{i,\bar{\sigma}}[c_{i,\sigma}^\dagger c_{j,\sigma},n_{\ell, \uparrow}]h_{j,\bar{\sigma}}n_{\ell,\downarrow}+n_{i,\bar{\sigma}}n_{\ell, \uparrow}[c_{i,\sigma}^\dagger c_{j,\sigma}n_{\ell,\downarrow}]h_{j,\bar{\sigma}}\right)
  \\ \nonumber &=& -tU_0\sum_{\langle i,j\rangle \sigma,\ell} \big{(} n_{i,\downarrow}^2c_{i,\uparrow}^\dagger c_{j,\uparrow}h_{j,\downarrow}-n_{i,\downarrow}c_{i,\uparrow}^\dagger c_{j,\uparrow}(h_{j,\downarrow}n_{j,\downarrow})\\ \nonumber &+&n_{i,\uparrow}^2c_{i,\downarrow}^\dagger c_{j,\downarrow}h_{j,\uparrow}-n_{i,\uparrow}c_{i,\downarrow}^\dagger c_{j,\downarrow}(h_{j,\uparrow}n_{j, \uparrow})\big{)}
 \end{eqnarray}
 using $n_{i,\sigma}^2=n_{i,\sigma}$ and $n_{j,\sigma}h_{j,\sigma}=0$ in the above we finally arrive at $[H_U, T_1] =U_0T_1$. 
 
\subsection{The Full Hamiltonian}

Similar to the last section, here we provide a rigorous derivation of the commutator $[H_0, T_{m,M,N}] =[H_U, T_{m,M,N}] +[H_V, T_{m,M,N}] = (mU_0 + (M-N)V_0)T_{m,M,N}$. In order to do this it is first useful to recall
\begin{equation}
\ {T}_{m,N_2,N_1} =\small{ \sum_{i,\delta,\sigma, \sigma'} \sum_{S[n_1]=N_1} \sum_{S[n_2]=N_2} O_i[n_2] ({T}_m)_{i,\delta,\sigma, \sigma'}O_{i+\delta}[n_1] }
\end{equation}
We can then derive the commutator expression by starting with 
\begin{eqnarray}
[H_U, T_{m,M,N}] &=& -tU_0\sum_{S[m]=M}\sum_{S[n]=N}\sum_{i,j,\ell,\sigma}[n_{\ell, \uparrow}n_{\ell, \downarrow},O_i[m]({T}_m)_{i,j,\sigma}O_j[n]]
\\ \nonumber &=& tU_0\sum_{S[m]=M}\sum_{S[n]=N}\sum_{i,j,\ell,\sigma}(O_i[m][({T}_m)_{i,j,\sigma}, n_{\ell, \uparrow}]O_j[n]n_{\ell, \downarrow}\\ \nonumber &+& O_i[m]n_{\ell, \uparrow}[({T}_m)_{i,j,\sigma},n_{\ell, \downarrow} ]O_j[n])
\end{eqnarray}  
From the section on the Hubbard model above it follows that
\begin{equation}
[({T}_m)_{i,j,\sigma}, n_{\ell, \uparrow}]n_{\ell, \downarrow} = -m\delta_{i, \ell} \delta_{\sigma,\uparrow}({T}_m)_{i,j,\uparrow}
\end{equation}
and 
\begin{equation}
n_{\ell, \uparrow}[({T}_m)_{i,j,\sigma}, n_{\ell, \downarrow}] = -m\delta_{i, \ell} \delta_{\sigma,\downarrow}({T}_m)_{i,j,\downarrow}
\end{equation}
thus
\begin{eqnarray}
[H_U, T_{m,M,N}]  &=& -mtU_0\sum_{S[m]=M}\sum_{S[n]=N}\sum_{i,j}O_i[m]({T}_m)_{i,j,\uparrow}O_j[n]+O_i[m]({T}_m)_{i,j,\downarrow}O_j[n]
 \nonumber \\ &=& mU_0T_{m,M,N}
\end{eqnarray}  
as desired. 

Now we consider $[H_V, T_{m,M,N}]$. To find this commutator it is useful to first write $H_V = V_0/2 \sum_{\ell,\alpha} n_{\ell,\alpha} \tilde{n}_\ell$ where $\tilde{n}_\ell = \sum_{\delta,\beta} n_{\ell,\beta}$ is the occupation number of all sites that are nearest neighbours to $\ell$. We then have
\begin{eqnarray}
[H_V, T_{m,M,N}]  &=& -\frac{V_0t}{2} \sum_{S[m]=M}\sum_{S[n]=N}\sum_{i,j,\ell,\sigma,\alpha}[n_{\ell, \alpha}\tilde{n}_{\ell},O_i[m]({T}_m)_{i,j,\sigma}O_j[n]]
\\ \nonumber &=&  \frac{V_0t}{2} \sum_{S[m]=M}\sum_{S[n]=N}\sum_{i,j,\ell,\sigma,\alpha}(O_i[m][({T}_m)_{i,j,\sigma}, n_{\ell, \alpha}]O_j[n]\tilde{n}_{\ell}\\ \nonumber&+&O_i[m]n_{\ell, \alpha}[({T}_m)_{i,j,\sigma},\tilde{n}_{\ell} ]O_j[n])
\end{eqnarray}
We now use the following
\begin{equation}
[({T}_m)_{i,j,\sigma}, n_{\ell,\alpha}] = \delta_{\sigma,\alpha}(\delta_{j,\ell}-\delta_{i, \ell}) ({T}_m)_{i,j,\sigma}
\end{equation}
as well as
\begin{equation}
[({T}_m)_{i,j,\sigma}, \tilde{n}_{\ell}]= \sum_{\delta} (\delta_{j, \ell+\delta} - \delta_{i,\ell+\delta}) ({T}_m)_{i,j,\sigma}
\end{equation}
to write
\begin{eqnarray}
[H_V, T_{m,M,N}]  &=&  
 \frac{V_0t}{2} \sum_{S[m]=M}\sum_{S[n]=N}\sum_{i,j,\sigma}(O_i[m]( T_m)_{i,j,\sigma}O_j[n]\tilde{n}_{j}\\ \nonumber &-&O_i[m]( T_m)_{i,j,\sigma}O_j[n]\tilde{n}_{i}) \\ \nonumber &+& \frac{Vt}{2} \sum_{S[m]=M}\sum_{S[n]=N}\sum_{i,j,\sigma,\alpha, \delta}(O_i[m]n_{j-\delta, \alpha}( T_m)_{i,j,\sigma}O_j[n]\\ \nonumber &-&O_i[m]n_{i-\delta, \alpha}( T_m)_{i,j,\sigma}O_j[n])
\\ \nonumber &=&  V_0t \sum_{S[m]=M}\sum_{S[n]=N}\sum_{i,j,\sigma}(O_i[m]( T_m)_{i,j,\sigma}O_j[n]\tilde{n}_{j}\\ \nonumber &-&O_i[m]\tilde{n}_{i}( T_m)_{i,j,\sigma}O_j[n]) \\ \nonumber &+& \frac{Vt}{2} \sum_{S[m]=M}\sum_{S[n]=N}\sum_{i,j,\sigma}(O_i[m][\tilde{n}_{j},( T_m)_{i,j,\sigma}]O_j[n]\\ \nonumber &-&O_i[m][( T_m)_{i,j,\sigma},\tilde{n}_{i}]O_j[n])
\\ \nonumber &=&  -V_0t \sum_{S[m]=M}\sum_{S[n]=N}\sum_{i,j,\sigma}\bigg{(}-O_i[m]( T_m)_{i,j,\sigma}O_j[n]\tilde{n}_{j}\\ \nonumber&+&O_i[m]\tilde{n}_{i}( T_m)_{i,j,\sigma}O_j[n]\bigg{)} 
\end{eqnarray}
Now we observe the following
\begin{eqnarray}
O_i[m]\tilde{n}_{i} &=& \prod_{\delta, \alpha} (m_{\delta,\alpha}n_{i+\delta,\alpha}+(1-m_{\delta,\alpha})h_{i+\delta,\alpha}) \sum_{\delta', \beta} n_{i+\delta', \beta}
\\ \nonumber &=&\sum_{\delta', \beta} \left(\prod_{(\delta, \alpha)\ne(\delta', \beta) } (m_{\delta,\alpha}n_{i+\delta,\alpha}+(1-m_{\delta,\alpha})h_{i+\delta,\alpha}) \right)\\ \nonumber &\times& (m_{\delta',\beta}n_{i+\delta',\beta}n_{i+\delta', \beta}+(1-m_{\delta',\beta})h_{i+\delta',\beta}n_{i+\delta', \beta})
\\ \nonumber &=&\sum_{\delta', \beta} \left(\prod_{(\delta, \alpha)\ne(\delta', \beta) } (m_{\delta,\alpha}n_{i+\delta,\alpha}+(1-m_{\delta,\alpha})h_{i+\delta,\alpha}) \right) m_{\delta',\beta}n_{i+\delta',\beta}
\end{eqnarray}
Note that in the summand above either $m_{\delta', \beta}=0$ and there is no contribution to the sum or $m_{\delta', \beta}=1$ and $n_{i+\delta', \beta}$ returns to its spot in the product over $(\delta,\alpha)\ne(\delta', \beta)$. Thus we have
\begin{eqnarray*}
O_i[m]\tilde{n}_{i} &=&\left(\sum_{\delta', \beta}m_{\delta',\beta}\right) \prod_{\delta, \alpha} (m_{\delta,\alpha}n_{i+\delta,\alpha}+(1-m_{\delta,\alpha})h_{i+\delta,\alpha}) = S[m]O_i[m]
\end{eqnarray*}
Similarly $\tilde{n}_{j}O_j[n]= S[n]O_j[n]$. This leads to
\begin{eqnarray}
\nonumber [H_V, T_{m,M,N}]   &=&  -V_0t \sum_{S[m]=M}\sum_{S[n]=N}\sum_{i,j,\sigma}\left(-S[n]O_i[m](\tilde{T}_m)_{i,j,\sigma}O_j[n]+S[m]O_i[m](\tilde{T}_m)_{i,j,\sigma}O_j[n]\right) 
\\ \nonumber &=&  -V_0t \sum_{S[m]=M}\sum_{S[n]=N}\sum_{i,j,\sigma}\left(-NO_i[m](\tilde{T}_m)_{i,j,\sigma}O_j[n]+MO_i[m](\tilde{T}_m)_{i,j,\sigma}O_j[n]\right) \\ &=&(M-N)V_0T_{m,M,N}
\end{eqnarray}
our desired result. 

\section{Solution of Commutator Equations} 

Here we give the solution to the commutator equations for the transformation $S$. Rather than presenting specific results for both models discussed in Chapter 4, we give a general description of a strong coupling expansion without specifying exactly what Hamiltonian it is specific to. 

Imagine we can write our Hamiltonian in two pieces $H=H_0+H_1$ where $H_0$ describes some (large) interaction effects and $H_1$ describes some hopping process. As was described in Chapter 4 we begin with the following transformation
\begin{equation}
c^\dagger_{i,\sigma} = e^{iS(\bar{c})} \bar{c}^\dagger_{i,\sigma}e^{-iS(\bar{c})}
\end{equation}
in such a way that the ``interaction" term, $H_0(\bar{c})$, is a constant of motion.  Applying the unitary transformation to our full Hamiltonian we write
\begin{equation}
H = e^{S(\bar{c})} (H_0(\bar{c})+H_1(\bar{c}))e^{-S(\bar{c})}\equiv H_0(\bar{c}) + H_1'(\bar{c})
\end{equation}
where $H_0(\bar{c})$ contains the ``large" terms in the Hamiltonian while $H_1$ contains the ``small" terms in the Hamiltonian. To be concrete, if we have $H_0=\chi\tilde{H}_0 $ and $H_1=\tau \tilde{H}_1$ we are interested in the limit ${\tau}/{\chi} \ll 1$ . In order for $H_0$ to be a constant of motion we must then have
\begin{equation}\label{condition}
[H_0(\bar{c}), H] = [H_0(\bar{c}), H_1'(\bar{c})] = 0. 
\end{equation}
To satisfy this condition we expand both $H_1'$ and $S(\bar{c})$ in a power series in the large parameter, $\chi$, contained in $H_0$ (later this will turn out to be $U_0$, the interaction strength). More concretely, we write
\begin{equation}
S(\bar{c}) = \sum_{n=1}^\infty \frac{1}{\chi^n}S_n(\bar{c}); \ \ \ \ \ \ H_1(\bar{c}) = \sum_{n=1}^\infty \frac{1}{\chi^{n-1}}H'_{1,n}(\bar{c}).
\end{equation}
 Our task is now to determine the contributions $S_n$ and $H_{1,n}'$ by insisting that Eq. (\ref{condition}) is satisfied to a given order in $1/\chi$, {\it i.e.}
 \begin{equation}
 [H_0(\bar{c}), H_{1,n}'(\bar{c})] = 0,
\end{equation}
for $n\ge1$. We can now find equations through which the first few $S_n$ are determined. We first note that
\begin{equation}
e^{S}Xe^{-S} = X+[S,X]+\frac{1}{2}[S,[S,X]]+...
\end{equation}
Using this expansion we have
\begin{eqnarray}
H&=& H_0(\bar{c})+H_1(\bar{c}) +[S,H_0(\bar{c})+H_1(\bar{c})] +\frac{1}{2} [S,[S,H_0(\bar{c})+H_1(\bar{c})]]+... \\ \nonumber &=& H_0(\bar{c})+\left(H_1(\bar{c}) +[S,H_0(\bar{c})] \right)+\left([S,H_1(\bar{c})]+\frac{1}{2} [S,[S,H_0(\bar{c})+H_1(\bar{c})]]\right)+...
\end{eqnarray}
We now let $H_0=\chi \tilde{H}_0$ and insert the expansion for $S$ to obtain
\begin{eqnarray}\label{effective1}
H&=&  H_0(\bar{c})+\left(H_1 +[S_1,\tilde{H}_0] \right)+\frac{1}{\chi}\left({[S_1,H_1]}+[S_2,\tilde{H}_0]+\frac{1}{2} [S_1,[S_1,\tilde{H}_0]]\right)+... \nonumber \\ 
\end{eqnarray}
From the above we can read off the first few terms in the expansion of $H_1'$, namely 
\begin{equation}
H'_{1,1} =H_1+[S_1,\tilde{H}_0] 
\end{equation}
and
\begin{equation}
H'_{2,1} =[S_1,H_1]+[S_2,\tilde{H}_0]+\frac{1}{2} [S_1,[S_1,\tilde{H}_0]]
\end{equation}

Our focus is now on solving for the first two terms in the expansion of $S$. $S_1$ must be found first and must be a solution to
\begin{equation}\label{s1}
0=[\tilde{H}_0,H_1+[S_1,\tilde{H}_0] ] ].
\end{equation}
Once $S_1$ is known we can solve for $S_2$ through
\begin{equation}\label{s2}
0=[\tilde{H}_0,[S_1,H_1]+[S_2,\tilde{H}_0]+\frac{1}{2} [S_1,[S_1,\tilde{H}_0]]].
\end{equation}
With $S_1$ and $S_2$ known we can substitute back into Eq. (\ref{effective1}) to obtain an effective Hamiltonian for the model. 

Our entire writing of an effective Hamiltonian relies on solutions to Equations (\ref{s1}) and (\ref{s2}). One convenient way to solve these conditions is to write $H_1$ as $H_1=\sum_n Y_n$ where the $Y_n$'s are given by 
\begin{equation}
[H_0,H_1]=\sum_{n} \epsilon_n Y_n
\end{equation} 
where $\{\epsilon_n\}$ are constants. This corresponds physically to decomposing the operator $H_1$ into pieces which change the energy of an eigenstate of $H_0$ by an amount $\epsilon_n$; formally $[H_0, Y_n] = \epsilon_nY_n$. The crux of this decomposition is then the ability to define the operators $Y_n$ in the first place. Assuming these operators can be defined we can use the expansion of $H_0$ in Eq. (\ref{s1}) to write
\begin{eqnarray}
&&0=\sum_n[\tilde{H}_0,Y_n]+[\tilde{H}_0,[S_1,\tilde{H}_0] ] \\ \nonumber &&\sum_n\frac{\epsilon_n}{\chi}Y_n=[\tilde{H}_0,[\tilde{H}_0, S_1] ].
\end{eqnarray}
The above equation can be solved by letting $S_1=\sum'_n\frac{\chi Y_n}{\epsilon_n}$ where the primed sum is over all $n$ such that $\epsilon_n\ne0$. We then use this expression for $S_1$ to write out $H_{1,1}'$ as
\begin{eqnarray}
H'_{1,1} &=&H_1 +[S_1,\tilde{H}_0] = \sum_n Y_n+\sum_n'\frac{\chi[Y_n,\tilde{H}_0] }{\epsilon_n} \\ \nonumber &=&\sum_n Y_n-\sum_n'Y_n = Y_0
\end{eqnarray} 
where we have defined the convention $\epsilon_0=0$. So to $\mathcal{O}(1)$ we have the effective Hamiltonian $H=H_0+Y_0+\mathcal{O}(1/\chi)$. As desired the leading order terms in the expansion $Y_0$ when acting by themselves do not change the energy of $H_0$. 

To get an effective Hamiltonian to $\mathcal{O}(1/\chi)$ we need to solve for the next term in the expansion of $S$. This is accomplished by solving Eq. (\ref{s2}), which is now possible with our knowledge of $S_1$. We have
\begin{eqnarray}
&&0=[\tilde{H}_0,[S_1,H_1]+[S_2,\tilde{H}_0]+\frac{1}{2} [S_1,[S_1,\tilde{H}_0]]] 
\\ \nonumber &&[\tilde{H}_0,[\tilde{H}_0,S_2]] =\frac{1}{2}[\tilde{H}_0, [S_1,[S_1,\tilde{H}_0]]] +[\tilde{H}_0,[S_1,H_1]]
\\ \nonumber &&[\tilde{H}_0,[\tilde{H}_0,S_2]] =\sum_n'\sum_m'\frac{\chi^2}{2\epsilon_n\epsilon_m}[\tilde{H}_0, [Y_n,[Y_m,\tilde{H}_0]]] +\sum_n'\sum_m\frac{\chi}{\epsilon_n}[\tilde{H}_0,[Y_n,Y_m]]
\\ \nonumber &&[\tilde{H}_0,[\tilde{H}_0,S_2]] =-\sum_n'\sum_m'\frac{\chi}{2\epsilon_n}[\tilde{H}_0, [Y_n,Y_m]] +\sum_n'\sum_m\frac{\chi}{\epsilon_n}[\tilde{H}_0,[Y_n,Y_m]]
\end{eqnarray}
Now from the Jacobi identity we know
\begin{eqnarray}\label{jacobi}
&&[\tilde{H}_0,[ {Y_m},Y_n]] + [Y_m,[Y_n,\tilde{H}_0]]+[Y_n,[\tilde{H}_0,Y_m]]=0 \\ \nonumber
&&\chi [\tilde{H}_0,[ {Y_m},Y_n]] -\epsilon_n [Y_m,Y_n]-\epsilon_m[Y_m,Y_n]=0
\\ \nonumber
&&[\tilde{H}_0,[ {Y_m},Y_n]] =\frac{(\epsilon_n +\epsilon_m)[Y_m,Y_n]}{\chi}
\end{eqnarray}
Using this in the above we find
\begin{eqnarray}
&&[\tilde{H}_0,[\tilde{H}_0,S_2]] =-\sum_n'\sum_m'\frac{(\epsilon_n +\epsilon_m)[Y_n,Y_m]}{2\epsilon_n}+\sum_n'\sum_m\frac{(\epsilon_n +\epsilon_m)[Y_n,Y_m]}{\epsilon_n}
\nonumber \\  &&[\tilde{H}_0,[\tilde{H}_0,S_2]] =\sum_n'\sum_m'\frac{(\epsilon_n +\epsilon_m)[Y_n,Y_m]}{2\epsilon_n}+\sum_n'[Y_n,Y_0]
\end{eqnarray}
Two applications of Eq. (\ref{jacobi}) gives
\begin{eqnarray}
&&[\tilde{H}_0,[\tilde{H}_0,[ {Y_m},Y_n]]] =\frac{(\epsilon_n +\epsilon_m)[\tilde{H_0},[Y_m,Y_n]]}{\chi}=\frac{(\epsilon_n +\epsilon_m)^2[Y_m,Y_n]}{\chi^2}
\end{eqnarray}
thus we see $S_2$ is given by
\begin{equation}
S_2 = \sum_{n,m\in \Lambda_{n,m} }\frac{\chi^2[Y_n,Y_m]}{2\epsilon_n(\epsilon_n +\epsilon_m)}+\sum_n'\frac{\chi^2}{\epsilon_n^2}[Y_n,Y_0]
\end{equation}
where $\Lambda_{n,m} = \{(n,m)| n\ne0, m\ne0, \epsilon_n\ne-\epsilon_m\}$. 

With the expression for $S_2$ known we can find the $\mathcal{O}(1/\chi)$ term in the effective Hamiltonian, we have
\begin{eqnarray}
H'_{2,1} &=&[S_1,H_1]+[S_2,\tilde{H}_0]+\frac{1}{2} [S_1,[S_1,\tilde{H}_0]]
\\ \nonumber &=&\sum_n'\sum_m\frac{\chi}{\epsilon_n}[Y_n,Y_m]+\sum_{n,m\in \Lambda_{n,m} }\frac{\chi^2[[Y_n,Y_m], \tilde{H}_0]}{2\epsilon_n(\epsilon_n +\epsilon_m)} \\ \nonumber &+&\sum_n'\frac{\chi^2}{\epsilon_n^2}[[Y_n,Y_0], \tilde{H}_0]+\sum_n'\sum_m'\frac{\chi^2}{2\epsilon_n\epsilon_m} [Y_n,[Y_m,\tilde{H}_0]]
\\ \nonumber &=&\sum_n'\sum'_m\frac{\chi}{2\epsilon_n}[Y_n,Y_m]-\sum_{n,m\in \Lambda_{n,m} }\frac{\chi}{2\epsilon_n}[Y_n,Y_m]
\end{eqnarray} 
With this we have an expression for an effective Hamiltonian valid to $\mathcal{O}(1/\chi^2)$, 
\begin{equation}\label{effective}
H = H_0 + Y_0 + \frac{1}{\chi}\left(\sum_n'\sum'_m\frac{\chi}{2\epsilon_n}[Y_n,Y_m]-\sum_{n,m\in \Lambda_{n,m} }\frac{\chi}{2\epsilon_n}[Y_n,Y_m]\right) + \mathcal{O}(1/\chi^2)
\end{equation}

The above general result readily simplifies to the results presented in Chapter 4 if the appropriate limits are taken. In particular, if we take $H_0=H_U$, $H_1=T$ and the $Y$ operators to be the $T_m$ with $\epsilon_m=mU_0$ then we get the strong coupling expansion of the Hubbard model. If we take $H_0=H_U+H_V$, $H_1=T+H_{SO}-H_{Z}$ and the $Y$ operators to be the $T_{m,N_1,N_2}$ with $\epsilon_{m,N_1,N_2}=mU_0+(N_1-N_2)V_0$ we obtain the strong coupling expansion of our full model Hamiltonian.  

\chapter{Calculation of Monte Carlo Local Energies}
This appendix will discuss how to evaluate the local energy expressions required in carrying out our Monte Carlo integration. Let us begin by writing
\begin{equation}
E(\rv^{2N_e}) = \frac{\langle \rv^{2N_e} | \tilde{H} |\tilde{\psi}_{Aux}\rangle}{\langle \rv^{2N_e} |\tilde{\psi}_{Aux}\rangle}
\end{equation}
where $ |\tilde{\psi}_{Aux}\rangle = P_G |\psi_{Aux}\rangle$ and $\tilde{H}$ is given in Equation (\ref{strongH}). Using the Hermiticity of $\tilde{H}$ it is far more intuitive (and useful) to think of $\tilde{H}$ acting to the left to alter the sample configuration than it is to think of it acting to the right. We are therefore interested in $\tilde{H}| \rv^{2N_e}   \rangle$. When acting on real space configurations the quartic interaction terms $H_U$ and $H_V$ give the simple results:
\begin{eqnarray}
&&H_U| \rv^{2N_e}   \rangle = \mathcal{D}U_0 | \rv^{2N_e}   \rangle, \ \ \ \ \ \  H_V | \rv^{2N_e}   \rangle = \mathcal{N}V_0| \rv^{2N_e}   \rangle
\end{eqnarray}
where $\mathcal{D}$ and $\mathcal{N}$ are, respectively, the number of double occupancies and the number of occupied nearest neighbour sites in the configuration $ \rv^{2N_e}$. Since we sample $ \rv^{2N_e}$ from a subspace of configurations with no double occupancies we will always have $\mathcal{D}=0$. 
Next we consider 
\begin{eqnarray}
&& \sum_M \tilde{T}_{0,M,M} | \rv^{2N_e}   \rangle=  \sum_{i,\delta,\sigma,M}  \sum_{S[m],S[n]=M} O_i[m] (\tilde{T}_0)_{i,\delta,\sigma}O_{i+\delta}[n]  | \rv^{2N_e}   \rangle
\end{eqnarray} 
We note that we must first consider a lattice site, $i+\delta$, occupied in the configuration $ \rv^{2N_e} $. $O_{i+\delta}[n]$ then operates on $| \rv^{2N_e}   \rangle$ and sets $M$ equal to the number of occupied nearest neighbours of the site $i$ in the configuration $\rv^{2N_e}$. We then have $ (\tilde{T}_m)_{i,\delta,\sigma}$ hop the electron at site $i+\delta$ to the site $i$. Knowing that $P_G$ will eventually act on our end result, we are only concerned with processes where the site $i$ is empty. Next  $O_i[m]$ acts, because $M$ has been fixed by  $O_{i+\delta}[n]$ the site $i$ must have the same number of nearest neighbours as $i+\delta$ to contribute. Let us define $nn_i$ as the number of nearest neighbours to site $i$. Let us also define $\eta( \rv^{2N_e},i)$ as a function that is one if $ \rv^{2N_e}$ has site $i$ occupied and zero otherwise. Then we have
\begin{eqnarray}
&& \sum_M \tilde{T}_{0,M,M} | \rv^{2N_e}   \rangle  =  \sum_{i,\delta,\sigma}  \mathcal{C}_0(i,\delta)(\tilde{T}_0)_{i,\delta,\sigma} | \rv^{2N_e}   \rangle
\end{eqnarray}
where 
\begin{equation}
\mathcal{C}_0(i,\delta)=\delta_{nn_{i+\delta},nn_i}\eta( \rv^{2N_e},i+\delta)[1-\eta( \rv^{2N_e},i)]
\end{equation}

Finally we consider 
\begin{eqnarray}
&& \sum_{m,M_1,N_1,M_2}' \frac{ \tilde{T}_{m,M_1, N_1}, \tilde{T}_{-m, M_2, M_1+M_2-N_1}}{(mU+(M_1-N_1)V)}| \rv^{2N_e}   \rangle\\  \nonumber  &=& (H'_{2,1}(m=1) +H'_{2,1}(m=-1)  +H'_{2,1}(m=0)) | \rv^{2N_e} \rangle
\end{eqnarray} 
This looks formidable but is simplified rather quickly. First, the $m=1$ terms above all give zero; $\tilde{T}_{-1, M,N}$ acting on $| \rv^{2N_e}\rangle $ where $ \rv^{2N_e} $ has no double occupancies gives zero. Next, the $m=-1$ terms are easily dealt with by remembering that $P_G$ will eventually be acting on our final configuration. For this reason $\tilde{T}_{-1,M_1, N_1} $ must remove the double occupancy $ \tilde{T}_{1, M_2, M_1+M_2-N_1}$ creates. 
Imposing conditions on nearest neighbours we obtain
\begin{eqnarray}
&& H'_{2,1}(m=-1)  | \rv^{2N_e}   \rangle =  \sum_{i,\delta, \delta',\sigma, \sigma'}\mathcal{C}_{1,-1}(i,\delta,\delta') (\tilde{T}_{-1})_{i-\delta,\delta',\sigma'}(\tilde{T}_1)_{i,\delta,\sigma} | \rv^{2N_e}   \rangle
\end{eqnarray}
where 
\begin{eqnarray}
&&\mathcal{C}_{1,-1}(i,\delta,\delta') = \frac{\delta_{nn_{i+\delta}, nn_{i-\delta'}}\eta( \rv^{2N_e},i+\delta)\eta( \rv^{2N_e},i)[1-\eta( \rv'^{2N_e},i-\delta')] }{V_0nn_{i+\delta}-V_0nn_{i}-U_0}
\end{eqnarray}
where $\rv'^{2N_e}$ is identical to $\rv$ except that the electron at $i+\delta$ has been moved to $i$. The $m=0$ contribution to the above is the most general. Following the reasoning developed so far, it is straightforward to show
\begin{eqnarray}
&& H'_{2,1}(m=0)  | \rv^{2N_e}   \rangle =  \sum_{i,i',\delta, \delta',\sigma, \sigma'}\mathcal{C}_{1,0}(i,i',\delta,\delta') (\tilde{T}_{0})_{i',\delta',\sigma'}(\tilde{T}_0)_{i,\delta,\sigma} | \rv^{2N_e}   \rangle
\end{eqnarray}
where 
\begin{eqnarray}
\mathcal{C}_{1,0}(i,i',\delta,\delta')  &=& \frac{\eta( \rv^{2N_e},i+\delta)\eta( \rv'^{2N_e},i'+\delta') [1-\eta( \rv'^{2N_e},i')][1-\eta( \rv^{2N_e},i)]}{V_0nn_{i+\delta}-V_0nn_{i}}\nonumber \\ &\times& \delta(nn_{i+\delta}+nn_{i'+\delta'}-nn_{i}-nn_{i'}) 
\end{eqnarray}
 and the above assumes $nn_{i}\ne nn_{i+\delta}$. Of course if $nn_{i}= nn_{i+\delta}$ then $\mathcal{C}_{1,0}(i,i',\delta,\delta')=0$. 

With all of these results spelled out the local energy is given as follows
\begin{eqnarray}
E(\rv^{2N_e}) &=& \mathcal{N}V_0 +\sum_{i,\delta,\sigma}  \mathcal{C}_0(i,\delta)\frac{\langle \rv^{2N_e} | (\tilde{T}_0)_{i,\delta,\sigma} |\tilde{\psi}_{Aux}\rangle}{\langle \rv^{2N_e} |\tilde{\psi}_{Aux}\rangle}+ \sum_{i,\delta, \delta',\sigma, \sigma'}\mathcal{C}_{1,-1}(i,\delta,\delta') \frac{\langle \rv^{2N_e} |  (\tilde{T}_{-1})_{i-\delta,\delta',\sigma'}(\tilde{T}_1)_{i,\delta,\sigma} |\tilde{\psi}_{Aux}\rangle}{\langle \rv^{2N_e} |\tilde{\psi}_{Aux}\rangle}  \nonumber \\ &+& \sum_{i,i',\delta, \delta',\sigma, \sigma'}\mathcal{C}_{1,0}(i,i',\delta,\delta') \frac{\langle \rv^{2N_e} |   (\tilde{T}_{0})_{i',\delta',\sigma'}(\tilde{T}_0)_{i,\delta,\sigma} |\tilde{\psi}_{Aux}\rangle}{\langle \rv^{2N_e} |\tilde{\psi}_{Aux}\rangle} + \mathcal{O}(U_0^{-2})
\end{eqnarray}
 The objects like $\langle \rv^{2N_e} | (\tilde{T}_m)_{i,\delta,\sigma} |\tilde{\psi}_{Aux}\rangle$ and $\langle \rv^{2N_e} |  (\tilde{T}_{0})_{i',\delta',\sigma'}(\tilde{T}_0)_{i,\delta,\sigma} |\tilde{\psi}_{Aux}\rangle$ correspond to one and two electron updates to $\rv^{2N_e}$ and so the ratio in each of the sums above becomes a ratio of Pfaffians, something that is easily calculated using the methods presented in Chapter 6. 
 
 From the above one can see that the majority of the calculation of the local energy is dedicated to imposing constraints on a given classical configuration $\rv^{2N_e}$. If these constraints are satisfied one can then use the methods of Chapter 6 to calculate the overlaps that contribute to $E(\rv^{2N_e})$.

\backmatter
\thispagestyle{plain} 
\bibliographystyle{apsrev}
\bibliography{topoSC}


\end{doublespace}
\end{document}